\documentclass[
reprint,
superscriptaddress,
amsmath,amssymb,
aps,
prx,
longbibliography,
]{revtex4-2}

\usepackage{graphicx}
\usepackage{dcolumn}
\usepackage{bm}

\usepackage{tcolorbox}
\usepackage{booktabs,tabularx,array}
\usepackage[colorlinks=true,linkcolor=blue,citecolor=blue,urlcolor=blue]{hyperref}

\begin{document}

\preprint{APS/123-QED}

\title{\textbf{Near-Wall Pathways of Anomalous Electron Transport in Hall Thrusters Revealed by 3D PIC Simulations} 
}%

\author{Zhe Liu}
\author{Zhongping Zhao}%

\author{Yinjian Zhao}
 \email{Contact author: zhaoyinjian@hit.edu.cn}
 
\affiliation{School of Energy Science and Engineering, Harbin Institute of Technology, Harbin 150001, People’s Republic of China}

\date{\today}

\begin{abstract}
Cross-field electron transport in Hall thrusters is widely attributed to high-frequency $E\times B$ instabilities, yet its net spatial pathway remains poorly resolved.
Here we perform instability-resolving three-dimensional particle-in-cell simulations of a Hall thruster using a boundary-faithful and highly integrated framework.
The model incorporates a realistic magnetic-field configuration, self-consistent dielectric wall charging, secondary electron emission, Monte Carlo ionization collisions, a self-consistent continuum neutral-gas evolution model, and an open near-plume outflow treatment.
From the strongly oscillatory three-dimensional fields, we extract the net instability-driven transport by time and azimuthal averaging of the correlation term $\langle n_e E_y\rangle$ and the corresponding effective perpendicular mobility.
The simulations reveal that anomalous electron transport is not distributed uniformly across the channel cross section.
Instead, it self-organizes into persistent near-wall pathways connected to the near-exit region.
By comparing conducting-wall, ceramic-wall-with-secondary-emission, and open-outflow closures, we show that the near-wall transport topology is robust, while the boundary treatment mainly redistributes the detailed strength of the pathway and its coupling to the exit and near-plume region.
These results demonstrate a previously unresolved spatial organization of instability-driven anomalous transport in Hall thrusters and highlight the unique role of 3D PIC simulations in revealing it.
\end{abstract}

\maketitle


\section{Introduction}

Hall effect thrusters (HETs) are among the most widely used electric propulsion devices\cite{Rafalskyi2021IodineEP,Levchenko2018SmartNanomaterialsEP, Ahedo2011PlasmasForSpacePropulsion, Mazouffre2016ElectricPropulsion, Lev2019ExpansionEP} for spacecraft station keeping, orbit raising, and deep-space missions because they provide high specific impulse at modest system complexity.
In a typical HET discharge, an axial electric field accelerates ions downstream\cite{DelaviereDelion2025DualTimeScale, Jorns2023FoundationsIonSources}, while electrons emitted from an external cathode must cross the magnetic field to sustain ionization, maintain current closure, and neutralize the exhaust plume
\cite{Boeuf2020RotatingSpokes, Andrenucci2003ScalingLaws, Zhurin1999ClosedDriftThrusters, GoebelKatz2008HallThrusters}.
The resulting cross-field electron transport therefore plays a central role in determining the discharge current, the location of the ionization and acceleration regions, the electron energy budget, plume properties, and ultimately the performance and efficiency of the thruster \cite{goebel_katz}.

A longstanding difficulty is that classical collisional transport, based on electron--neutral scattering alone, generally underpredicts the level of cross-field electron mobility inferred from experiments and required by predictive models \cite{lafleur_2016_pop1,lafleur_2016_pop2,janes_lowder_1966}.
This discrepancy has motivated extensive work on instability-driven anomalous transport\cite{Coche2014AzimuthalAxialPIC, Charoy2021ITTEDIInteraction, Janhunen2018ECDI2D, Petronio2021MTSIExB, Sengupta2020ModeTransitionsEDI}, especially in relation to the electron drift instability (EDI) and related $E\times B$ modes that arise naturally in strongly magnetized crossed-field plasmas \cite{forslund_1970_prl,forslund_1971_prl,boeuf_2013_prl,tsikata_2015_prl,koshkarov_2019_prl, Croes2017EDI2DPIC, Taccogna2019NumericalStudiesEDI, Boeuf2014RotatingStructures, Janhunen2018NonlinearStructures, Katz2018NearPlumePIC, Asadi2019ECDIHallThruster, Charoy2020ComparisonKineticPIC}.
Recent high-level studies have significantly deepened the field by clarifying the growth and saturation of the EDI, its nonlinear spectral transfer, and the direct experimental inference of anomalous diffusion profiles \cite{brown_jorns_2023_prl, roberts_jorns_2024_prl}.
However, an important question remains insufficiently resolved: beyond identifying the instability and estimating an effective transport coefficient, where in the discharge does the net instability-driven anomalous electron transport actually flow?
In other words, the spatial topology of the transport pathway itself remains far less established than the existence of the instability that drives it.

This question is intrinsically spatial and inherently three-dimensional.
The EDI develops through azimuthal dynamics, yet it is simultaneously coupled to axial ionization and acceleration, as well as to radial wall and sheath interaction\cite{mcdonald_gallimore_2011,ellison_raitses_fisch_2012, Sheehan2013KineticTheory}.
As a result, the transport cannot, in general, be inferred reliably from axisymmetric or reduced descriptions that suppress the azimuthal degree of freedom or average it out by construction \cite{villafana_2023_pop,ducrocq_2018_pop,Chen_POP,escobar_ahedo_2014,escobar_ahedo_2015}.
Moreover, HETs are wall-bounded plasmas.
Wall charging, sheath formation, and secondary electron emission (SEE) modify the near-wall electric field, the electron energy distribution, and the local instability environment, so they are not merely implementation details of a simulation boundary but part of the transport physics itself \cite{kaganovich_2012_prl, chabert_2025_prl, Chen2025ThreeDimensionalHelical}.
At the same time, the discharge plume expands into an effectively unbounded vacuum, so any finite simulation domain must also treat the outflow in a way that does not artificially clamp the far-field potential or introduce a nonphysical current sink \cite{andrews_2022_open_boundary}.
For these reasons, a credible answer to the spatial organization of anomalous transport requires not only a fully kinetic and instability-resolving model, but also a boundary treatment that does not predetermine the result.

\begin{figure*}[htbp]
    \centering
    \includegraphics[width=0.8\linewidth]{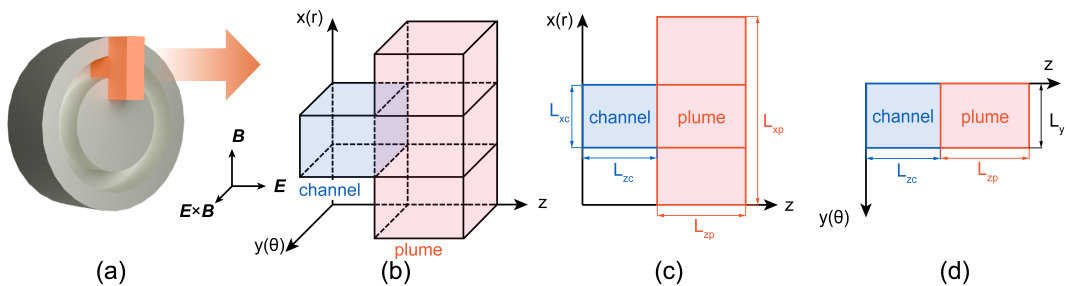}
    \caption{
    Physical configuration and computational domain.
    (a) Schematic of the annular Hall thruster and the modeled azimuthal sector.
    (b) Three-dimensional decomposition of the simulation domain into the channel region (blue) and plume region (red) in the $(x,y,z)\equiv(r,\theta,z)$ coordinate system, with the nominal $\mathbf{E}$, $\mathbf{B}$, and $\mathbf{E}\!\times\!\mathbf{B}$ directions indicated.
    (c) Meridional $(x,z)$ view defining the channel size $(L_{xc},L_{zc})$ and plume extension $(L_{xp},L_{zp})$.
    (d) Azimuthal $(y,z)$ view showing the simulated sector length $L_y$ with periodic boundary conditions in $y$.
    }
    \label{fig:simulation domain}
\end{figure*}

Particle-in-cell (PIC) simulation is the natural framework for this problem because it can evolve the plasma self-consistently without prescribing an empirical anomalous mobility and can resolve the non-Maxwellian kinetics, sheath physics, and high-frequency field fluctuations associated with EDI-driven transport \cite{boeuf_tutorial,lafleur_2018_comparison,garrigues_2018_pop}.
Yet obtaining physically meaningful transport information from three-dimensional PIC simulations of Hall thrusters remains exceptionally demanding.
The simulation must simultaneously capture the magnetic topology, wall response, SEE, ionization dynamics, near-plume expansion, and long-time nonlinear instability development, while also maintaining sufficient particle statistics to extract averaged transport diagnostics from strongly oscillatory fields.
As emphasized in recent methodological discussions, credible 3D Hall-EDI campaigns require not only raw computational power but also careful choices of initialization, boundary closure, collision and neutral modeling, averaging strategy, and convergence assessment\cite{Xie_2024,Xie_2025,Chen_POP,IEPC-2025-063}.
Consequently, there remains a strong need for a transport-resolving, convergence-aware 3D PIC study that is both physically revealing and practically instructive.

The present work addresses this need.
Its primary physical result is that the net instability-driven anomalous electron transport in a Hall thruster is not distributed uniformly across the channel cross section.
Instead, after time and azimuthal averaging of the fully three-dimensional oscillatory fields, the transport emerges as persistent near-wall pathways connected to the near-exit region.
This spatial transport topology is extracted from a 3D PIC framework that incorporates a realistic magnetic-field configuration, self-consistent dielectric wall charging, SEE, Monte Carlo ionization collisions, self-consistent continuum neutral-gas evolution, and an open outflow treatment for the near plume.
By comparing conducting-wall, ceramic-wall-with-SEE, and open-outflow closures, we show that the near-wall transport topology is robust, while the boundary treatment mainly redistributes the detailed strength of the pathway and its coupling to the exit and near-plume region.
In addition to this physics result, the paper also serves as a detailed reference for long-time, transport-resolving 3D PIC Hall-thruster simulations, documenting the practical requirements and numerical sensitivities needed to make such transport claims credible.

The remainder of the paper is organized as follows.
Section~\ref{sec:sim_model_setup} presents the simulation model and numerical setup, including the dielectric-wall, secondary-electron-emission (SEE), and open-outflow treatments, together with the computational geometry, simulation cases, initialization strategy, applied magnetic field, and neutral-gas modeling.
Section~\ref{sec:results} presents the simulation results, beginning with the slow global evolution and time-averaged discharge structure, and then examining the near-wall pathways of anomalous electron transport, the instantaneous three-dimensional EDI structures underlying them, and their spectral characteristics.
Section~\ref{sec:numerical} assesses the numerical effects associated with the timestep, grid resolution, and plume-domain size.
Finally, Section~\ref{sec:conclusion} discusses the physical implications of these findings and summarizes the main conclusions of the present work for future three-dimensional PIC studies of anomalous transport in Hall thrusters.

\begin{figure*}[htbp]
    \centering
    \includegraphics[width=0.8\linewidth]{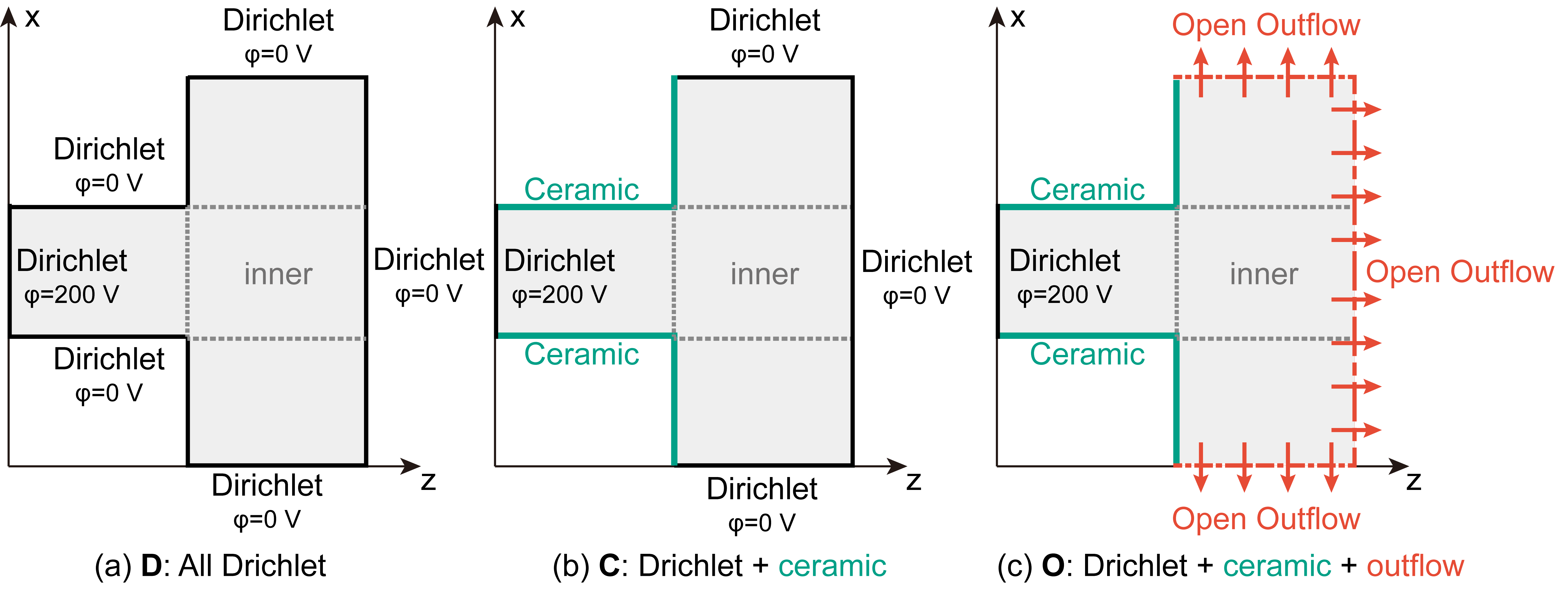}
\caption{
Summary of the simulation cases and their computational cost.
Here, $\Delta t$ is the PIC time step, $t_{\mathrm{run}}$ is the total simulated physical time,
$t_{\mathrm{wall}}$ is the measured wall-clock runtime on the target machine,
$\bar{t}_{\mathrm{step}}$ is the average wall-clock time per time step,
and $N_{i,\mathrm{peak}}$ is the peak number of macro-ions.
}
    \label{fig:fig_BC}
\end{figure*}

\section{Simulation Model and Numerical Setup}
\label{sec:sim_model_setup}

\subsection{Physical Configuration and Computational Domain}
\label{subsec:physical_configuration_domain}

Fig.~\ref{fig:simulation domain} summarizes the physical configuration and computational domain adopted in this work.
A reduced azimuthal sector of an annular Hall thruster is modeled in a three-dimensional Cartesian coordinate system $(x,y,z)$,
which corresponds locally to the radial, azimuthal, and axial directions $(r,\theta,z)$, respectively.
With this local slab approximation, the azimuthal curvature of the full annulus is neglected,
as in many previous Hall-thruster PIC studies \cite{Charoy2019,Villafana2021,villafana_2023_pop,Chen_POP}.
This approximation retains the essential crossed-field $\mathbf{E}\!\times\!\mathbf{B}$ dynamics
while making long-time, full-3D kinetic simulations computationally tractable.

The computational domain is decomposed into two connected regions:
an acceleration-channel region and a downstream plume region.
As illustrated in Fig.~\ref{fig:simulation domain}(b) and (c),
the channel occupies the axial interval $z\le z_{\mathrm{exit}}$
and the annular radial span $x\in[x_{\min},x_{\max}]$,
with channel width $L_{xc}=x_{\max}-x_{\min}$ and channel length $L_{zc}=z_{\mathrm{exit}}$.
Downstream of the exit plane, the plume region extends to
$z_{\max}=z_{\mathrm{exit}}+L_{zp}$ in the axial direction
and to the full radial range $x\in[0,L_{xp}]$
to accommodate near-field expansion of the discharge plume.
Accordingly, the total computational box has dimensions
$L_x\times L_y\times L_z$,
where $L_x=L_{xp}$ and $L_z=z_{\max}$.

The azimuthal direction is represented by a finite sector of length $L_y$ with periodic boundary conditions,
as shown in Fig.~\ref{fig:simulation domain}(d).
This treatment provides a computationally efficient approximation to the full annulus
while preserving the dominant azimuthal instability dynamics.
The nominal directions of the electric field, magnetic field, and electron $\mathbf{E}\!\times\!\mathbf{B}$ drift
are also indicated in Fig.~\ref{fig:simulation domain}(b).
Throughout this work, positive $z$ denotes the downstream direction,
$x$ denotes the radial direction across the annular gap,
and $y$ denotes the azimuthal direction.

It should be emphasized that $x=0$ in the present local Cartesian representation
does not correspond to the thruster axis.
Instead, the simulated domain represents a radial segment extracted from a finite-radius annular device.
The actual geometric radius used to construct the applied magnetic field will be introduced in
Sec.~\ref{subsec:background_magnetic_field}.
This distinction is important when interpreting the radial extent of the plume region and the exported magnetic-field configuration.

The baseline computational domain used in this study is
$L_x\times L_y\times L_z = 25.6\times 6.4\times 25.6~\mathrm{mm}$.
It is discretized by a uniform Cartesian mesh with
$N_x\times N_y\times N_z = 256\times 64\times 256$,
corresponding to
$\Delta x=\Delta y=\Delta z=0.1~\mathrm{mm}$.
Modified domain sizes and mesh resolutions considered for sensitivity studies
will be summarized together with the simulation cases in the next subsection.

\begin{table}[t]
\centering
\caption{
Summary of simulation cases, computational cost, and computing resources.
$\Delta t$ is the PIC timestep; $t_{\mathrm{run}}$ is the total simulated time; $t_{\mathrm{wall}}$ is the measured wall-clock time; $\bar{t}_{\mathrm{step}}$ is the time per step; and $N_{i,\mathrm{peak}}$ is the peak number of macro-ions.
Cases~D, $D_{2\Delta t}$, and $O_{\mathrm{loadC}}$ used 192 MPI ranks on AMD EPYC~9965; Case~C used 128 MPI ranks on EPYC~9965; Cases~O and $O_{\mathrm{bigger}}$ used 128 MPI ranks on EPYC~9754; and Case~$D_{\mathrm{finer}}$ used 512 MPI ranks on four EPYC~9754 processors.
}
\label{tab:case_summary}
\begin{tabularx}{\linewidth}{@{}l r c c c c@{}}
\toprule
Case & Meaning(BCs) & $t_{\mathrm{run}}$ ($\mu$s) & $t_{\mathrm{wall}}$(d) & $\overline{t}_{\mathrm{step}}$(s) & $N_{i,\mathrm{peak}}$ \\
\midrule
D &
All Dirichlet BCs &
40.22 &
30.94 &
0.3323&
1.51E8 \\
D$_{2\Delta t}$ &
Doubled $\Delta t$ &
48.33 &
14.76 &
0.2639&
7.87E7 \\
D$_{\text{finer}}$ &
Halved cell size &
2.48 &
41.82 &
3.6462&
6.15E8 \\
C &
Adding ceramic BCs &
41.36 &
33.25 &
0.3473&
1.16E8 \\
O &
Adding outflow BCs &
28.52 &
38.61 &
0.5850&
1.32E8 \\
O$_{\text{loadC}}$ &
Load C to start &
22.32 &
17.38 &
0.3363&
9.33E7\\
O$_{\text{bigger}}$ &
Bigger plume region &
7.75 &
30.08 &
1.6756&
2.98E8 \\
\bottomrule
\end{tabularx}
\end{table}

\subsection{Boundary Treatments and Simulation Cases}
\label{subsec:boundary_treatments_cases}

To isolate the physical impact of boundary modeling on the 3D PIC solution,
we organize the simulations as a hierarchy of cases in which the boundary treatment is progressively improved from a commonly used baseline.
The corresponding boundary-condition configurations are summarized in Fig.~\ref{fig:fig_BC},
and the considered cases together with their computational cost are listed in Tab.~\ref{tab:case_summary}.

\subsubsection{Case hierarchy and purpose}
\label{subsubsec:case_hierarchy}

We first consider a baseline case, denoted Case D,
in which Dirichlet electrostatic boundaries are applied on the non-periodic surfaces.
Specifically, except for the anode surface where the fixed anode potential is imposed,
all other outer boundaries are set to a fixed zero potential,
and particles reaching these boundaries are absorbed.
This boundary treatment is the same as that adopted in our previous works \cite{Xie_2024,Xie_2025,Chen_POP}
and is also common in many earlier Hall-thruster PIC simulations \cite{Villafana2021,villafana_2023_pop}.
Case D therefore serves as the reference for assessing how more realistic wall and outflow treatments modify the discharge evolution and instability characteristics.

Based on Case D, two additional cases are introduced to examine numerical sensitivity.
Case D$_{2\Delta t}$ uses the same geometry and boundary conditions as Case D but doubles the PIC timestep,
in order to assess timestep effects on the EDI dynamics.
Case D$_{\text{finer}}$ uses the same physical domain but halves the grid spacing in all directions and correspondingly halves the timestep.
This refined case is intended to probe grid-resolution sensitivity during the early stage of the discharge,
although its computational cost prevents a long-time run.
The corresponding magnetic-field export and mesh will be introduced later in Sec.~\ref{subsec:background_magnetic_field}.

A more realistic wall treatment is then introduced in Case C,
where the discharge-channel walls are modeled as dielectric ceramic boundaries rather than fixed-potential conducting proxies,
as illustrated in Fig.~\ref{fig:fig_BC}(b).
The thruster exit surface is also treated as ceramic.
This case is designed to examine how self-consistent wall charging and secondary electron emission modify the near-wall plasma state and the resulting instability activity.

Finally, because the simulated plume covers only a near-field region and the potential on the truncated plume boundary has generally not relaxed to zero,
an open-outflow treatment is introduced in Case O,
as shown in Fig.~\ref{fig:fig_BC}(c).
Case O therefore combines dielectric channel walls with an open electrostatic and particle outflow treatment in the plume region.
To reduce the cost of reaching a late-time quasi-steady state with this more expensive configuration,
an additional Case O$_{\text{loadC}}$ is initialized by loading the solution from Case C at a later time,
so that cases O and O$_{\text{loadC}}$ can be compared to assess consistency.
Moreover, because the open-boundary approximation may be less effective for a near-plume truncation than in a magnetic-nozzle configuration where the method was originally proposed \cite{andrews_2022_open_boundary},
another case with a larger downstream plume region is considered and labeled O$_{\text{bigger}}$.
Its magnetic-field export and extended domain will also be introduced in Sec.~\ref{subsec:background_magnetic_field}.

\subsubsection{Main boundary-treatment framework}
\label{subsubsec:boundary_framework}

The boundary treatments introduced above affect both the electrostatic field solve and the particle dynamics.
In the main text, we retain only the physical framework and the key governing relations,
while the discrete implementation details are deferred to Apx.~\ref{apx:dielectric_bc_discretization},
Apx.~\ref{apx:see_sampling},
and Apx.~\ref{apx:open_bc_discretization}.

\paragraph{Dielectric channel walls.}
The discharge-channel walls are modeled as dielectric (ceramic) insulators that accumulate free surface charge, consistent with the long-recognized importance of ceramic-wall plasma interaction, secondary electron emission, and near-wall transport effects in Hall thrusters \cite{ahedo_2003_wall,barral_2003_wall}.
In contrast to a conducting-wall proxy, the wall response is not imposed by prescribing the potential.
Instead, it enters the electrostatic problem through the normal electric field at the plasma-facing wall,
\begin{equation}
E_n \;=\; -\left.\frac{\partial \phi}{\partial n}\right|_{\mathrm{wall}},
\label{eq:dielectric_neumann_main}
\end{equation}
where the boundary field is obtained self-consistently from the evolving surface charge density $\sigma(\mathbf{x}_w,t)$.
For a wall surface element $f$ with area $A_f$,
the surface charge is advanced from the net collected particle charge as
\begin{equation}
\sigma_f^{\,n+1}
=
\sigma_f^{\,n}
+
\frac{1}{A_f}\sum_{p\in \mathcal{H}_f} q_p\,w_p ,
\label{eq:sigma_update_main}
\end{equation}
where $\mathcal{H}_f$ is the set of macroparticles interacting with the surface element during the time step, $q_p$ denotes the particle charge,
and $w_p$ denotes the macroparticle weight.
The boundary normal field then follows from Gauss's law,
\begin{equation}
E_{bc}^{\min}=\frac{\sigma^{\min}}{\epsilon_0},
\qquad
E_{bc}^{\max}=-\frac{\sigma^{\max}}{\epsilon_0},
\label{eq:en_from_sigma_main}
\end{equation}
which closes the field problem while allowing the wall charging state to evolve self-consistently with the plasma.

\paragraph{Secondary electron emission.}
Secondary electron emission (SEE) is included as a wall-mediated electron source coupled to the dielectric charging process.
In the present model, SEE is applied only to incident electrons.
When an electron macroparticle strikes a wall, its incident kinetic energy
\begin{equation}
E_{\mathrm{inc}}=\frac{1}{2}m_e\left|\mathbf{v}_{\mathrm{inc}}\right|^2
\end{equation}
is used to determine the emission yield through a linear ramp with saturation
\cite{Tavant_2018},
\begin{equation}
\sigma(E_{\mathrm{inc}})
=
\min\!\left(
\sigma_{\max},
\;
\sigma_0+\left(1-\sigma_0\right)\frac{E_{\mathrm{inc}}}{E_{\mathrm{SEE}}}
\right).
\label{eq:see_yield_linear_main}
\end{equation}
The continuous yield is converted into an integer number of emitted secondary macroparticles using stochastic sampling,
and the associated net charge transferred to the wall is
\begin{equation}
q_{\mathrm{wall}} = (1-N)\,q_e\,w_p ,
\label{eq:see_net_charge_main}
\end{equation}
where $N$ is the sampled number of emitted secondaries.
In this way, SEE modifies the wall-current balance and feeds back directly into the dielectric charging through Eq.~\eqref{eq:sigma_update_main}.
The emitted secondaries are launched into the plasma half-space with a low-energy distribution characterized by an effective temperature $T_{\mathrm{SEE}}$.
The detailed stochastic-emission and velocity-sampling procedure is described in Apx.~\ref{apx:see_sampling}.

\paragraph{Open plume outflow.}
Because the physical plume is unbounded whereas the numerical domain must be truncated,
imposing a fixed Dirichlet condition on the downstream and radial plume boundaries can introduce artificial field distortion and a strong dependence on the truncation location.
To better approximate the influence of the exterior region,
we employ an open electrostatic boundary of Robin type following a far-field monopole approximation \cite{andrews_2022_open_boundary},
\begin{equation}
\left.\frac{\partial \phi}{\partial n}\right|_{\mathbf{x}_b}
+\kappa_b\big(\phi(\mathbf{x}_b)-\phi_\infty\big)\approx 0,
\qquad
\kappa_b=\frac{\hat{\mathbf{n}}_b\cdot\mathbf{r}_b}{\mathbf{r}_b\cdot\mathbf{r}_b},
\label{eq:open_robin_cont_main}
\end{equation}
where $\mathbf{r}_b=\mathbf{x}_b-\mathbf{x}_0$ and $\phi_\infty$ is the far-field reference potential.
This boundary reduces the clamping effect associated with a fixed-potential truncation while remaining compatible with the cell-centered Poisson solve.

In addition to the field boundary condition, a particle outflow treatment is required.
Heavy species reaching the open plume faces are absorbed and removed from the simulation.
For electrons, however, we apply an energy-selective transmission rule so that electrostatically confined low-energy electrons are not artificially drained through the domain truncation.
Electrons reaching an open face are transmitted only if their total kinetic energy satisfies
\begin{equation}
\frac{1}{2}m_e |\mathbf{v}|^2 \;\ge\; e\,\max\!\left(\phi_P-\phi_\infty,\,0\right),
\label{eq:electron_outflow_energy_criterion_main}
\end{equation}
where $\phi_P$ is the interior-cell potential adjacent to the boundary.
Otherwise, the electron is specularly reflected at the open face.
This treatment mitigates artificial electron loss at the truncation and reduces the sensitivity of the near-plume solution to the boundary placement.
The discrete Robin implementation and its incorporation into the Poisson stencil are given in Apx.~\ref{apx:open_bc_discretization}.

\subsection{Baseline Numerical and Physical Parameters}
\label{subsec:baseline_parameters}

The baseline numerical and physical parameters adopted in this study are summarized in Tabs.~\ref{tab:numerical_para} and \ref{tab:physical_para}.
Unless otherwise stated, these values correspond to the baseline configuration used in Case D and inherited by cases C and O.
Case-specific changes in timestep, mesh resolution, and plume-domain extent have been summarized previously in Tab.~\ref{tab:case_summary}.

For the baseline setup, the computational domain introduced in Sec.~\ref{subsec:physical_configuration_domain}
is discretized by a uniform Cartesian mesh with
$N_x\times N_y\times N_z = 256\times 64\times 256$,
corresponding to
$\Delta x=\Delta y=\Delta z=0.1~\mathrm{mm}$.
Time integration is performed with a fixed timestep
$\Delta t=5~\mathrm{ps}$,
and the electrostatic potential is obtained from the Poisson solver with a convergence tolerance of
$\epsilon=10^{-6}$.
The baseline macro-particle weight is
$w_p\approx 8597$.
In the azimuthal direction, periodic boundary conditions are imposed.

Several auxiliary cases adopt modified numerical parameters.
Case D$_{2\Delta t}$ uses a doubled timestep,
$\Delta t=10~\mathrm{ps}$,
to examine timestep sensitivity.
Case D$_{\text{finer}}$ retains the same physical domain as Case D
but uses
$N_x\times N_y\times N_z=512\times 128\times 512$,
with
$\Delta x=\Delta y=\Delta z=0.05~\mathrm{mm}$,
$\Delta t=2.5~\mathrm{ps}$,
and a reduced macro-particle weight
$w_p\approx 1064$.
Case O$_{\text{bigger}}$ keeps the baseline cell size of
$0.1~\mathrm{mm}$
but enlarges the computational box to
$N_x\times N_y\times N_z=384\times 64\times 384$
to assess sensitivity to the downstream plume extent.

The default physical and model inputs are listed in Tab.~\ref{tab:physical_para}.
The plasma consists of electrons and singly charged xenon ions,
with ion-to-electron mass ratio
$m_i/m_e=2.3933\times10^{5}$.
The anode potential is fixed at
$\phi_a=200~\mathrm{V}$,
and when the open-outflow treatment is enabled,
the far-field reference potential is set to
$\phi_\infty=0~\mathrm{V}$.
The reference point used in the Robin outflow model is
$\mathbf{r}_{0,\mathrm{out}}=(12.85,\,3.25,\,7.30)~\mathrm{mm}$.

Electrons are injected through the cathode/injection model using a fixed current
$I_{\mathrm{inj}}=0.177~\mathrm{A}$.
The injected cathode electrons are prescribed with a thermal speed
$v_{e,t}=2297~\mathrm{km/s}$
and a drift speed
$u_{e,x}=-2297~\mathrm{km/s}$.
The injection plane is offset by 10 cells from the $x$-maximum boundary,
and the injection location is randomly sampled in the $y$--$z$ plane for each injected macroparticle.

Collisions are treated using the MCC module.
In the present work, electron-impact ionization of xenon is included,
with ionization threshold
$E_{\mathrm{ion}}=12.14~\mathrm{eV}$.
When dielectric walls are enabled,
secondary electron emission is characterized by the parameter set
$(\sigma_{\max},\sigma_0,E_{\mathrm{SEE}},T_{\mathrm{SEE}})
=
(2.9,\,0.5,\,50~\mathrm{eV},\,2~\mathrm{eV})$.
These quantities provide the default model inputs for the wall-emission treatment introduced in Sec.~\ref{subsec:boundary_treatments_cases}.

\begin{table}[htbp]
\centering
\caption{Baseline numerical setup for the reference simulation configuration.}
\label{tab:numerical_para}
\small
\setlength{\tabcolsep}{6pt}
\begin{tabularx}{\linewidth}{@{}>{\raggedright\arraybackslash}X >{$}l<{$} l@{}}
\toprule
\textbf{Parameter} & \textbf{Symbol} & \textbf{Value} \\
\midrule

\multicolumn{3}{@{}l}{\textbf{Grid}}\\
Grid size (cells) & N_x\times N_y\times N_z & \mbox{$256\!\times\!64\!\times\!256$} \\
Domain size & L_x\times L_y\times L_z & \mbox{$25.6\!\times\!6.4\!\times\!25.6~\mathrm{mm}$} \\
Grid spacing & \Delta x=\Delta y=\Delta z & \mbox{$0.1~\mathrm{mm}$} \\

\midrule
\multicolumn{3}{@{}l}{\textbf{Time integration and solver}}\\
Time step & \Delta t & \mbox{$5~\mathrm{ps}$} \\
Poisson solver tolerance & \epsilon & \mbox{$10^{-6}$} \\
Macro-particle weight & w_p & \mbox{$8597$} \\

\bottomrule
\end{tabularx}
\end{table}

\begin{table}[htbp]
\centering
\caption{Baseline physical and model parameters for the reference simulation configuration.}
\label{tab:physical_para}
\begin{tabular}{p{0.56\linewidth} p{0.12\linewidth} p{0.26\linewidth}}
\toprule
\textbf{Parameter} & \textbf{Symbol} & \textbf{Value} \\
\midrule
\multicolumn{3}{l}{\textbf{Plasma species}}\\
Species & -- & electrons + Xe$^{+}$ \\
Ion-to-electron mass ratio & $m_i/m_e$ & $2.3933\times10^{5}$ \\

\midrule
\multicolumn{3}{l}{\textbf{Electrostatic boundary parameters}}\\
Anode potential & $\phi_a$ & $200~\mathrm{V}$ \\
Far-field potential & $\phi_\infty$ & $0~\mathrm{V}$ \\
Outflow reference point (mm) & $\mathbf{r}_{0,\mathrm{out}}$ & $(12.85,\,3.25,\,7.30)$ \\

\midrule
\multicolumn{3}{l}{\textbf{Cathode / injection model}}\\
Cathode current & $I_{\mathrm{inj}}$ & $0.177~\mathrm{A}$ \\
Cathode electron thermal speed & $v_{e,t}$ & $2297~\mathrm{km/s}$ \\
Cathode electron drift speed (x) & $u_{e,x}$ & $-2297~\mathrm{km/s}$ \\

\midrule
\multicolumn{3}{l}{\textbf{Collisions and neutrals}}\\
Electron-neutral process & -- & ionization \\
Ionization threshold & $E_{\mathrm{ion}}$ & $12.14~\mathrm{eV}$ \\

\midrule
\multicolumn{3}{l}{\textbf{Secondary electron emission}}\\
Maximum yield & $\sigma_{\max}$ & $2.9$ \\
Baseline parameter & $\sigma_{0}$ & $0.5$ \\
Characteristic energy & $E_{\mathrm{SEE}}$ & $50~\mathrm{eV}$ \\
Secondary electron temperature & $T_{\mathrm{SEE}}$ & $2~\mathrm{eV}$ \\
\bottomrule
\end{tabular}
\end{table}

\subsection{Initial Plasma Distribution and Particle Loading}
\label{subsec:init_plasma}


To reduce the long transient associated with a spatially uniform plasma initialization, as noted in previous studies \cite{Xie_2024,Xie_2025,Chen_POP}, and to provide a stable and repeatable starting point for discharge development, we prescribe a quasineutral seed plasma at $t=0$ in the channel and near-field plume regions. The initial electron and singly ionized xenon ($\mathrm{Xe}^{+}$) number densities are set equal, $n_e(x,y,z,0)=n_i(x,y,z,0)\equiv n_{\mathrm{seed}}(x,y,z)$.

The seed plasma is assigned a peak number density of $n_{\mathrm{pk}}=7.5\times10^{17}\ \mathrm{m^{-3}}$ and is constructed as a separable analytic profile in the radial, azimuthal, and axial directions,
\begin{align}
n_{\mathrm{seed}}(x,y,z)
&= n_{\mathrm{pk}}\,F_x(x)\,F_y(y)\,F_z(z)\,M(x,z),
\label{eq:nseed}
\\
M(x,z)
&=
\begin{cases}
H(x-x_{\mathrm{i}})\,H(x_{\mathrm{o}}-x), & 0\le z<z_{\mathrm{ex}},\\
1, & z\ge z_{\mathrm{ex}},
\end{cases}
\label{eq:mask}
\end{align}
where $H(\cdot)$ is the Heaviside step function,
$z_{\mathrm{ex}}=7.2~\mathrm{mm}$ denotes the channel exit plane,
and the annular channel gap is bounded by
$[x_{\mathrm{i}},x_{\mathrm{o}}]=[6.4,\,19.2]~\mathrm{mm}$.
Thus, for $z<z_{\mathrm{ex}}$ the seed plasma is loaded only inside the annular channel,
whereas for $z\ge z_{\mathrm{ex}}$ it extends across the full radial span
to provide a continuous plasma background in the near-field plume.

To ensure smoothness and strict positivity,
the radial and axial shape functions include a finite floor parameter $\eta\in(0,1)$,
which may be interpreted as a background-to-peak density ratio.
In addition, a weak azimuthal modulation is imposed to break perfect symmetry in a controlled manner.
The shape functions are defined as
\begin{align}
F_x(x) &= \eta + (1-\eta)\left[\sin\!\left(\pi\frac{x-x_{\mathrm{i}}}{x_{\mathrm{o}}-x_{\mathrm{i}}}\right)\right]_+,
\label{eq:Fr}\\
F_y(y) &= (1-\delta)+\delta\,\sin\!\left(\frac{2\pi m\,y}{L_y}\right),
\label{eq:Fy}\\
F_z(z) &= \eta + (1-\eta)\exp\!\left[-\left(\frac{z-z_0}{\sigma_z}\right)^2\right],
\label{eq:Fz}
\end{align}
where $[\cdot]_+\equiv\max(\cdot,0)$,
$L_y$ is the periodic azimuthal length,
and $(z_0,\sigma_z)=(6.0~\mathrm{mm},\,3.0~\mathrm{mm})$ define the center and characteristic width of the axial envelope.
In this work, $\eta=0.2$, $\delta=0.1$, and $m=7$ are used.
With these choices,
$F_x,F_z\in[\eta,1]$ and $F_y\in[1-2\delta,\,1]$ remain strictly positive.
The resulting azimuthal peak-to-valley variation is $2\delta$ (about $20\%$),
which is sufficient to seed symmetry breaking without externally imposing the subsequent nonlinear dynamics.
Because the simulation is performed in an open domain with continuous particle loss and replenishment through transport, injection, and ionization, this weak initial azimuthal modulation does not constrain the plasma to oscillate persistently in the imposed seed mode; instead, the initially prescribed pattern is rapidly refreshed, and the subsequent oscillations are established self-consistently by the physical instability dynamics of the system.

The analytic profile above is then realized in the PIC solver by Monte Carlo particle loading.
For each species, the total number of real particles represented by the initial seed is obtained from the discrete volume integral of $n_{\mathrm{seed}}$ over the mesh.
With a cell volume of $\Delta V=\Delta x\,\Delta y\,\Delta z=(0.1~\mathrm{mm})^3=10^{-12}\ \mathrm{m^3}$, the corresponding total number of real particles per species is $N_0=4.298\times10^{11}$.
The seed plasma is represented by $N_{\mathrm{mp}}=5.0\times10^{7}$ initialized macroparticles per species, corresponding to a macroparticle weight of $w_p=N_0/N_{\mathrm{mp}}=8.597\times10^{3}$.
At the peak density,
the number of real particles per cell is
$n_{\mathrm{pk}}\Delta V=7.5\times 10^{5}$,
corresponding to
$n_{\mathrm{pk}}\Delta V/w_p \approx 87$
macroparticles per cell in order of magnitude.
This keeps the initial particle noise at a controllable level while avoiding the much larger computational cost that would result from initializing the entire domain with a uniformly large number of particles per cell.

In practice, the target distribution $n_{\mathrm{seed}}(x,y,z)$ is realized by reading a pre-sampled set of
$n_{\mathrm{load}}=10^{6}$ particle positions from an external file.
Each MPI rank retains only the samples that fall inside its local subdomain and obtains $n$ positions.
To reach the desired total particle number while keeping the I/O and pre-sampling cost moderate, the retained positions are replicated $n_{\mathrm{copy}}=50$ times for each species, so that each rank initializes $n_{\mathrm{load}}\,n_{\mathrm{copy}}$ macroparticles per species and the total number of initialized macroparticles becomes $N_{\mathrm{mp}}=n_{\mathrm{load}}\,n_{\mathrm{copy}}=5.0\times10^{7}$.
To enforce exact discrete charge neutrality at $t=0$,
macro-electrons and macro-ions are initialized with the same sampled positions.

Particle velocities are assigned independently after the spatial loading.
For each species $s$ and each velocity component $k\in\{x,y,z\}$,
a Gaussian random variate is generated through the Box--Muller transform,
\begin{equation}
v_{k,s}=u_{k,s}+v_{\mathrm{th},s}\sqrt{-2\ln r_1}\cos(2\pi r_2),
\quad r_1,r_2\sim \mathcal{U}(0,1),
\label{eq:init_boxmuller}
\end{equation}
which corresponds to a drifting Maxwellian with mean drift $u_{k,s}$ and thermal speed
\begin{equation}
v_{\mathrm{th},s}=\sqrt{\frac{eT_s}{m_s}}.
\label{eq:vth_def}
\end{equation}
In this work, the initial temperatures are $T_{e0}=30~\mathrm{eV}$ and $T_{i0}=1.5~\mathrm{eV}$, corresponding to thermal speeds of $v_{\mathrm{th},e}=2.2970546\times10^{6}\ \mathrm{m/s}$ and $v_{\mathrm{th},i}=1.0499183\times10^{3}\ \mathrm{m/s}$, respectively.

The initial electron drift is set to zero.
For ions, the transverse drifts are also set to zero,
whereas the axial ion drift is prescribed as a smooth $z$-dependent profile to reduce the initial transient and accelerate the early approach to the discharge state,
\begin{equation}
v_{z,i}(z)=\frac{1}{2}\left[
a+\frac{b-a}{1+(z/c)^d}
\right],
\label{eq:init_ion_vz_profile}
\end{equation}
where $a=1.6575\times10^{4}$, $b=-1.5208\times10^{3}$, $c=9.1~\mathrm{mm}$, and $d=4.4528$.
Accordingly, electrons are initialized with an isotropic Maxwellian,
while ions have Maxwellian transverse velocities $(v_{x,i},v_{y,i})$ and an axial drift $v_{z,i}$ prescribed by Eq.~\eqref{eq:init_ion_vz_profile}.

\begin{figure}[htbp]
    \centering
    \includegraphics[width=0.9\linewidth]{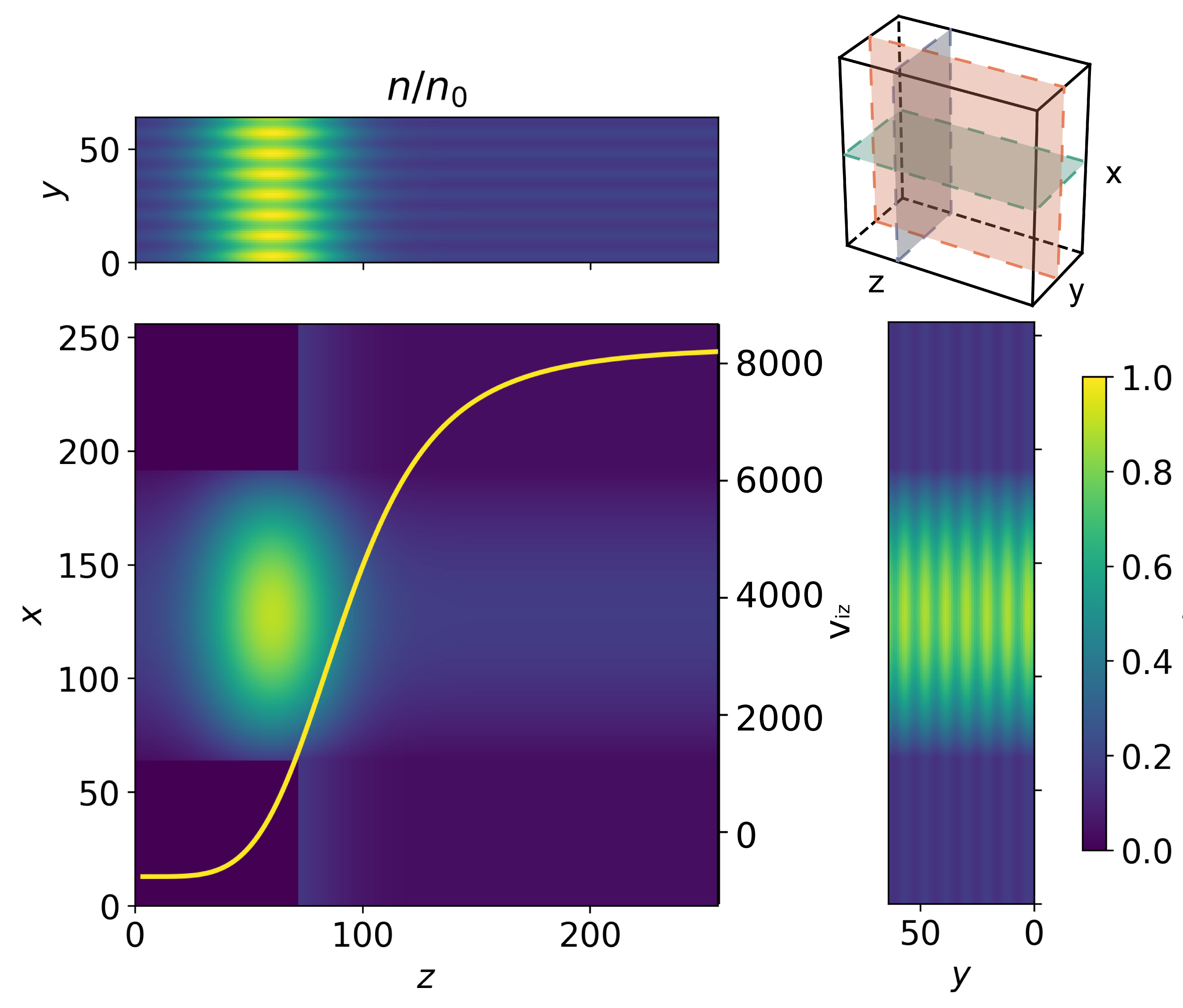}
    \caption{
    Prescribed initial plasma loading.
    Normalized seed density $n_{\mathrm{seed}}/n_0$ is shown on three slices:
    $x$--$y$ at $i_z=72$,
    $x$--$z$ at $i_y=32$,
    and $y$--$z$ at $i_x=128$,
    where $i_x$, $i_y$, and $i_z$ denote grid indices.
    The yellow curve in the $x$--$z$ plane indicates the imposed initial ion axial drift profile $v_{z,i}$ (m/s).
    The inset shows the slice locations in the 3D computational domain.
    }
\label{fig:init_den}
\end{figure}

Fig.~\ref{fig:init_den} summarizes the prescribed initialization.
It shows the normalized seed density $n_{\mathrm{seed}}/n_0$ on three slices
($x$--$y$ at $i_z=72$,
$x$--$z$ at $i_y=32$,
and $y$--$z$ at $i_x=128$,
where $i_x,i_y,i_z$ denote grid indices).
The imposed initial ion axial drift profile $v_{z,i}(z)$ is overlaid on the $x$--$z$ slice
to illustrate how the seed plasma and ion pre-acceleration are arranged within the domain.
The inset indicates the relative positions of the slices in the 3D computational box.

\subsection{Background Magnetic Field}
\label{subsec:background_magnetic_field}

The applied background magnetic field used in this study is derived from the magnetic-circuit design of a Hall thruster prototype.
An axisymmetric magnetostatic model is constructed in Finite Element Method Magnetics (FEMM) \cite{meeker2015femm},
in which the magnetic properties of the ferromagnetic components and surrounding media are specified,
together with the coil current and number of turns.
The resulting magnetostatic solution provides the imposed background magnetic field for the PIC simulations.

Because the plasma solver adopts a local Cartesian slab representation,
the exported magnetic field is mapped onto the computational $(x,z)$ plane,
where $x$ corresponds locally to the radial direction and $z$ to the axial direction.
The magnetic field therefore contains only radial and axial components in the present model.
This construction is consistent with the local annular-sector approximation introduced in Sec.~\ref{subsec:physical_configuration_domain}.
In FEMM, the reference Hall thruster geometry is defined such that the radius from the thruster axis to the channel center is $57.5~\mathrm{mm}$.
Accordingly, the simulated Cartesian domain represents a local radial segment extracted from a finite-radius annular device rather than a full cylindrical cross section.

\begin{figure}[htbp]
    \centering
    \includegraphics[width=0.98\linewidth]{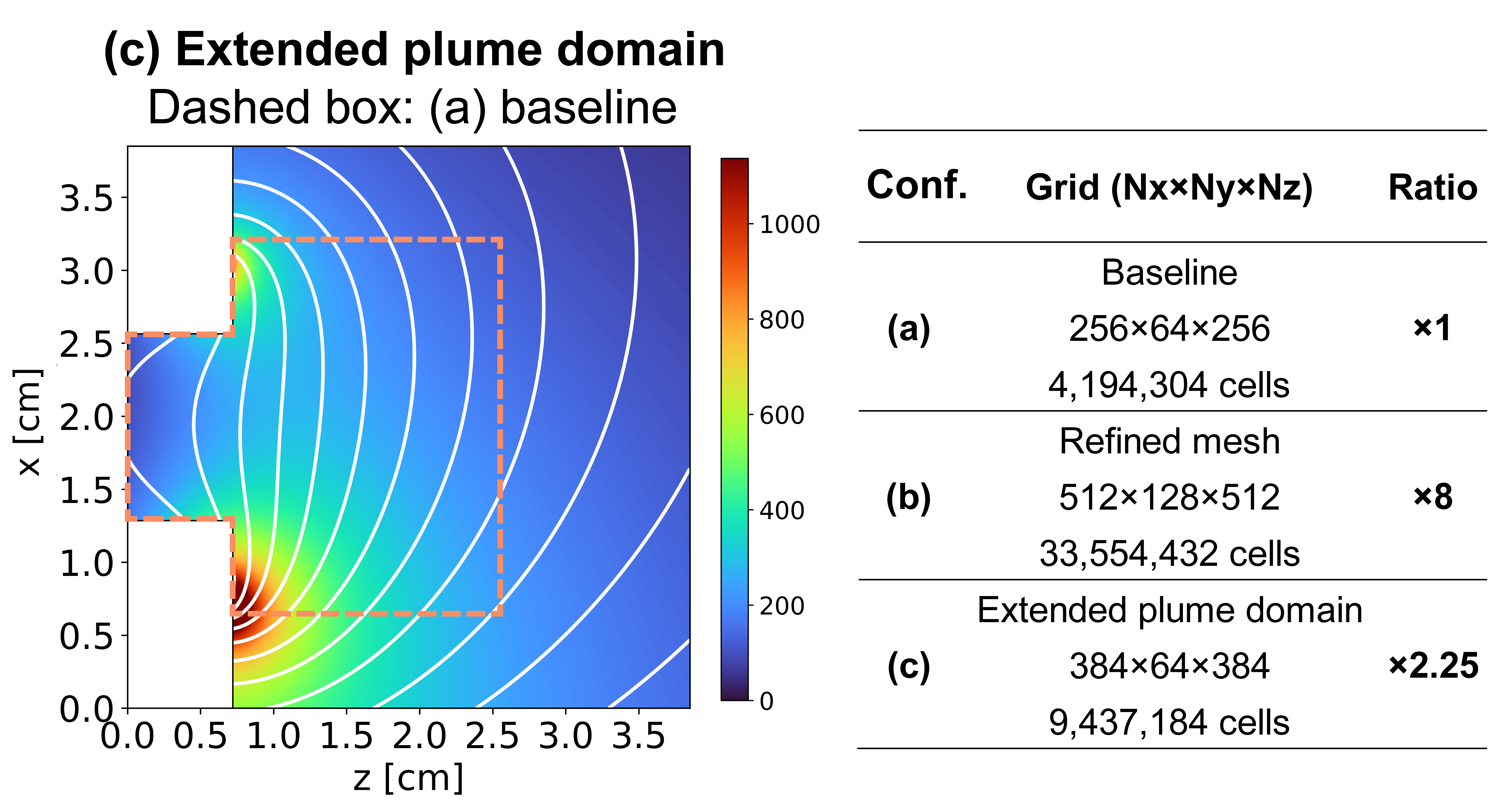}
    \caption{
    Background magnetic-field configurations in the $x$--$z$ plane.
    Colors show the magnetic-field magnitude $|\mathbf{B}|$ (Gauss), and white contours denote the magnetic vector potential.
    (a) Baseline export used for cases D, D$_{2\Delta t}$, C, O, and O$_{\text{loadC}}$.
    (b) Refined export over the same physical region, used for Case D$_{\text{finer}}$.
    (c) Extended plume-domain export, used for Case O$_{\text{bigger}}$; the dashed box indicates the baseline-domain extent.
    The inset summarizes the total cell count and relative cell-count ratio for the three configurations.
    }
\label{fig:init_B}
\end{figure}

Three magnetic-field exports are prepared, as shown in Fig.~\ref{fig:init_B}.
Configuration (a) corresponds to the baseline computational domain and mesh spacing.
Configuration (b) covers the same physical region but uses half the export spacing in both directions,
thereby providing the higher-resolution magnetic field required by Case D$_{\text{finer}}$.
Configuration (c) retains the baseline export spacing but extends farther downstream into the plume region,
which is required for Case O$_{\text{bigger}}$.
Thus, the three magnetic-field configurations are directly associated with the simulation cases summarized in Tab.~\ref{tab:case_summary}.

Fig.~\ref{fig:init_B} presents the magnetic-field magnitude $|\mathbf{B}|$ together with contours of the magnetic vector potential on the $x$--$z$ plane.
The baseline configuration in Fig.~\ref{fig:init_B}(a) is used for cases D, D$_{2\Delta t}$, C, O, and O$_{\text{loadC}}$.
The refined export in Fig.~\ref{fig:init_B}(b) is used for Case D$_{\text{finer}}$,
and the extended export in Fig.~\ref{fig:init_B}(c) is used for Case O$_{\text{bigger}}$.
In this way, changes in mesh resolution or plume-domain size remain consistent with the corresponding magnetic-field representation adopted in each case.

\subsection{Neutral Background Initialization, Continuum Evolution, and Verification}
\label{subsec:neutral_background}

To initialize the neutral propellant background prior to plasma evolution,
we first generate a quasi-steady neutral field using a collisionless free-molecular particle method.
The resulting time-averaged neutral density and mean velocity fields are then exported to the coupled 3D PIC solver.
During the coupled plasma simulation, however, neutrals are no longer evolved as macroparticles.
Instead, the neutral number density is advanced using a reduced-cost continuum continuity solver,
while the neutral velocity field is prescribed from the pre-processing stage.
This hybrid treatment greatly reduces the computational cost relative to a fully kinetic neutral description,
while retaining the dominant neutral transport and depletion physics relevant to the discharge evolution.

\subsubsection{Free-molecular neutral pre-processing}
\label{subsubsec:neutral_freemolecular}

In the pre-processing stage, neutral dynamics are decoupled from the plasma and neutral--neutral collisions are neglected.
Neutral macroparticles therefore undergo collisionless free streaming between boundary interactions.
The neutral simulation is performed in the same local Cartesian coordinate system $(x,y,z)$ used by the plasma solver,
with $x\equiv r$, $y\equiv \theta$, and $z$ the axial direction.
Diagnostic quantities are obtained by azimuthally averaging particle statistics and projecting them onto a two-dimensional $(x,z)$ grid,
which is subsequently used by the coupled plasma simulation.

The neutral simulation domain spans
$x\in[0,L_{xp}]$ and $z\in[0,z_{\max}]$,
where
$L_{xp}=25.6~\mathrm{mm}$ and $z_{\max}=25.6~\mathrm{mm}$.
The physical discharge channel occupies
$x\in[x_{\min},x_{\max}]$ for $z\le z_{\mathrm{exit}}$,
with
$x_{\min}=6.4~\mathrm{mm}$,
$x_{\max}=19.2~\mathrm{mm}$,
and
$z_{\mathrm{exit}}=7.2~\mathrm{mm}$,
while the azimuthal extent is
$y\in[0,L_y]$
with periodic boundary conditions.

Neutral particles are continuously injected from the anode plane at $z=0$.
At each time step,
$N_{\mathrm{new}}=2000$ neutral macroparticles are introduced with $(x,y)$ sampled uniformly in the injection window
$x\in[x_{1,\mathrm{inj}},x_{2,\mathrm{inj}}]=[7.68,\,17.92]~\mathrm{mm}$
and
$y\in[0,L_y]$.
Their velocities follow a drifting Maxwellian distribution:
a thermal spread corresponding to the gas temperature
$T_{\mathrm{gas}}=573.15~\mathrm{K}$
and xenon atomic mass
$m_{\mathrm{gas}}=2.18\times 10^{-25}~\mathrm{kg}$
is applied to all three components,
while a constant axial drift
$u_{\mathrm{drift},z}=1000~\mathrm{m/s}$
is superimposed on the axial velocity to impose the mean propellant throughflow.
Backflow samples with $v_z<0$ are allowed;
such particles simply exit through the upstream boundary and are absorbed.
The imposed drift speed $u_{\mathrm{drift},z}=1000~\mathrm{m/s}$ is about three times higher than the nominal physical value and is adopted deliberately, following our previous studies\cite{IEPC-2025-063}, as a modeling choice to increase the breathing-mode frequency and thereby make the cross-frequency interaction between the low-frequency breathing mode and the high-frequency EDI more tractable in a single simulation.

Particle trajectories are advanced explicitly with
\begin{equation}
\mathbf{x}^{n+1}=\mathbf{x}^{n}+\mathbf{v}^{n}\Delta t ,
\label{eq:neutral_push}
\end{equation}
and particles are removed once they leave the computational domain in either $x$ or $z$.
This provides a simple open-boundary treatment for the pre-processing stage without reinjection from the far field.

Within the channel region ($z\le z_{\mathrm{exit}}$),
interactions with the inner and outer radial walls are modeled by a mixed specular--diffuse reflection law.
When a particle crosses a wall,
its position is shifted slightly back into the domain to avoid repeated crossings caused by finite time stepping.
Its post-collision velocity is then determined probabilistically:
with probability $1-\sigma_t$, specular reflection is applied by reversing only the wall-normal velocity component;
with probability $\sigma_t$, diffuse reflection is applied by re-emitting the particle from the wall with a Maxwellian distribution at the wall temperature
$T_{\mathrm{wall}}=773.15~\mathrm{K}$,
with the normal velocity directed back into the domain.
In this work,
$\sigma_t=0.7$ is used.

After an initial transient filling stage,
particle statistics are accumulated over a later sampling window to reduce Monte Carlo noise and obtain quasi-steady mean fields.
The neutral binning grid is cell-centered with
$N_x\times N_z=256\times 256$
and
$\Delta x=\Delta z=0.1~\mathrm{mm}$.
The neutral-particle time step is
$\Delta t=0.125~\mu\mathrm{s}$,
and the total run length is
$N_{\mathrm{step}}=1000$ steps,
corresponding to
$125~\mu\mathrm{s}$.
Sampling is performed over the last 500 steps only,
so that the exported fields are not contaminated by the initial filling transient.

Because a finite collisionless domain without ionization
may otherwise accumulate an unrealistically large far-field neutral population,
the time-averaged neutral density is attenuated downstream using an exponential taper,
\begin{equation}
f(z)=
\begin{cases}
1, & z \le z_1, \\[2pt]
\exp\!\left(-\dfrac{z-z_1}{L}\right), & z > z_1 ,
\end{cases}
\qquad
L = \frac{z_2-z_1}{\ln(1/\epsilon)} ,
\label{eq:neutral_taper}
\end{equation}
with
$z_1=0.8\,z_{\mathrm{exit}}=5.76~\mathrm{mm}$,
$z_2=z_{\max}=25.6~\mathrm{mm}$,
and
$\epsilon=0.1$,
which gives
$L=8.62~\mathrm{mm}$.
The tapered density is then written as
\begin{equation}
n_{n0}(x,z)\leftarrow n_{n0}(x,z)\,f(z).
\end{equation}
The final outputs of the pre-processing stage are the time-averaged neutral density
$n_{n0}(x,z)$
and mean velocity fields
$(u_{x0},u_{z0})(x,z)$,
which are used to initialize and drive the continuum neutral model in the coupled PIC simulations.

\subsubsection{Continuum neutral evolution in the coupled PIC solver}
\label{subsubsec:neutral_continuum_pic}

Evolving neutrals as macroparticles throughout the full 3D PIC simulation would add substantial computational cost.
Therefore, during the coupled plasma run,
the neutral number density is advanced using a continuum continuity equation,
while the neutral velocity field is prescribed from the free-molecular pre-processing stage.
The governing equation is
\begin{equation}
\frac{\partial n_n}{\partial t}+\nabla\cdot\!\left(n_n\mathbf{u}_n\right)
= -\dot{n}_{\mathrm{ion}},
\label{eq:neutral_continuity_ion_3d}
\end{equation}
where
$\mathbf{u}_n=(u_x,u_y,u_z)$
is the neutral mean velocity and
$\dot{n}_{\mathrm{ion}}$
is the neutral consumption rate due to electron-impact ionization.
In the coupled simulation,
$\dot{n}_{\mathrm{ion}}$ is computed self-consistently from the plasma state,
and the removed neutrals are converted into newly created ions and electrons by the plasma module.
The coupling is therefore two-way:
the prescribed neutral advection field transports the neutrals,
while plasma ionization depletes them in space and time.

Equation~\eqref{eq:neutral_continuity_ion_3d} is discretized on the cell-centered Cartesian mesh using an explicit finite-volume method.
Denoting the cell-averaged neutral density by $n^n_{i,j,k}$,
the update reads
\begin{equation}
\begin{aligned}
n^{n+1}_{i,j,k}
&=
n^{n}_{i,j,k}
-\frac{\Delta t}{\Delta x}
\left(F^{x}_{i+\frac12,j,k}-F^{x}_{i-\frac12,j,k}\right) \\
&\quad
-\frac{\Delta t}{\Delta y}
\left(F^{y}_{i,j+\frac12,k}-F^{y}_{i,j-\frac12,k}\right)\\
&\quad
-\frac{\Delta t}{\Delta z}
\left(F^{z}_{i,j,k+\frac12}-F^{z}_{i,j,k-\frac12}\right) \\
&\quad
-\Delta t\,\dot{n}_{\mathrm{ion},i,j,k},
\end{aligned}
\label{eq:neutral_fv_update}
\end{equation}
where
$(F^x,F^y,F^z)$
are the numerical mass fluxes through the cell faces.

The face fluxes are evaluated using the Lax--Friedrichs (Rusanov) form.
For example, at an $x$-face,
\begin{equation}
\begin{aligned}
F^{x}_{i+\frac12,j,k}
&=
\frac12\Big(u_{x,L}\,n_L + u_{x,R}\,n_R\Big)
-\frac12\,\alpha_x\,(n_R-n_L), \\
\alpha_x
&=
\max\!\left(|u_{x,L}|,|u_{x,R}|\right),
\end{aligned}
\label{eq:neutral_lf_flux_x}
\end{equation}
where
$(n_L,u_{x,L})$
and
$(n_R,u_{x,R})$
are taken from the two cells adjacent to the face.
The $y$- and $z$-direction fluxes are defined analogously.
This scheme is robust for advection-dominated transport and introduces sufficient numerical dissipation to suppress spurious oscillations near steep gradients.

The neutral update is performed once per PIC time step using
\begin{equation}
\Delta t_n=\Delta t_{\mathrm{PIC}} .
\label{eq:neutral_dt_equals_pic}
\end{equation}
Because the PIC time step is already constrained by plasma dynamics,
it is much smaller than the neutral advection CFL limit in the present simulations.
This keeps the neutral and plasma evolution synchronized without requiring neutral subcycling.

In the azimuthal direction, periodic boundary conditions are applied.
At the anode plane, a prescribed inflow neutral density profile is imposed from the free-molecular pre-processing.
At the plume outflow boundaries, zero-gradient conditions are used.
Solid channel walls are treated as impermeable for the continuum neutral transport,
so that the normal neutral mass flux vanishes at wall-adjacent faces.
These boundary conditions ensure consistent mass transport while preventing unphysical neutral penetration into solid regions.

\subsubsection{Verification against the free-molecular solution}
\label{subsubsec:neutral_verification}

\begin{figure}[!htbp]
    \centering
    \includegraphics[width=0.95\linewidth]{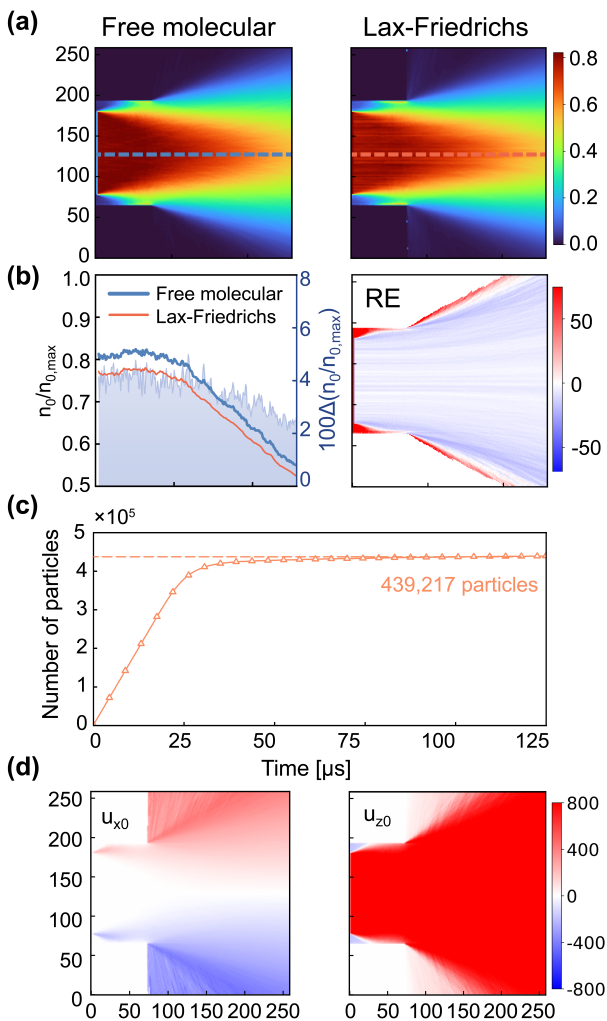}
    \caption{
    Verification of the continuum neutral model against the free-molecular pre-processing.
    (a) Normalized neutral density $n_{0}/n_{0,\max}$ from the free-molecular method (left) and the finite-volume Lax--Friedrichs solution (right); the dashed line indicates the centerline used for the one-dimensional comparison.
    (b) Left: centerline profiles of $n_{0}/n_{0,\max}$ from the two methods; the shaded region denotes the pointwise difference.
    Right: relative-error map
    $RE=100\,(n_{0,F}-n_{0,LF})/n_{0,F}$,
    evaluated only in cells with $n_{0,F}/n_{0,\max}>0.01$.
    (c) Time evolution of the total number of neutral macroparticles in the free-molecular simulation, saturating at $4.39\times 10^{5}$.
    (d) Mean velocity fields $u_{x0}$ and $u_{z0}$ (m/s) obtained from the free-molecular run and used as prescribed advection velocities in the coupled continuum neutral solver.
    }
    \label{fig:neutral_fields}
\end{figure}

Before plasma coupling is activated,
the continuum neutral solver is verified by setting
$\dot{n}_{\mathrm{ion}}=0$
and evolving Eq.~\eqref{eq:neutral_continuity_ion_3d}
with the pre-processed velocity field and inlet profile.
Fig.~\ref{fig:neutral_fields} compares the normalized neutral density obtained from the free-molecular pre-processing and from the finite-volume Lax--Friedrichs solver.
The agreement shows that,
when driven by the particle-derived mean transport field,
the continuum solver reproduces the quasi-steady neutral distribution with only modest discrepancies.

The free-molecular simulation reaches a statistically steady neutral population of
$4.39\times 10^{5}$ macroparticles
for the chosen injection rate and averaging window.
The corresponding mean velocity fields
$u_{x0}$ and $u_{z0}$
are also shown in Fig.~\ref{fig:neutral_fields}
and are subsequently used as the prescribed advection field in the coupled continuum neutral model.
This verification supports the use of the reduced-cost continuum neutral treatment in the plasma simulations,
where the ionization sink term is enabled to capture self-consistent neutral depletion at far lower computational cost than a fully kinetic neutral description.

\section{Simulation Results}
\label{sec:results}

With the simulation model and numerical setup established, we now turn to the main results.
To present them clearly, we proceed from the slow global discharge evolution to the quasi-steady averaged structure, and then to the instability-driven anomalous transport pathways, the underlying instantaneous three-dimensional EDI dynamics, and their spectral characteristics.

\begin{figure}[htbp]
    \centering
    \includegraphics[width=1\linewidth]{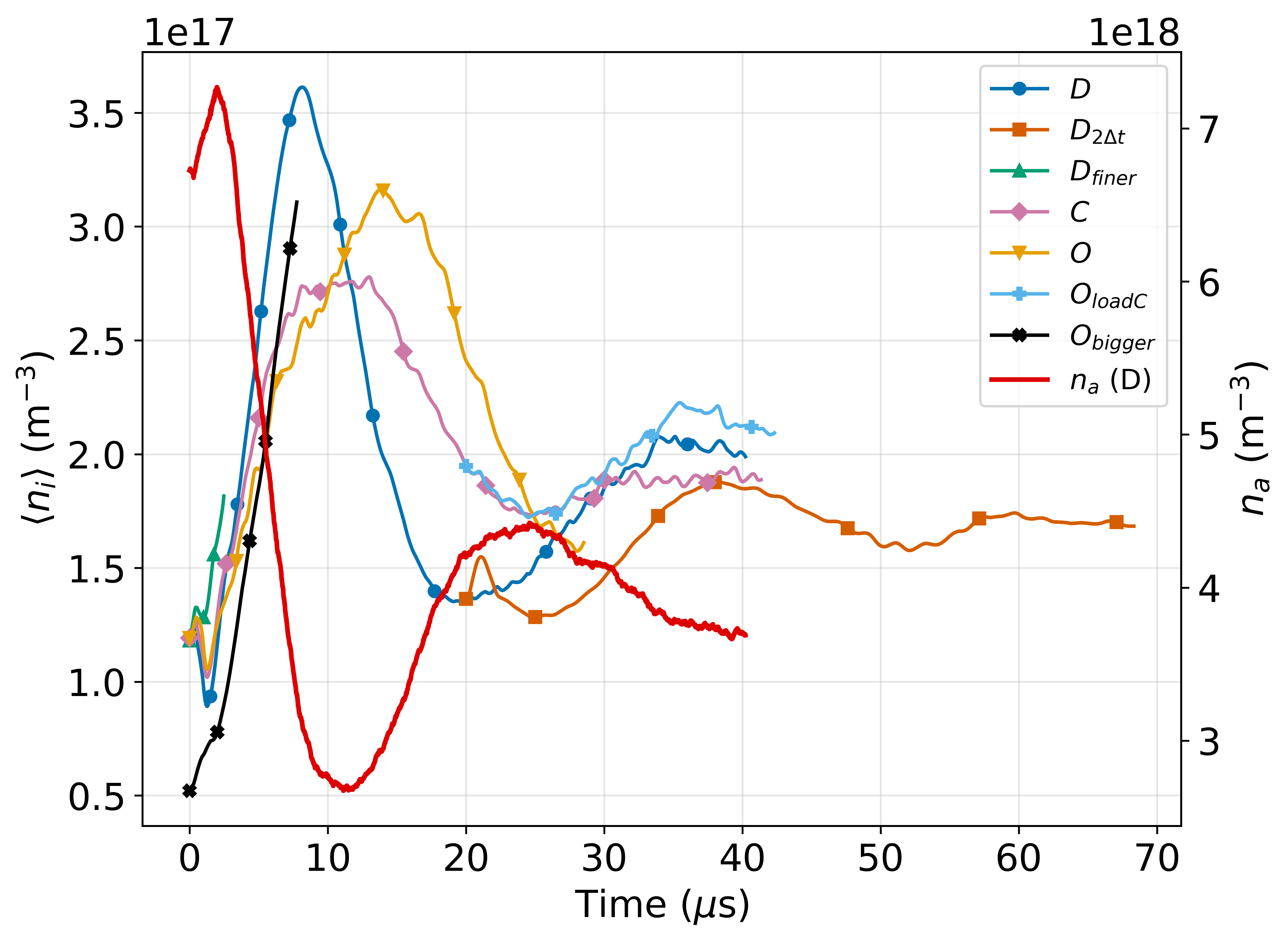}
    \caption{Time evolution of the spatially averaged ion number density, $\langle n_i\rangle$, for all cases (left axis). The neutral number density $n_a$ at the grid indices $(128,32,80)$ in Case~D is shown on the right axis.}
    \label{fig:ni_t_cases}
\end{figure}

\subsection{Slow Global Evolution and Selection of the Quasi-Steady Analysis Window}
\label{sec:analysis_window}

We begin by distinguishing the slow, global discharge evolution from the much faster instability dynamics that are the main focus of this work.
Fig.~\ref{fig:ni_t_cases} shows the time evolution of the spatially averaged ion number density, $\langle n_i\rangle$, for all simulation cases.
Nearly all cases exhibit an early transient in which $\langle n_i\rangle$ first rises rapidly to a peak, then decreases, and finally evolves toward a later recovery stage.
For Case~D, for example, $\langle n_i\rangle$ reaches approximately $3.6\times 10^{17}~\mathrm{m^{-3}}$ at $t\approx 10~\mu\mathrm{s}$, decreases to approximately $1.4\times 10^{17}~\mathrm{m^{-3}}$ by $t\approx 20~\mu\mathrm{s}$, and then increases more gradually to approximately $2.0\times 10^{17}~\mathrm{m^{-3}}$ at $t\approx 35~\mu\mathrm{s}$.
The same figure also overlays the neutral number density, $n_a$, at a representative monitoring point $(128,32,80)$ in Case~D.
As $\langle n_i\rangle$ increases during the initial ionization stage, the local neutral density decreases, whereas during the subsequent reduction of $\langle n_i\rangle$, the local neutral density shows a partial recovery.
This broadly anti-correlated evolution is consistent with neutral depletion and refilling during the slow breathing-like modulation of the discharge.

Although the detailed amplitudes and timings differ from case to case, Cases~D, C, O, and O$_{\text{loadC}}$ all display the same qualitative rise--decay--recovery trend.
This indicates that the low-frequency global evolution is a robust feature of the self-consistent discharge development, rather than a peculiarity of one specific boundary treatment.
At the same time, Fig.~\ref{fig:ni_t_cases} shows that changing the boundary treatment does modify the macroscopic trajectory of $\langle n_i\rangle(t)$, including the magnitude of the initial overshoot, the depth of the subsequent depletion, and the level approached at later times.
These differences confirm that the wall and outflow models affect the slowly varying background state on which the high-frequency instability develops.

For the present purpose, however, the key point is that the slow evolution of $\langle n_i\rangle$ occurs on a time scale of order $10~\mu\mathrm{s}$, whereas the EDI evolves much more rapidly.
Therefore, strict convergence of the global quantity $\langle n_i\rangle(t)$ is not required before meaningful instability diagnostics can be extracted.
Instead, what is needed is a time interval during which the high-frequency EDI has become established and its spectral content and spatial organization vary only weakly over the averaging window.
In other words, the relevant requirement for the following transport analysis is a \emph{local quasi-steady state} of the instability, even if the slowly varying discharge envelope still evolves in the background.

This distinction is particularly important for Case~D.
Because of computational cost, Case~D was terminated at $t=40.22~\mu\mathrm{s}$ before the low-frequency evolution became fully stationary.
Nevertheless, by the end of the run the rapid initial overshoot has long passed, the subsequent recovery has slowed substantially, and the discharge structure has become much more repeatable than during the earlier transient stages.
Case~D$_{2\Delta t}$, which restarts from the Case~D solution at $t=20~\mu\mathrm{s}$ and uses a doubled time step, further suggests that the low-frequency oscillation amplitude continues to decrease at later times and approaches a narrower range of variation.
We therefore interpret the late-time interval of Case~D as sufficiently mature for extracting representative EDI and transport diagnostics, while recognizing that it does not yet correspond to a fully converged global steady state.

A similar argument applies to Cases~C and O.
Case~C extends to $t=41.36~\mu\mathrm{s}$ and is closer to a global quasi-steady state in the sense of $\langle n_i\rangle(t)$.
Case~O reaches a shorter physical time because of its higher computational cost, whereas the restarted Case~O$_{\text{loadC}}$ begins from a later-time state loaded from Case~C and therefore provides access to a comparably mature discharge stage under the outflow boundary treatment.
As a result, Cases~D, C, and O$_{\text{loadC}}$ together provide the most appropriate set for comparing late-time instability structure and anomalous electron transport under different boundary closures.

Guided by these considerations, the following sections use a representative late-time window around $t\approx 40~\mu\mathrm{s}$ as the primary analysis stage for the baseline transport diagnostics.
For Case~D, the field maps shown below at $t=40~\mu\mathrm{s}$ are therefore interpreted as representative of a late, locally quasi-steady instability state.
For Case~C and Case~O$_{\text{loadC}}$, the corresponding late-time windows are chosen to match this stage as closely as possible in the same spectral and structural sense.
This common analysis window is not intended to imply that the three cases have identical macroscopic histories; rather, it provides a consistent basis for isolating and comparing the instability-driven transport pathway after the early transient has decayed.

Finally, the short and computationally expensive runs D$_{\mathrm{finer}}$ and O$_{\mathrm{bigger}}$ remain valuable even though they do not reach the same late-time window.
They will be used later only for targeted robustness checks during the early stage, namely to assess the sensitivity of the EDI characteristics to grid refinement and plume-domain enlargement over the time interval available to those cases.

Having identified the slow global modulation and the representative late-time analysis window, we next examine the time-averaged spatial structure of the discharge at several stages of the evolution.
This will clarify how the background density, potential, acceleration, heating, and azimuthal drift are organized before turning to the instability-resolved transport pathway itself.

\begin{figure*}[htbp]
    \centering
    \includegraphics[width=0.32\linewidth]{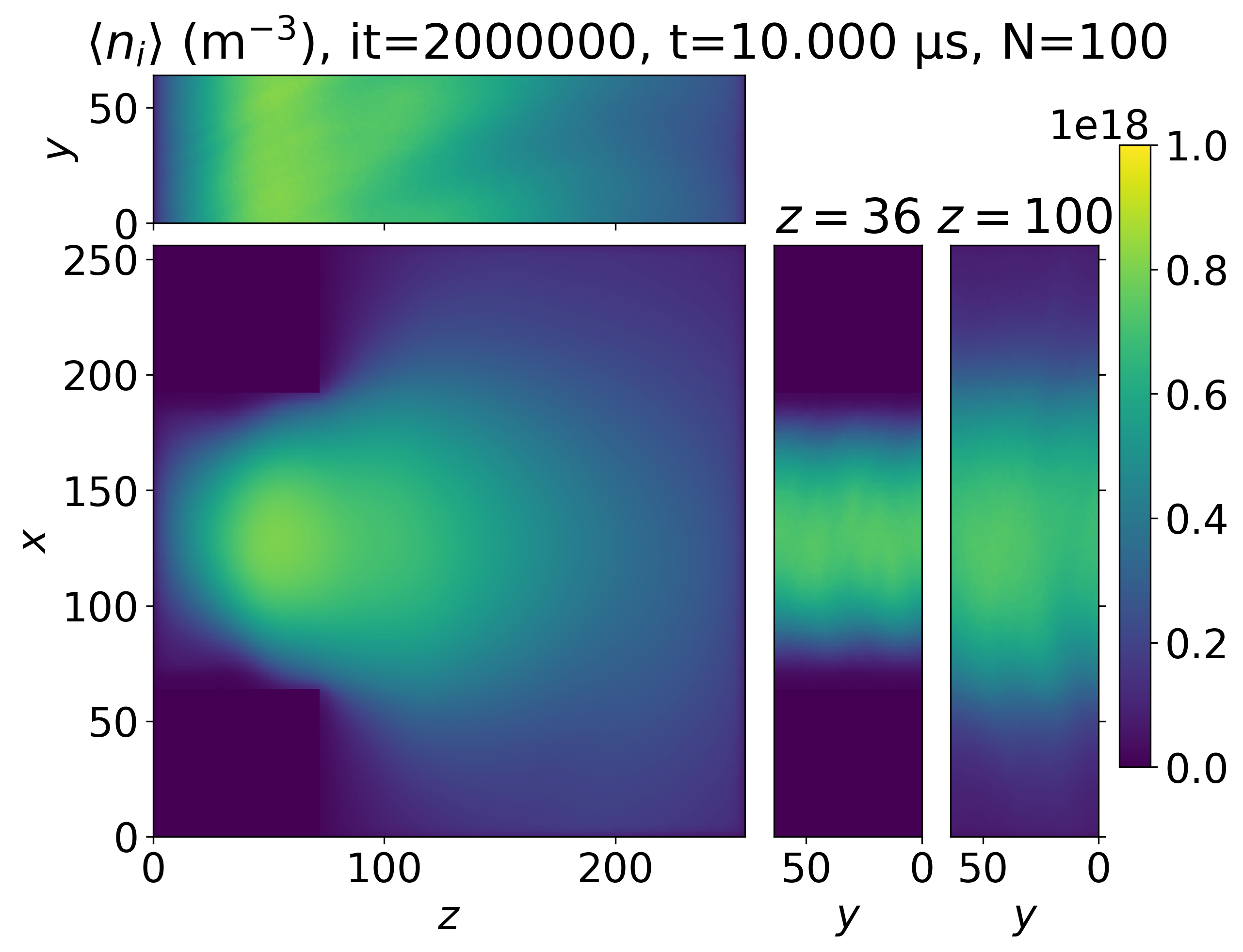}
    \includegraphics[width=0.32\linewidth]{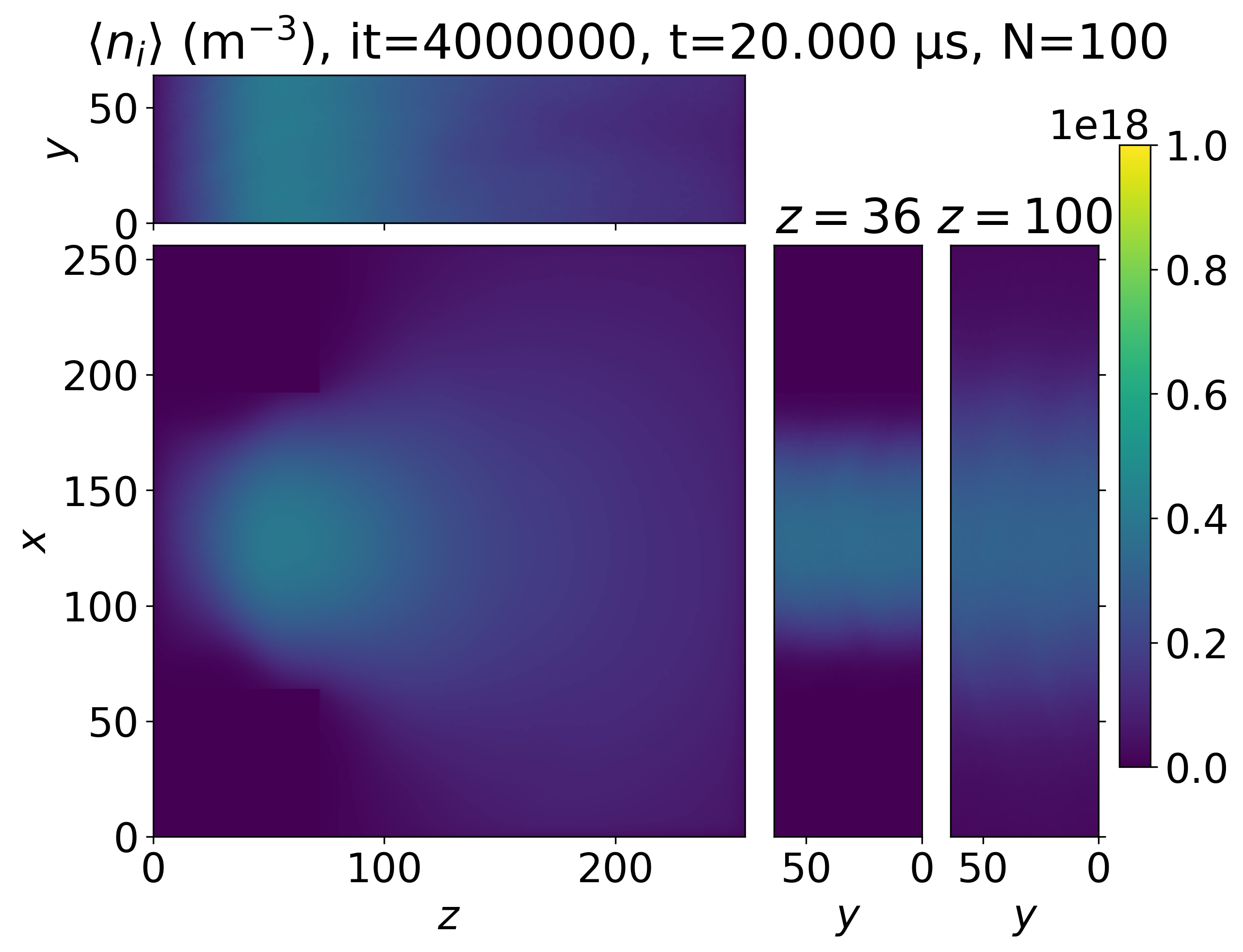}
    \includegraphics[width=0.32\linewidth]{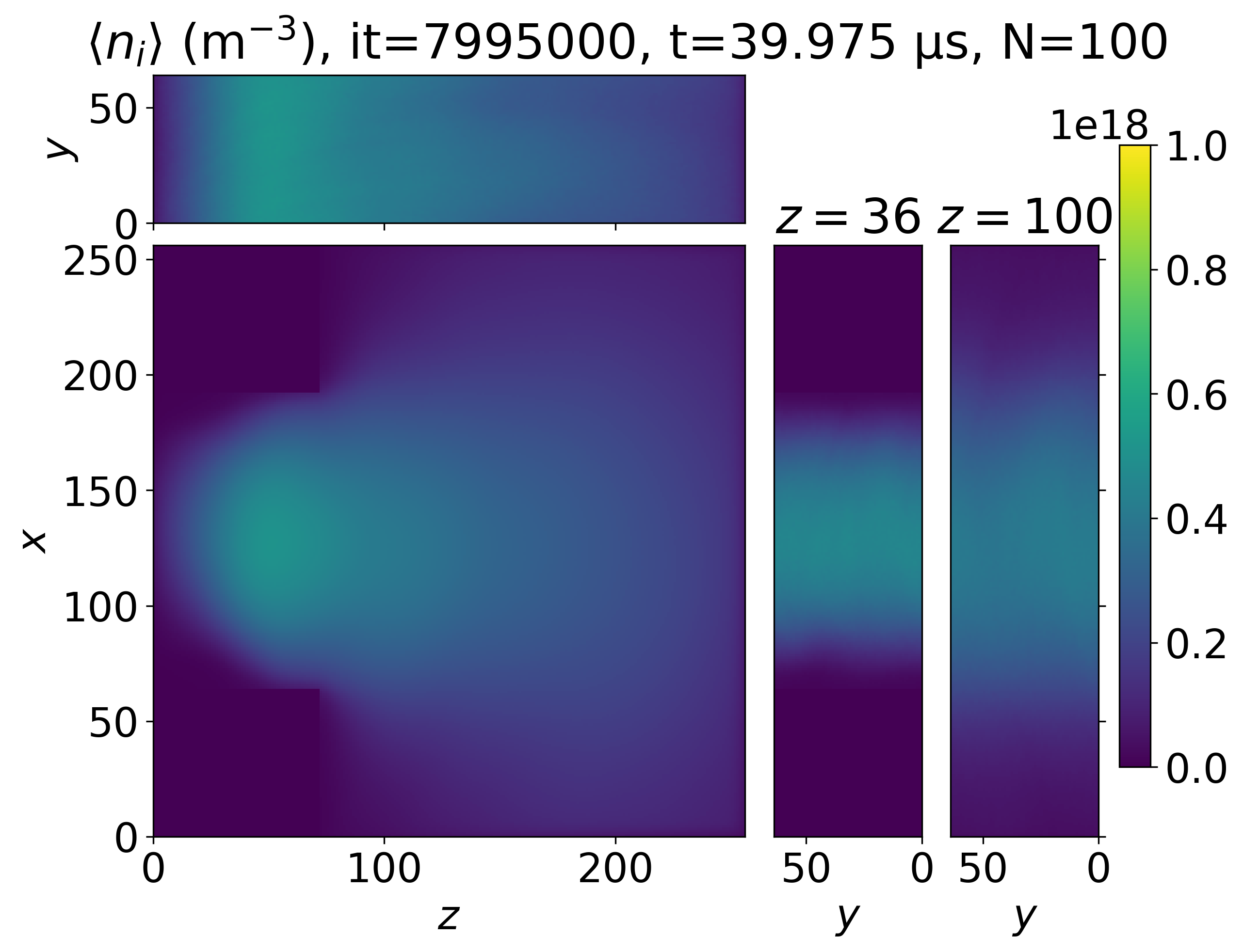}
    \includegraphics[width=0.32\linewidth]{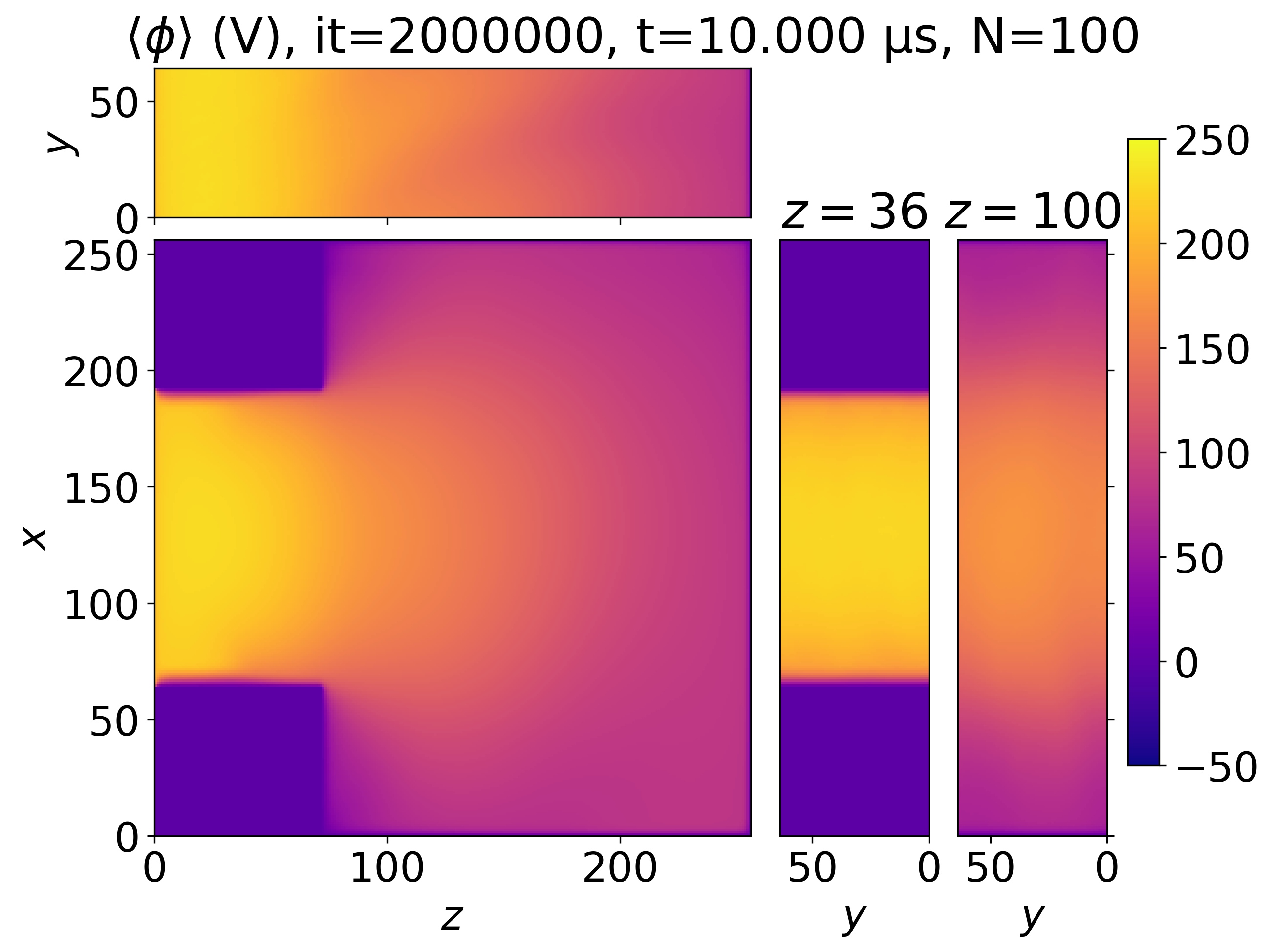}
    \includegraphics[width=0.32\linewidth]{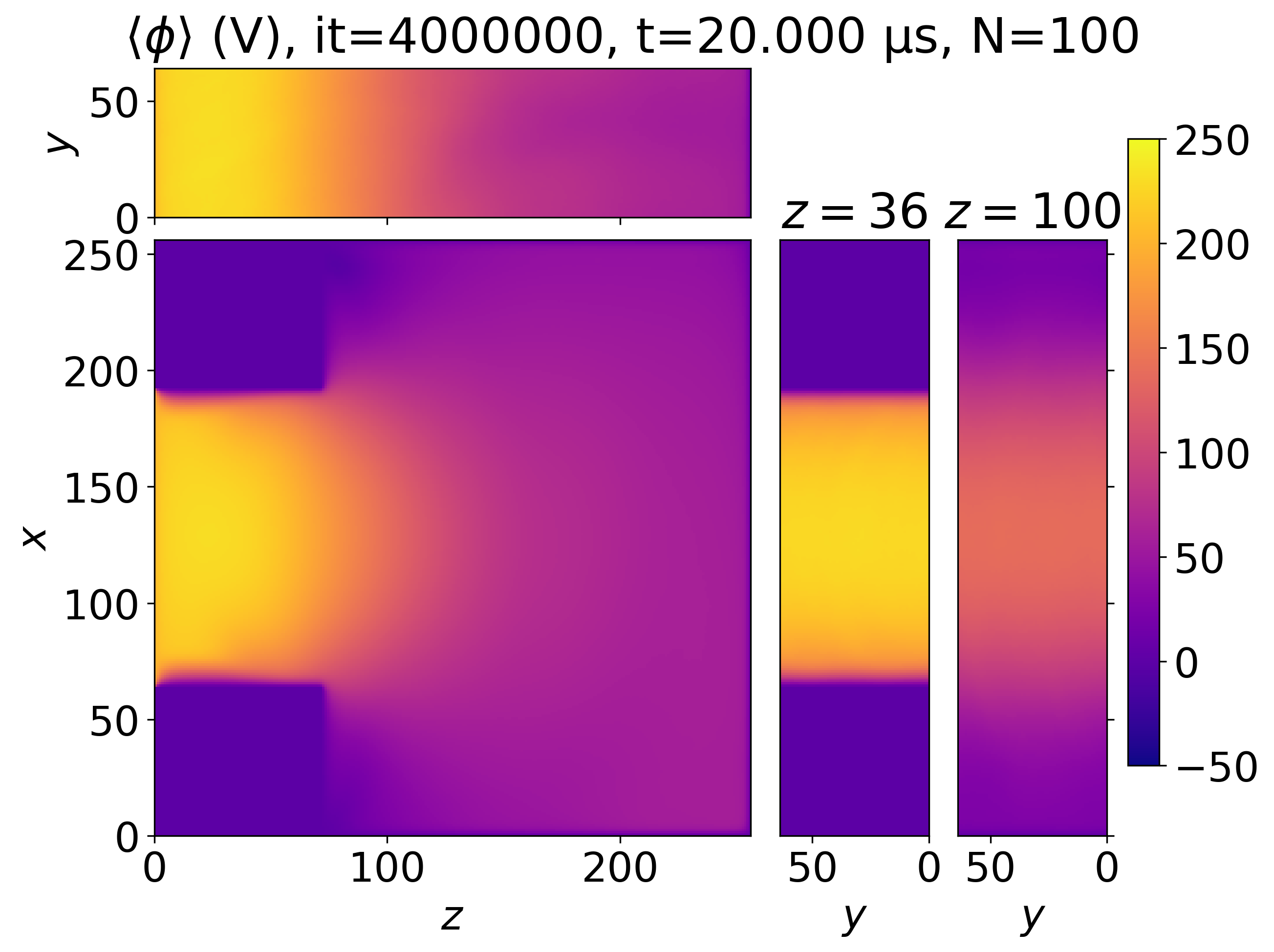}
    \includegraphics[width=0.32\linewidth]{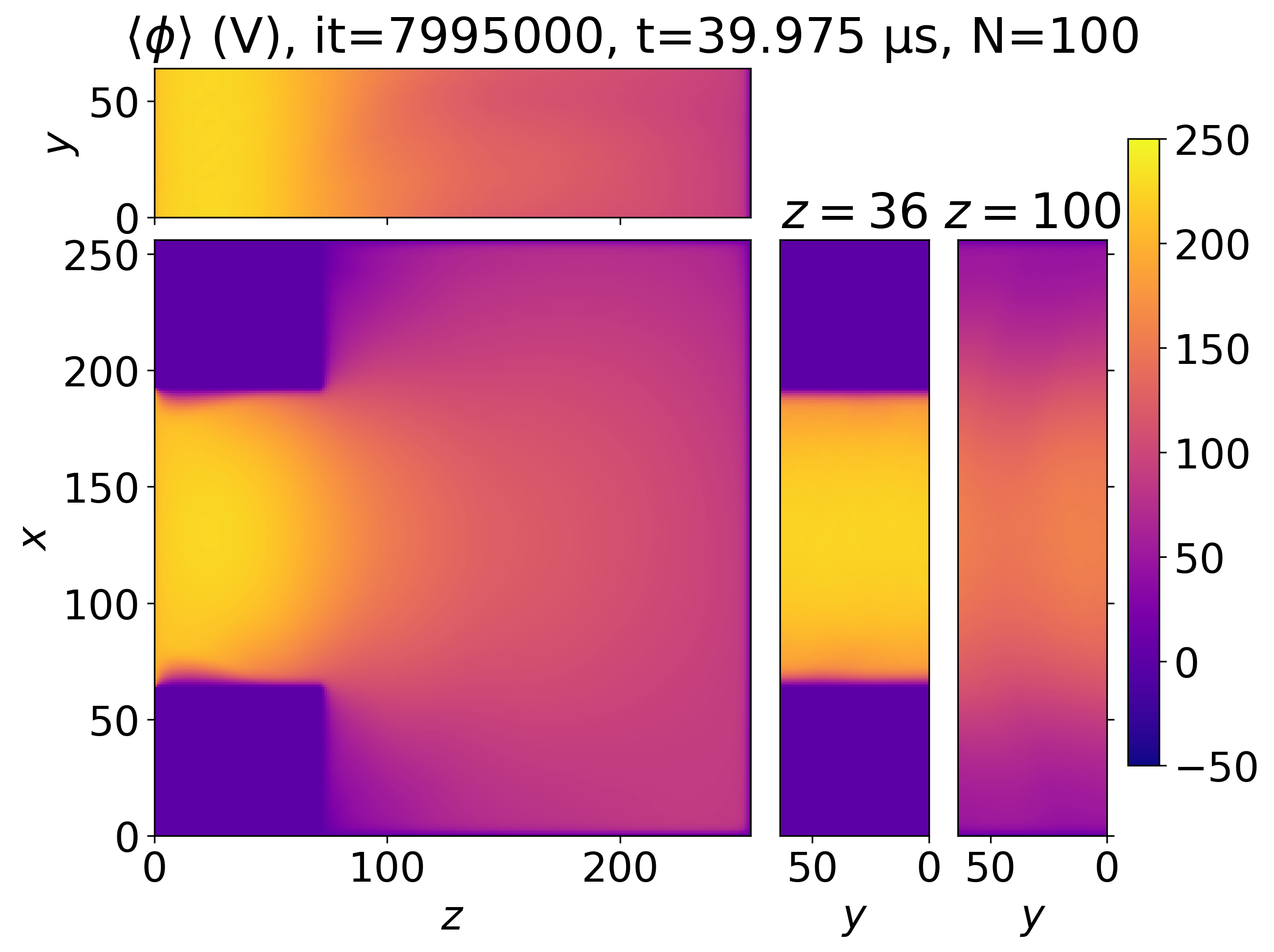}
    \includegraphics[width=0.32\linewidth]{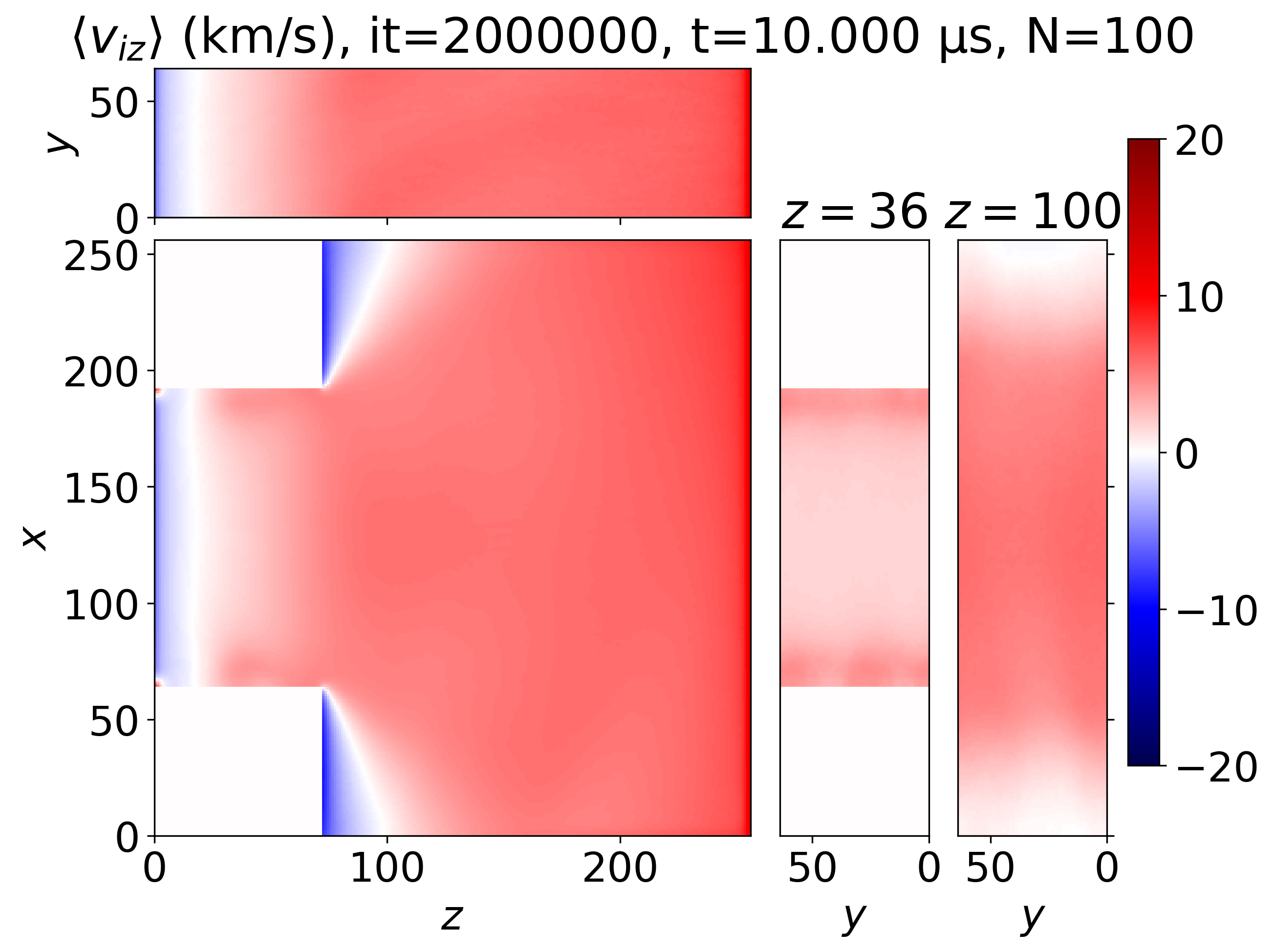}
    \includegraphics[width=0.32\linewidth]{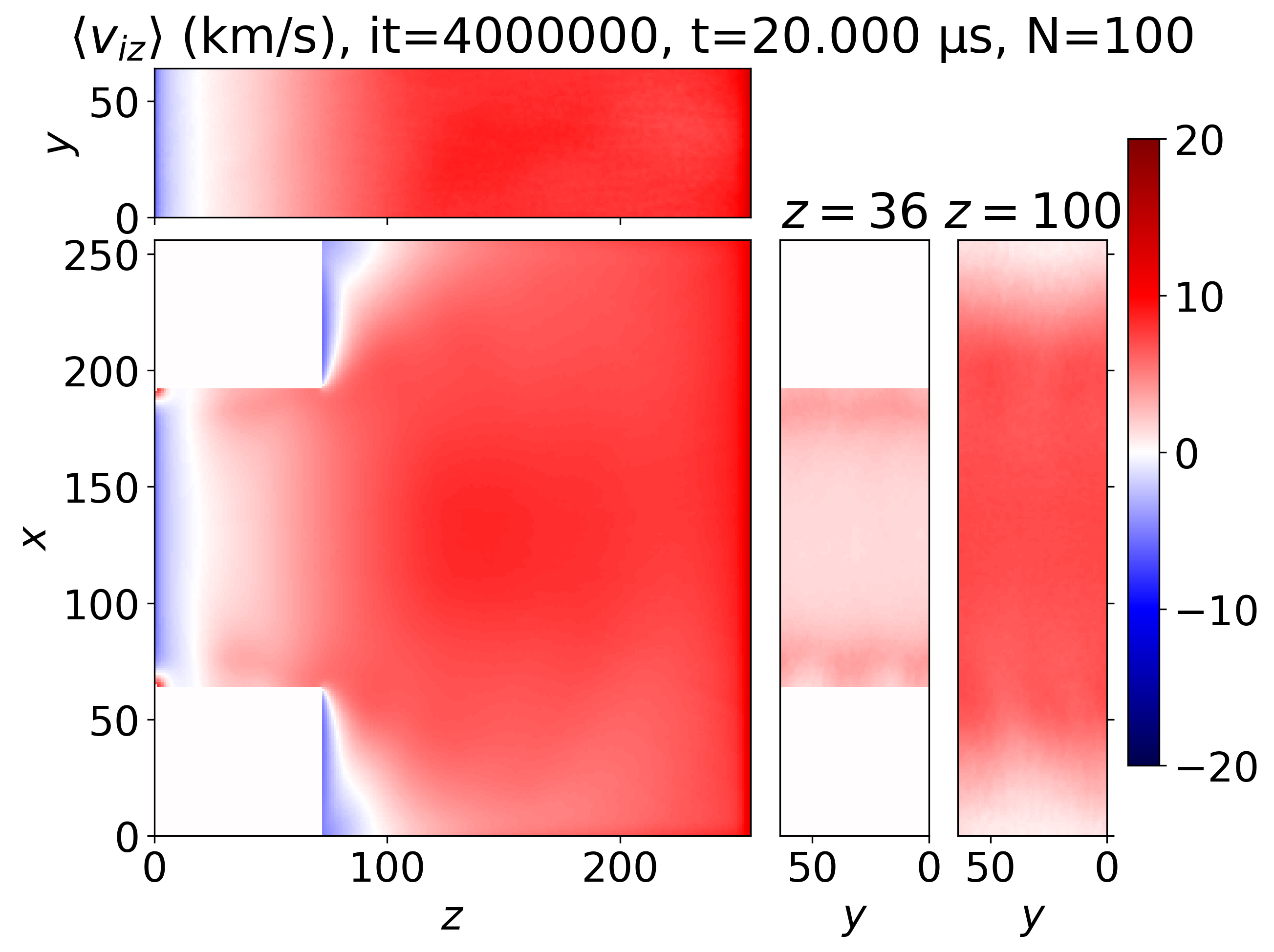}
    \includegraphics[width=0.32\linewidth]{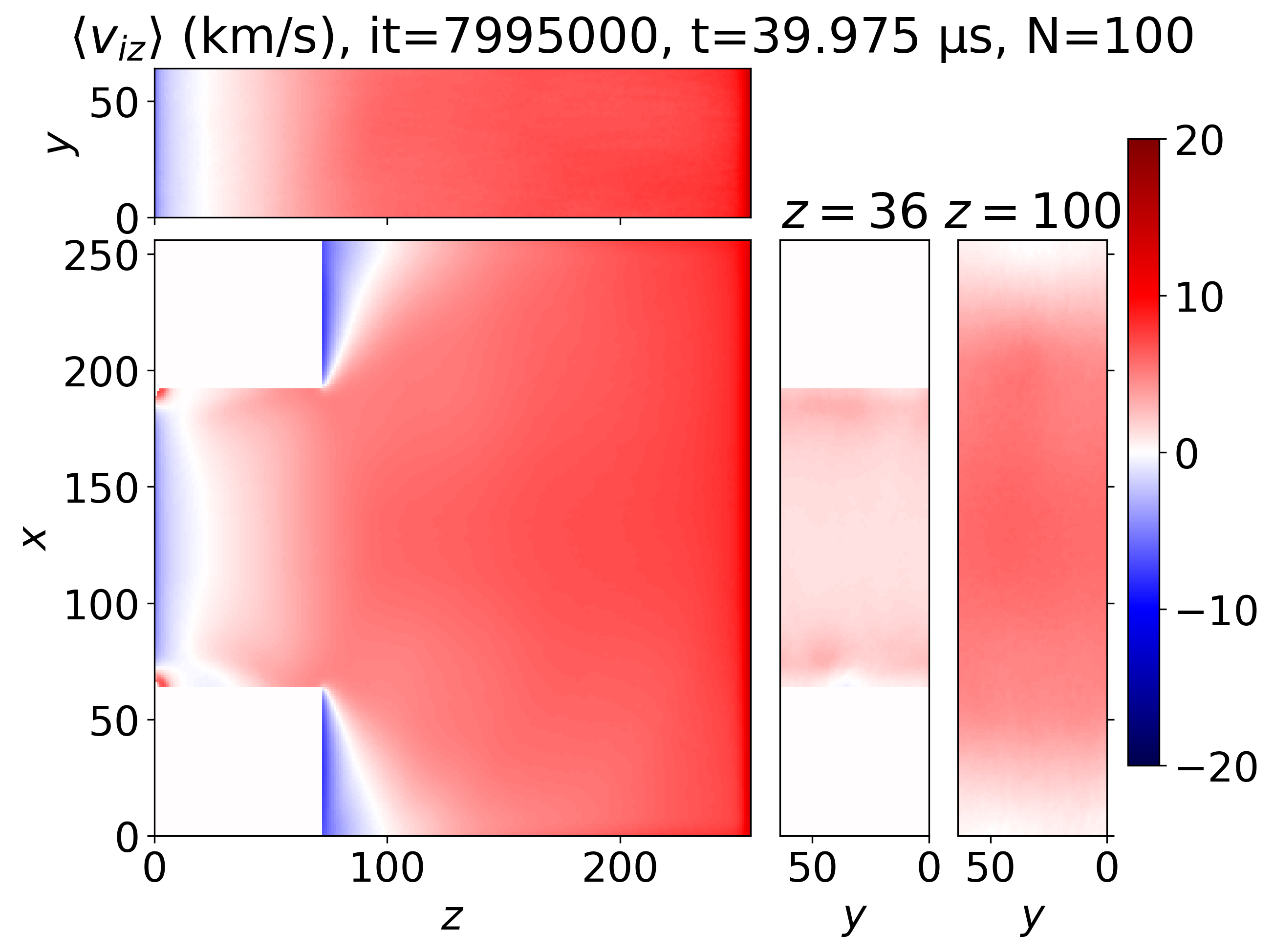}
    \includegraphics[width=0.32\linewidth]{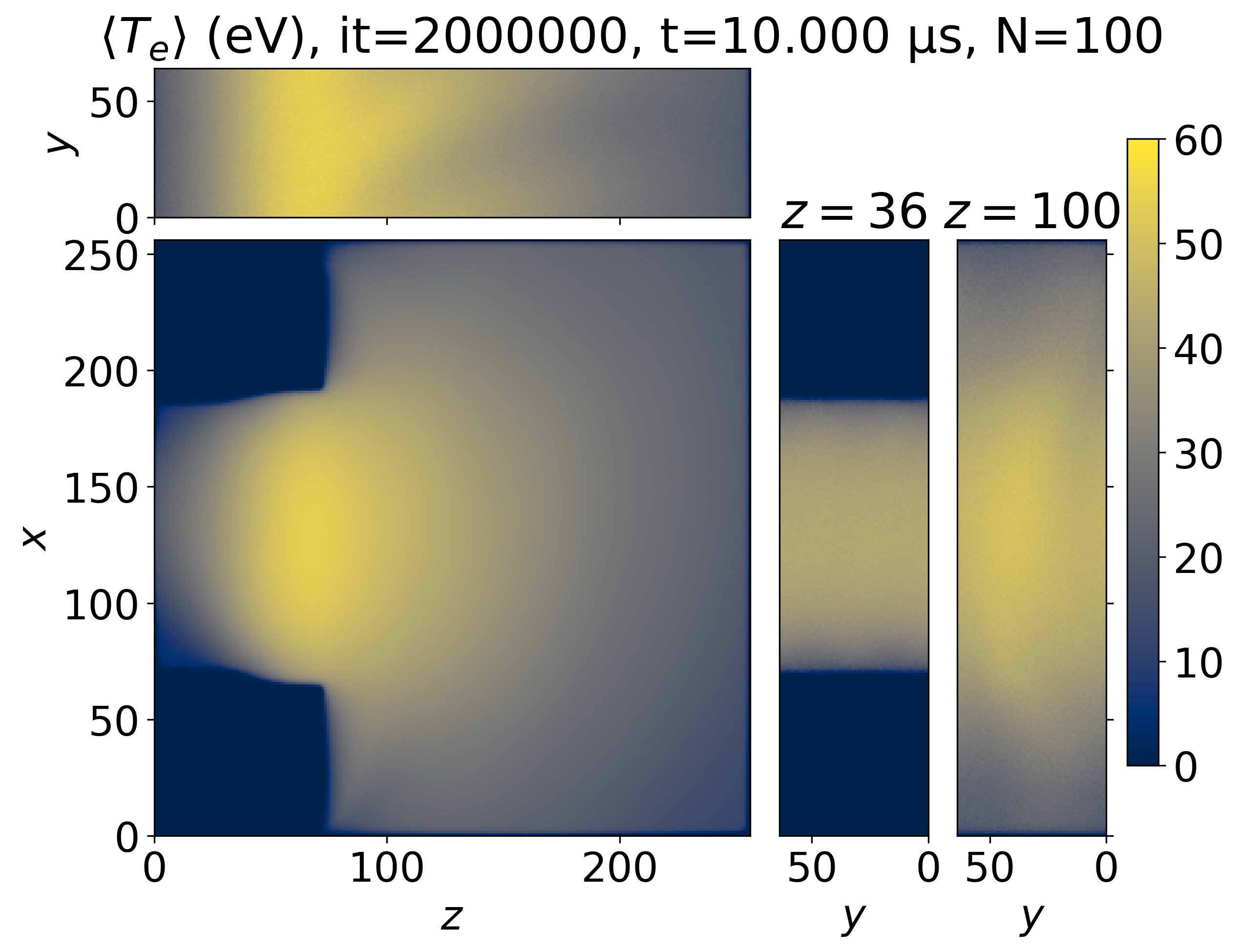}
    \includegraphics[width=0.32\linewidth]{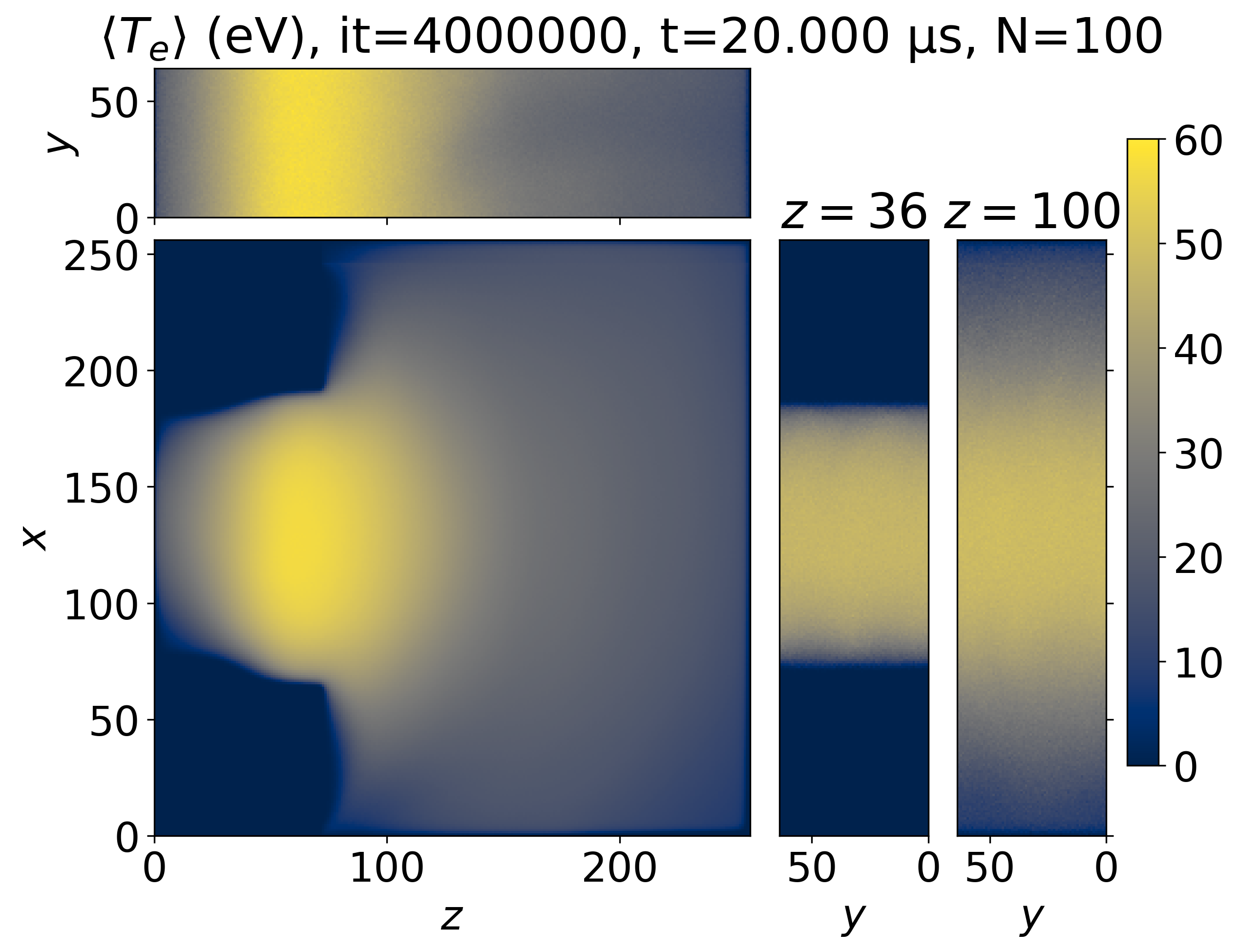}
    \includegraphics[width=0.32\linewidth]{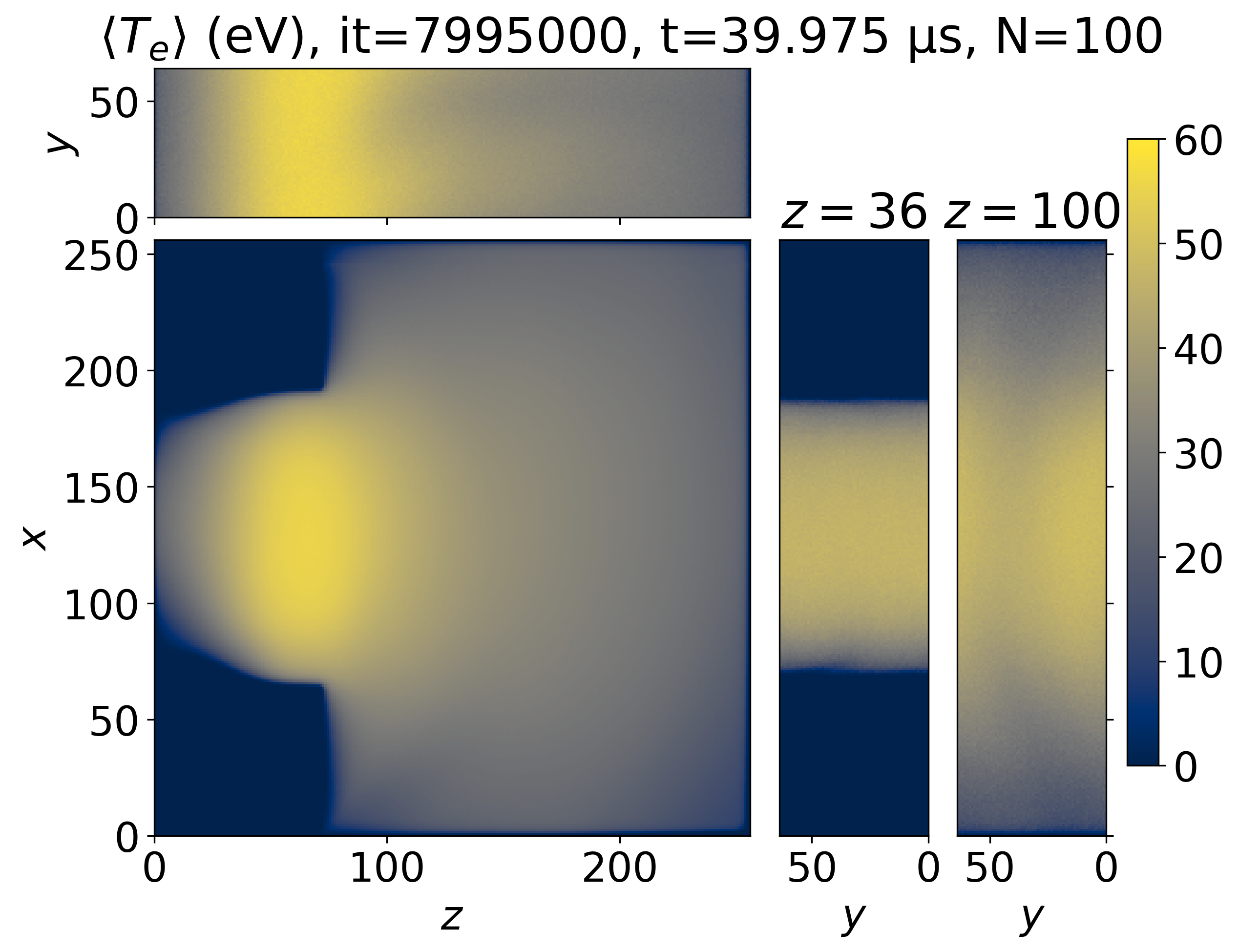}
    \includegraphics[width=0.32\linewidth]{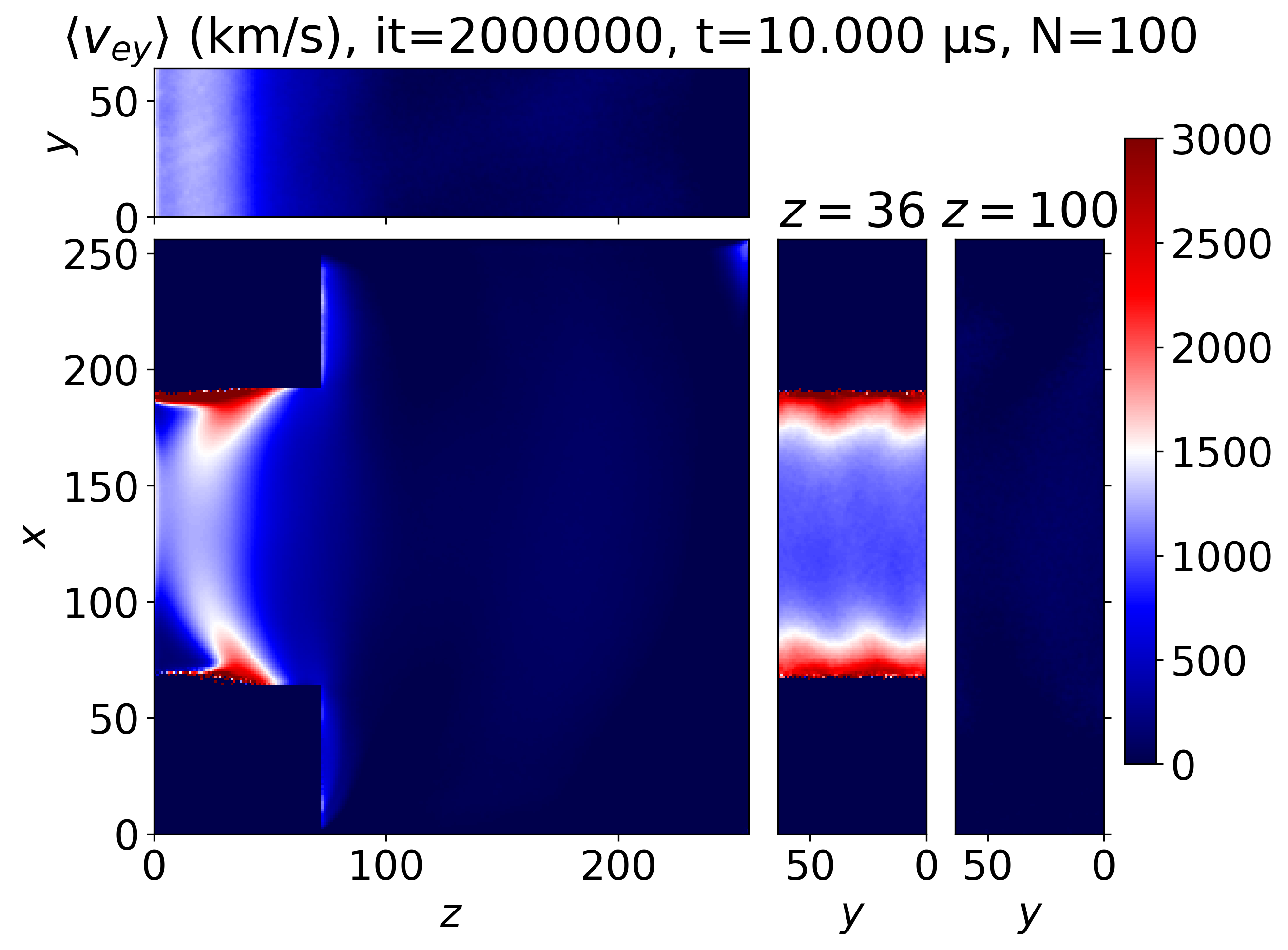}
    \includegraphics[width=0.32\linewidth]{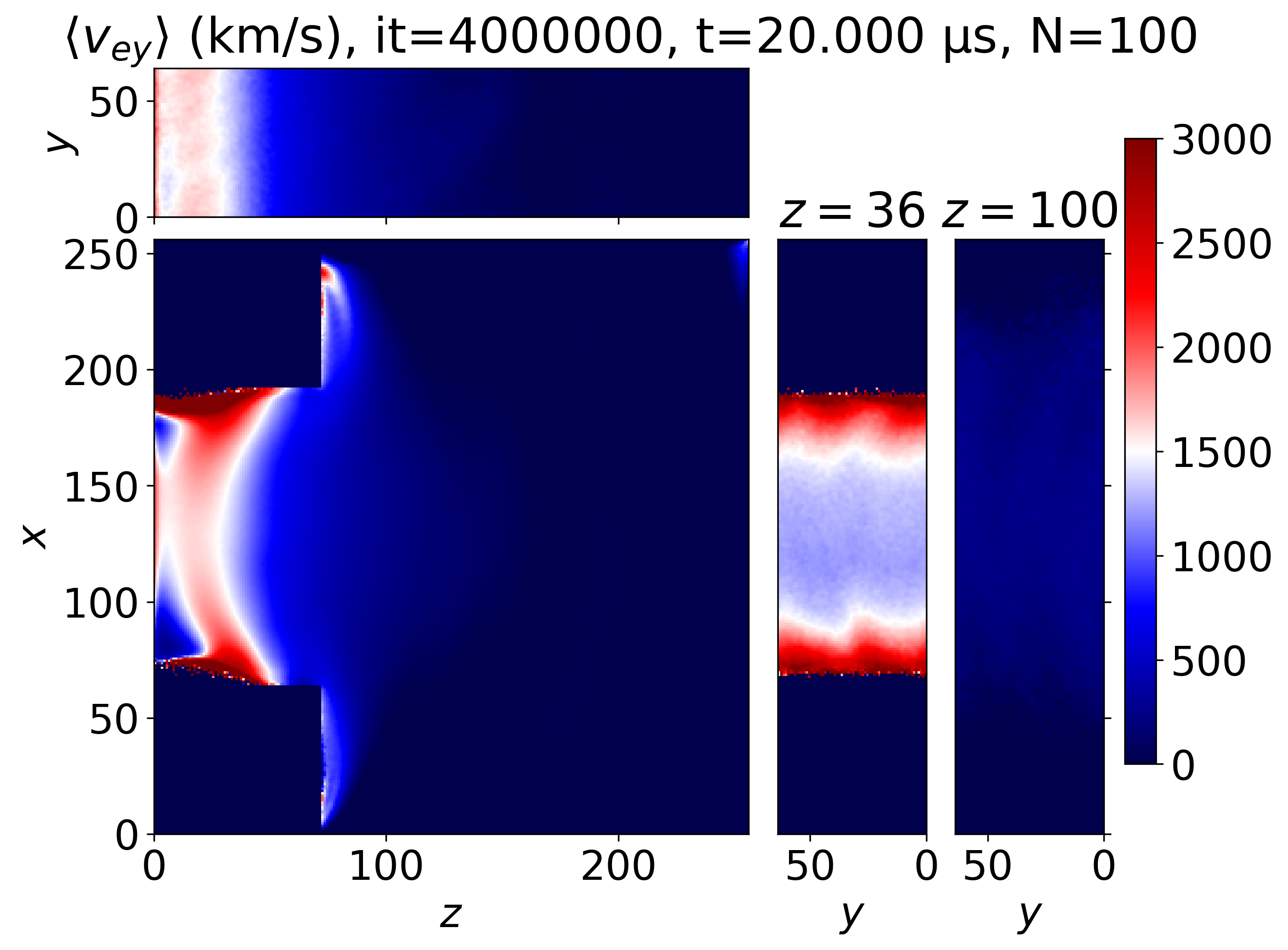}
    \includegraphics[width=0.32\linewidth]{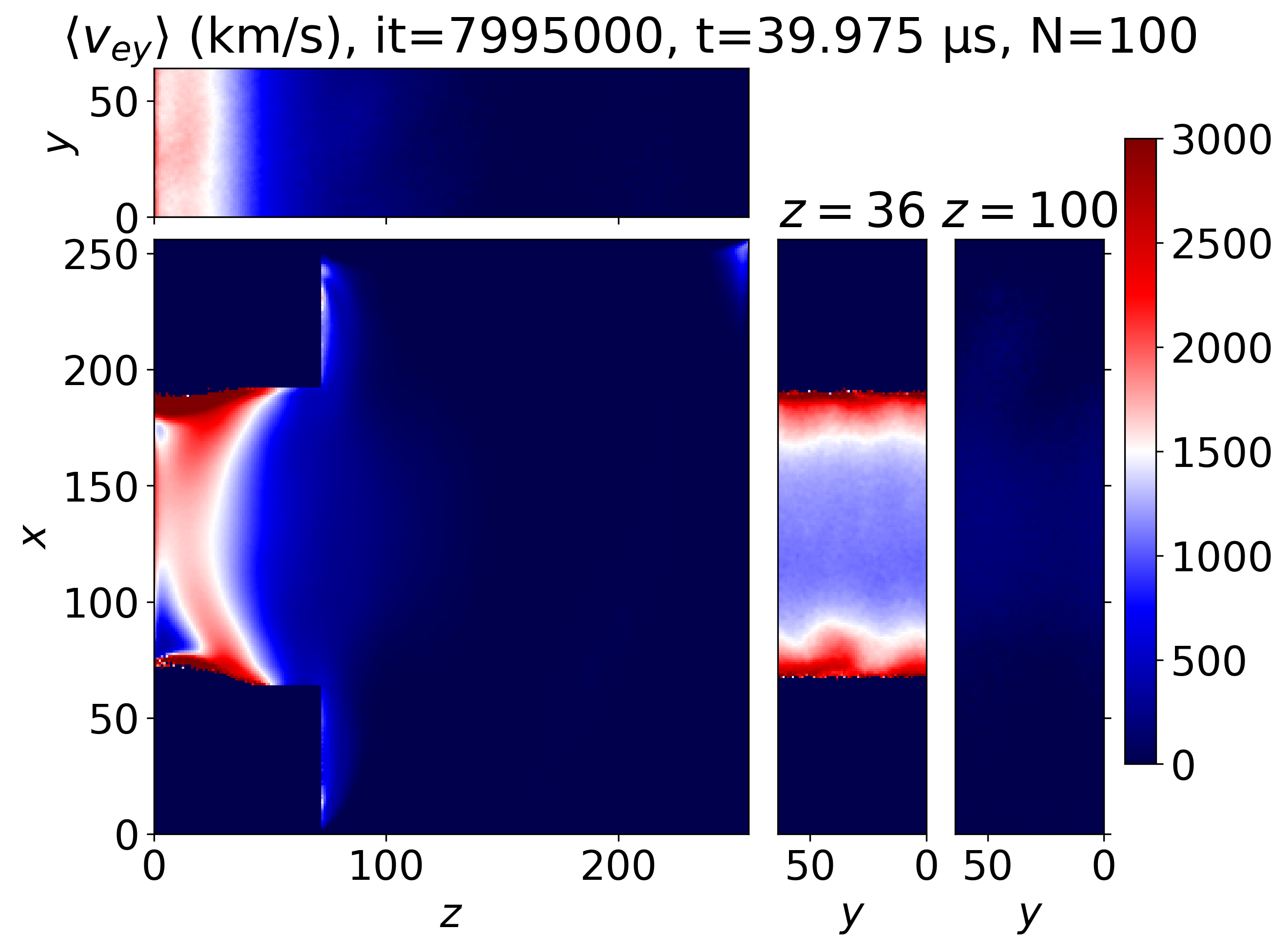}
    \caption{
Time-averaged background fields of Case~D at three representative stages of the slow global evolution: $t=10$, 20, and $40~\mu\mathrm{s}$ (left to right).
From top to bottom, the panels show ion number density $n_i$, electric potential $\phi$, axial ion velocity $v_{iz}$, electron temperature $T_e$, and azimuthal electron velocity $v_{ey}$.
Each snapshot includes mid-plane $z$--$y$ and $z$--$x$ cuts, together with $y$--$x$ slices at $z=36$ (inside the channel) and $z=100$ (near-exit plume).
The fields are averaged using $N=100$ samples
at every 1000 time steps; the $z$--$x$ panels are additionally averaged over the periodic $y$ direction.
Coordinates are reported in grid-index units.
}
\label{fig:Tripanel_tavg}
\end{figure*}

\subsection{Time-Averaged Field Geometry}
\label{sec:avg_geometry}

Having identified the slow global modulation and the representative late-time analysis window in Sec.~\ref{sec:analysis_window}, we next examine the time-averaged background discharge structure.
The purpose of this subsection is not yet to quantify anomalous transport itself, but rather to establish the slowly varying field geometry on which the high-frequency instability develops.
In particular, Fig.~\ref{fig:Tripanel_tavg} shows that the discharge does not evolve as a spatially uniform state over the breathing-like cycle.
Instead, the bulk plasma density, potential drop, ion acceleration, and electron heating remain organized around the downstream half of the channel and the near-exit region, whereas the mean azimuthal electron drift follows a different spatial distribution and is strongest near the walls in the upstream half of the channel.
Taken together, these fields define the background configuration within which the later instability-driven transport pathway emerges.
In the following, Case~D is used as the representative example, since the other cases exhibit broadly similar time-averaged macroscopic field distributions.

Fig.~\ref{fig:Tripanel_tavg} presents time-averaged snapshots of Case~D at three representative stages of the slow evolution: $t=10~\mu\mathrm{s}$, corresponding to the early high-density stage; $t=20~\mu\mathrm{s}$, corresponding to the depleted stage; and $t=40~\mu\mathrm{s}$, corresponding to the late, locally quasi-steady stage used as the main reference below.
From top to bottom, the figure shows the ion number density $n_i$, electric potential $\phi$, axial ion velocity $v_{iz}$, electron temperature $T_e$, and azimuthal electron velocity $v_{ey}$.
The fields are displayed using a tripanel layout consisting of a $z$--$y$ cut at the mid-plane in $x$, a $z$--$x$ cut at the mid-plane in $y$, and two $y$--$x$ slices at $z=36$ and 100, representing an in-channel plane and a near-exit plume plane, respectively.
These maps reveal how the slow breathing-like evolution modulates the overall discharge level while leaving the main spatial skeleton of the discharge largely unchanged.

The ion-density maps show that the discharge is sustained by a dense plasma core located in the downstream half of the channel, which then expands into the near-field plume after exiting the thruster.
Across the three stages, the absolute density level follows the slow modulation identified in Fig.~\ref{fig:ni_t_cases}: the plasma is strongest at $t=10~\mu\mathrm{s}$, becomes depleted at $t=20~\mu\mathrm{s}$, and recovers to an intermediate, more repeatable state by $t=40~\mu\mathrm{s}$.
However, despite this global modulation, the spatial organization remains similar.

The potential maps show that the dominant axial potential drop is localized around the thruster exit and extends into the near-field plume.
This indicates that the main acceleration zone is not fully confined inside the channel, but instead straddles the exit region\cite{hargus_cappelli_2001,chaplin_2018_hermes}.
The overall potential level varies over the slow evolution, being elevated at $t=10~\mu\mathrm{s}$ and reduced at $t=20~\mu\mathrm{s}$, before becoming more stable by $t=40~\mu\mathrm{s}$.
Nevertheless, the location of the principal potential gradient remains anchored near the exit.
This persistent localization of the axial electric field is important for the later analysis, because it sets the background field geometry experienced by the electrons.

The axial ion velocity $v_{iz}$ further confirms the role of this near-exit potential structure.
A high-$\langle v_{iz}\rangle$ jet emerges downstream of the strongest axial potential drop and persists into the near-field plume, whereas the upstream channel remains comparatively weakly accelerated.
As expected, the overall magnitude of $\langle v_{iz}\rangle$ varies with the slow discharge state, being higher at $t=10~\mu\mathrm{s}$ and lower at $t=20~\mu\mathrm{s}$, but its spatial pattern remains largely tied to the same near-exit region.


\begin{figure*}[!htbp]
    \centering
    \includegraphics[width=0.8\linewidth]{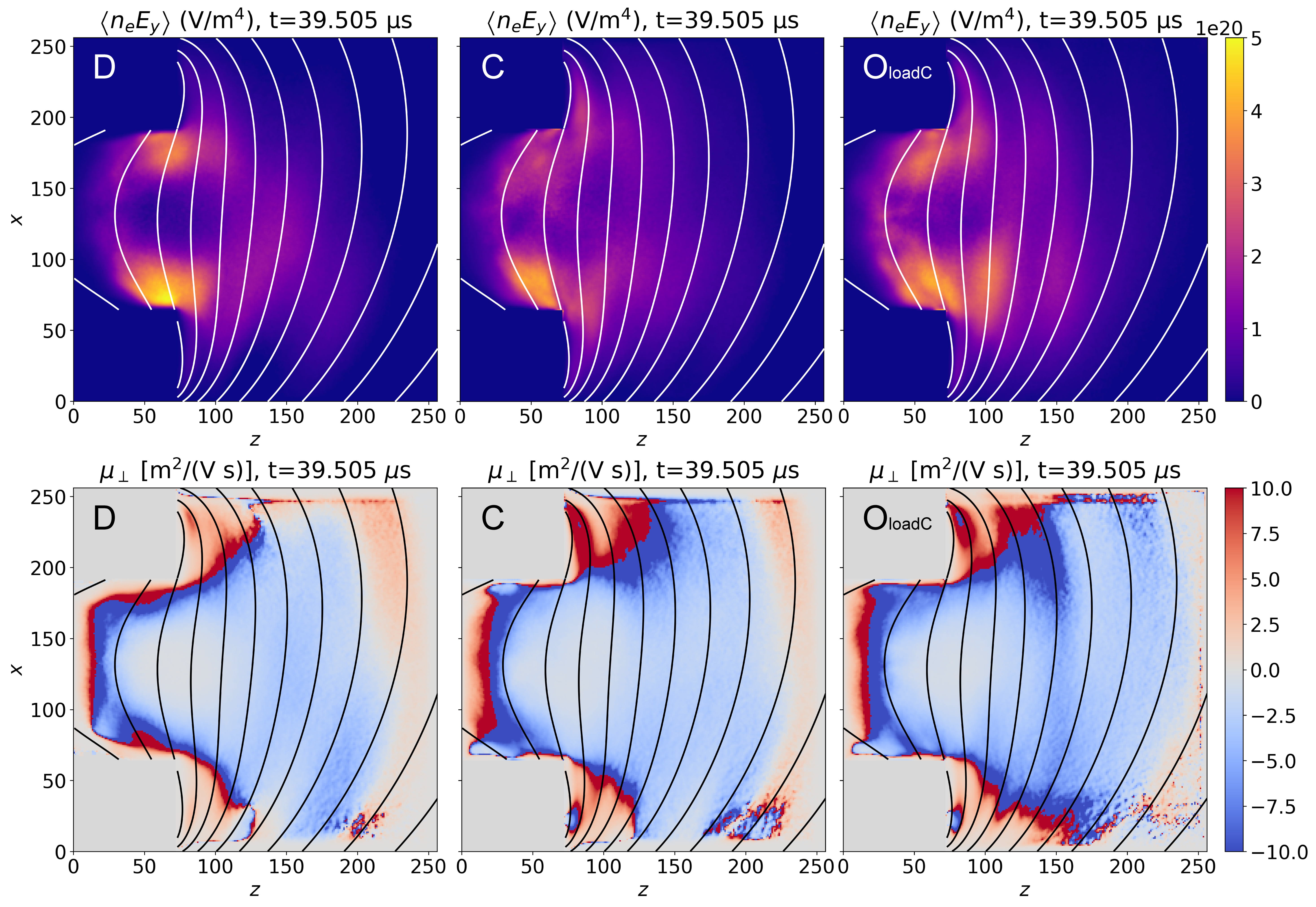}
\caption{
Mapping of the net anomalous electron transport pathway on the $z$--$x$ plane at a representative late-time stage for Cases~D, C, and O$_{\mathrm{loadC}}$ (left to right).
Top row: time- and azimuthally-averaged correlation term $\langle n_e E_y\rangle$.
Bottom row: corresponding effective perpendicular mobility $\mu_{\perp}$ defined by Eqs.~\eqref{eq:muperp_def}--\eqref{eq:Eperp_def}.
The averages use 200 samples over the late-time analysis window to filter out the rapid EDI oscillations and reveal the persistent instability-driven contribution to cross-field electron transport.
Coordinates are reported in grid-index units.
Negative $\mu_{\perp}$ indicates electron transport directed toward the anode across the magnetic field under the present sign convention.
}
    \label{fig:nE}
\end{figure*}

The electron-temperature maps exhibit a broadly similar spatial structure throughout the three stages.
A hot-electron region is concentrated in the downstream channel and near the exit, where the potential gradient is strongest.
Downstream of the exit, $T_e$ remains elevated in the near-field plume and then decreases gradually with axial distance as the plasma expands.
Compared with the larger modulation seen in $n_i$ and $\phi$, the temporal variation of $T_e$ is more moderate in amplitude, but it exhibits an opposite trend relative to the ion density.
Specifically, when the plasma density is high, the electron temperature becomes somewhat lower, consistent with enhanced ionization activity that consumes electron energy.
By contrast, during the low-density stage, $T_e$ becomes higher because ionization is weaker and the discharge is effectively waiting for neutral replenishment.
As the neutral population recovers and the plasma density rises again, the electron temperature correspondingly decreases.
These results indicate that, although the main electron-heating zone remains anchored near the downstream channel and exit region, its intensity responds dynamically to the slower breathing-like cycle through the competition between electron heating, ionization loss, and neutral replenishment.

The azimuthal electron velocity maps exhibit a distinct spatial organization from the fields discussed above.
The maps in Fig.~\ref{fig:Tripanel_tavg} show that $\langle v_{ey}\rangle$ is strongest in the upstream half of the channel rather than in the downstream near-exit region.
It is also enhanced near the inner and outer walls, so that the mean azimuthal drift is organized into wall-adjacent regions within the channel.
This behavior is consistent with the fact that a dominant contribution to $\langle v_{ey}\rangle$ comes from the $\mathbf{E}\times\mathbf{B}$ drift.
With $\mathbf{E}=-\nabla\phi$ and $\mathbf{B}=(B_x,\,0,\,B_z)$, the azimuthal component of the drift can be written approximately as
\begin{equation}
v_{E,y}\approx \frac{(\mathbf{E}\times\mathbf{B})_y}{B^2}
=\frac{E_zB_x-E_xB_z}{B^2}.
\label{eq:vEy_ExB_components}
\end{equation}
Near the walls, the second term, $-E_xB_z/B^2$, can become important.
In Case~D, the wall-adjacent sheath produces a substantial normal electric field $E_x$, while the fringe magnetic field near the exit introduces a finite $B_z$ component.
As a result, the mean azimuthal drift is redistributed into wall-adjacent regions.
Because $E_x$ changes sign across opposite walls, this second term need not enhance the signed drift in the same direction on both sides; rather, its main effect is to reorganize the drift structure near the walls.

Over the three stages shown in Fig.~\ref{fig:Tripanel_tavg}, the temporal variation of $\langle v_{ey}\rangle$ differs from the more obvious breathing-cycle modulation seen in the other fields.
In particular, the mean azimuthal drift at $t=10~\mu\mathrm{s}$ is not the strongest.
By $t=20~\mu\mathrm{s}$, $\langle v_{ey}\rangle$ in the channel-center region becomes noticeably stronger than at $t=10~\mu\mathrm{s}$, while the wall-adjacent high-drift layers remain present.
The distribution at $t=40~\mu\mathrm{s}$ is broadly similar to that at $t=20~\mu\mathrm{s}$, indicating that the mean azimuthal drift evolves mainly through a redistribution of intensity within the channel rather than through a simple monotonic modulation.
These time-averaged maps therefore show that the mean azimuthal electron drift has a distinct temporal behavior from the other averaged fields, even though its overall spatial organization remains robust.

We next turn to the corresponding transport diagnostics and show that, once the rapid oscillations are averaged out, the net anomalous electron transport is organized into persistent near-wall pathways rather than being distributed uniformly across the channel cross section.

\subsection{Mapping Near-Wall Pathways of Anomalous Electron Transport}
\label{sec:transport_pathway}

Having established in Sec.~\ref{sec:avg_geometry} that the time-averaged discharge structure provides a nonuniform background field geometry for the instability, we now turn to the central result of this work: the spatial mapping of the net anomalous electron transport pathway.
Because the instantaneous fields are strongly oscillatory, the pathway cannot be inferred directly from a single snapshot.
Instead, it must be extracted from a time- and azimuthally-averaged diagnostic that filters out the rapid EDI oscillations while preserving their nonzero net contribution to cross-field transport.
For this purpose, Fig.~\ref{fig:nE} presents the correlation term $\langle n_e E_y\rangle$ on the $z$--$x$ plane for Cases~D, C, and O$_{\text{loadC}}$ at a representative late-time stage, using 200 samples over the analysis window.
This quantity provides a direct measure of the net instability-driven correlation between density and azimuthal electric-field fluctuations and therefore serves as a compact map of where the anomalous transport is most strongly organized.

To interpret the same result in a more familiar transport form, the bottom row of Fig.~\ref{fig:nE} also shows the corresponding effective perpendicular mobility $\mu_{\perp}$, defined as
\begin{equation}
\mu_{\perp}\equiv
-\frac{\langle n_e E_y\rangle}
{B\,\langle n_e\rangle\,\langle E_{\perp}\rangle},
\label{eq:muperp_def}
\end{equation}
with
\begin{equation}
E_{\perp}\equiv \frac{B_x E_z - B_z E_x}{B},
\qquad
B=\sqrt{B_x^2+B_z^2}.
\label{eq:Eperp_def}
\end{equation}
Under the present sign convention, negative $\mu_{\perp}$ corresponds to electron transport directed toward the anode across the magnetic field.
The purpose of introducing $\mu_{\perp}$ here is not to replace the fluctuation correlation itself, but to help interpret the same spatial structure in terms of an effective transport coefficient that can be compared more easily with common Hall-thruster transport language.

The most important result in Fig.~\ref{fig:nE} is that all three cases exhibit the same global transport topology.
In the top row, the strongest $\langle n_e E_y\rangle$ does not fill the channel cross section uniformly, nor is it centered in the channel core.
Instead, it is concentrated in two band-like regions located adjacent to the inner and outer walls, slightly upstream of and around the exit region.
These two bands define persistent near-wall pathways of anomalous electron transport.
A broader but weaker region of non-negligible correlation extends into the near-plume, yet the dominant transport signature remains clearly wall-adjacent.


The same conclusion is reinforced by the bottom-row maps of $\mu_{\perp}$.
It should be noted, however, that $\mu_{\perp}$ may become unrealistically large in regions where $\langle n_e\rangle$ and $\langle E_{\perp}\rangle$ are both very small, especially inside the near-wall sheath, while the correlation term $\langle n_e E_y\rangle$ is not negligible.
In such regions, the ratio used to define $\mu_{\perp}$ is strongly magnified and no longer provides a quantitatively reliable measure of transport.
Accordingly, these extreme values are clipped in the color scale.
Outside these regions, although the normalization changes the local contrast and introduces additional dependence on the mean density and background-field geometry, the strongest negative $\mu_{\perp}$ remains concentrated in the same near-wall regions.
The mobility maps therefore do not introduce a different picture; rather, they confirm in a transport-coefficient form that the dominant cross-field conduction is organized into near-wall pathways.
In all three cases, the inner-wall side tends to show a stronger transport signature than the outer-wall side, suggesting that the wall-adjacent transport is asymmetric, likely because of the radial gradient in the background magnetic field, even though the two-pathway topology itself remains robust.

A key implication of Fig.~\ref{fig:nE} is that the near-wall transport pathway is not an artifact of one particular boundary treatment.
Case~D uses conducting Dirichlet walls, Case~C introduces ceramic dielectric walls with self-consistent charging and SEE, and Case~O$_{\text{loadC}}$ further incorporates the outflow treatment.
Despite these substantial changes in boundary closure, the same near-wall transport topology persists.
Therefore, the present 3D PIC results indicate that near-wall localization is a robust property of the instability-driven anomalous transport itself, rather than a special consequence of the conducting-wall approximation.

The differences among the three cases are mainly quantitative and are concentrated near the exit and in the near-plume.
Comparing Cases~D and C, the principal change is a redistribution of both $\langle n_e E_y\rangle$ and $\mu_{\perp}$ near the wall surface at and slightly downstream of the thruster exit.
The overall near-wall pathway inside the channel remains similar, but the ceramic wall treatment modifies how the transport couples to the exit and downstream plasma.
Case~O$_{\text{loadC}}$ remains close to Case~C in the channel interior, while displaying a somewhat stronger transport signature in the near-plume.
This suggests that the open outflow treatment primarily affects the downstream extension and strength of the pathway, rather than creating or eliminating the pathway itself.


\begin{figure*}[htbp]
    \centering
    \includegraphics[width=0.32\linewidth]{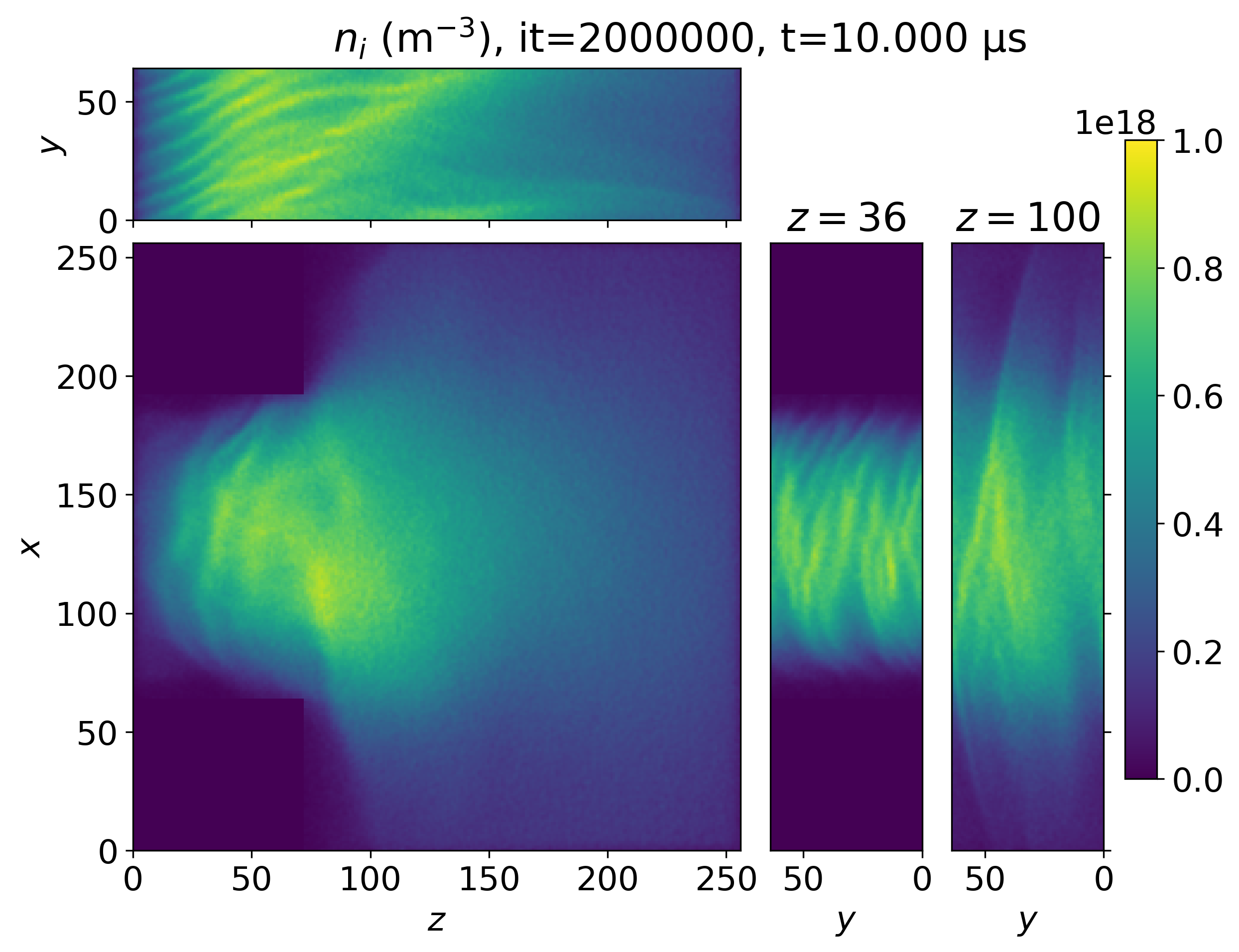}
    \includegraphics[width=0.32\linewidth]{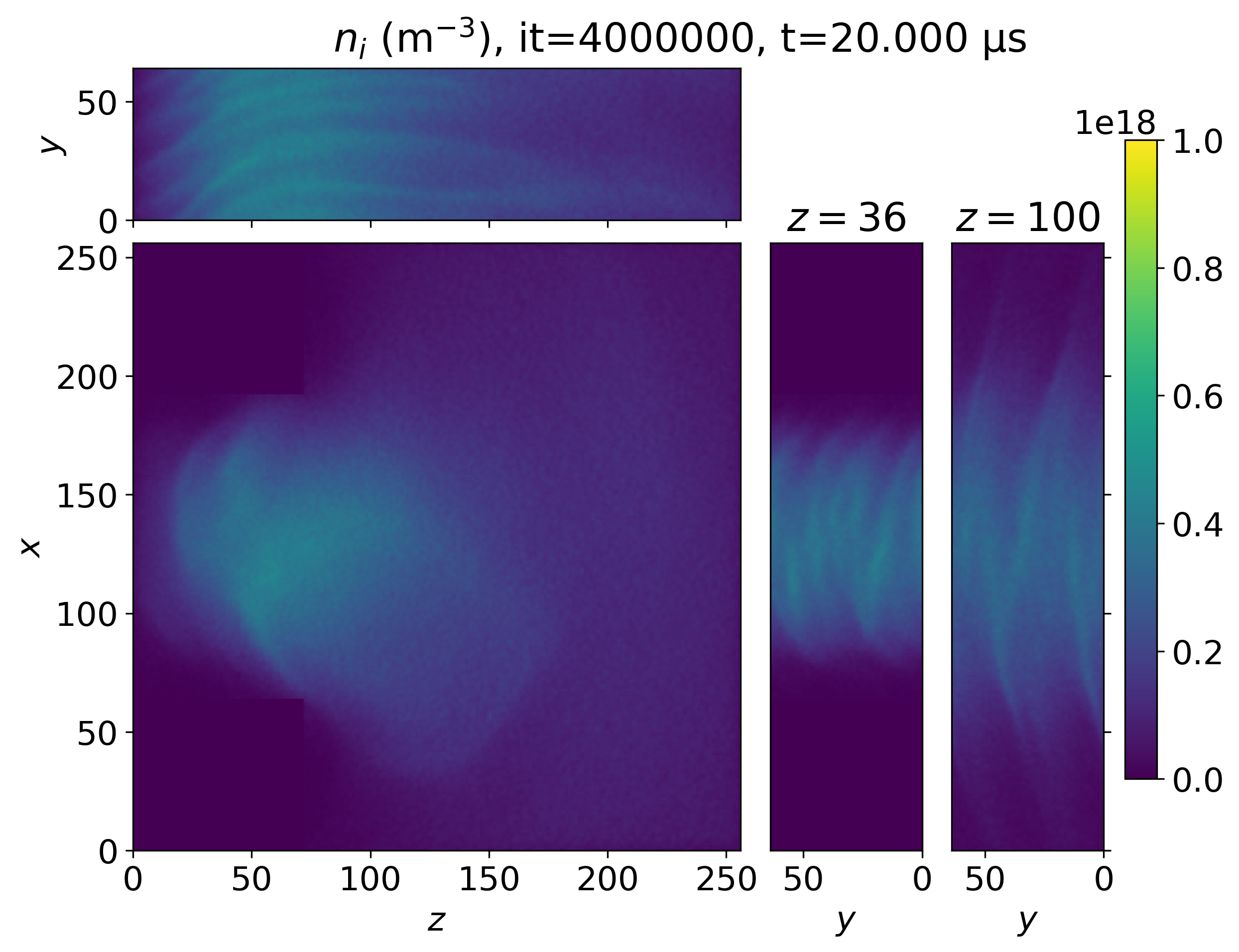}
    \includegraphics[width=0.32\linewidth]{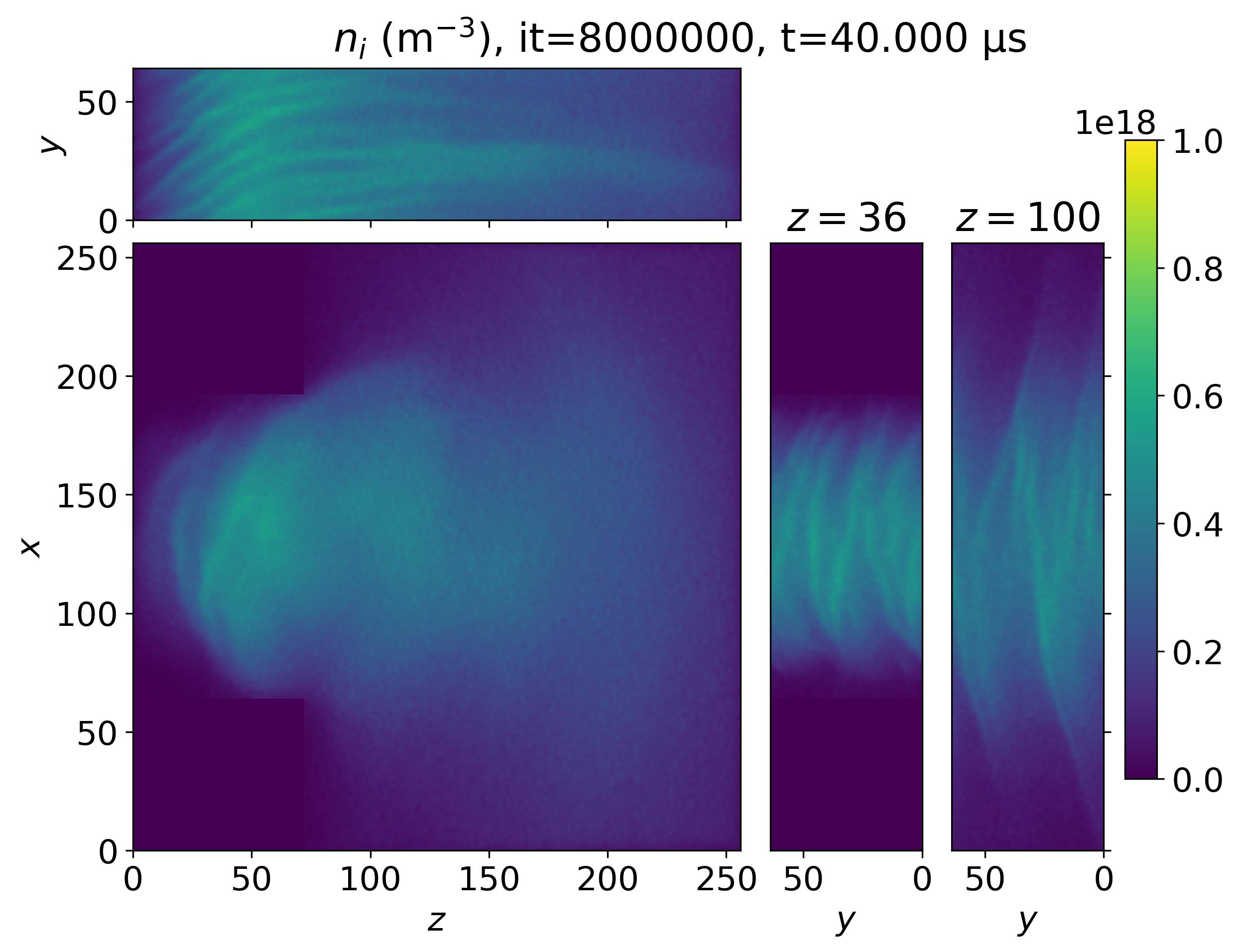}
    \includegraphics[width=0.32\linewidth]{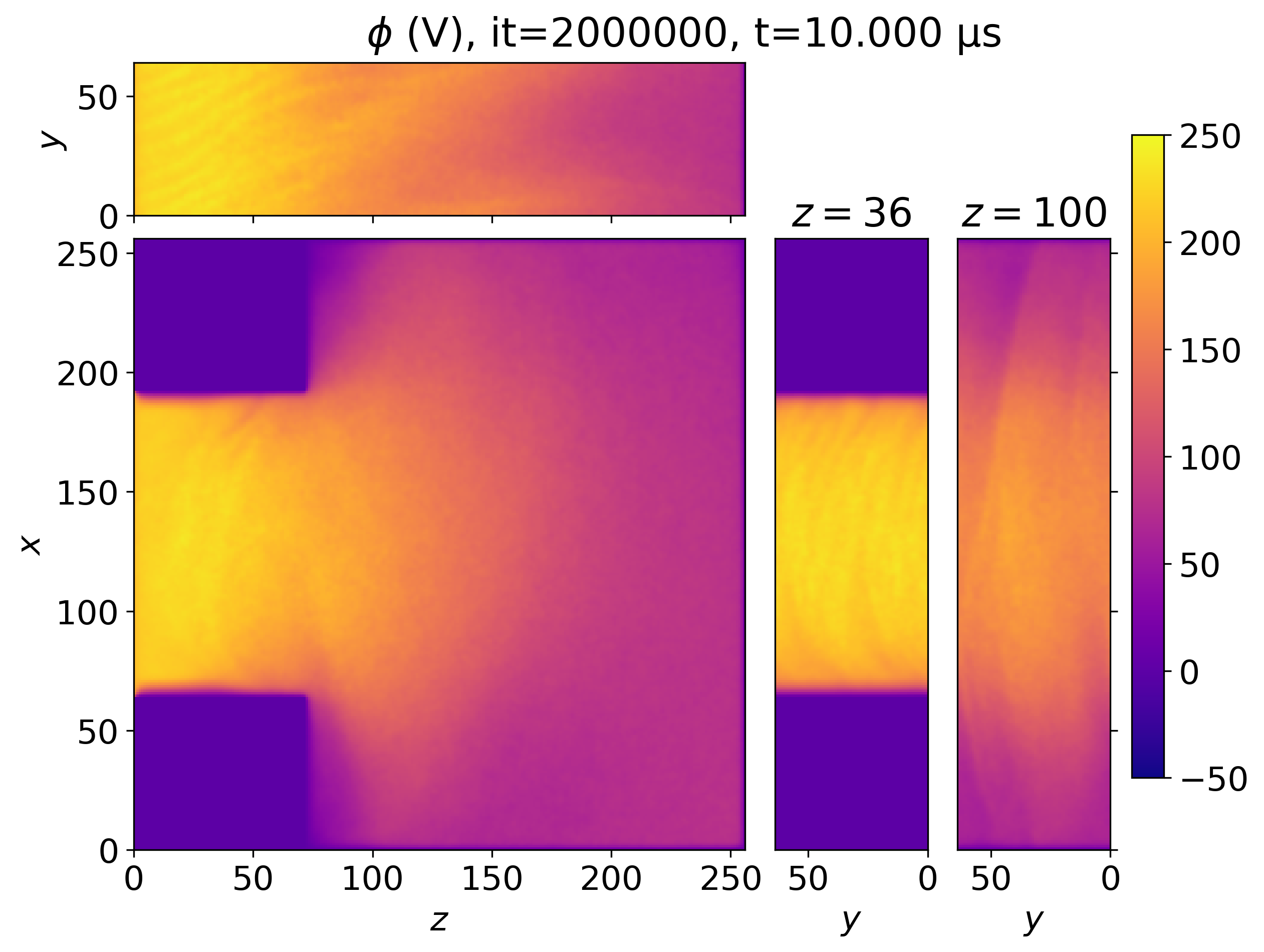}
    \includegraphics[width=0.32\linewidth]{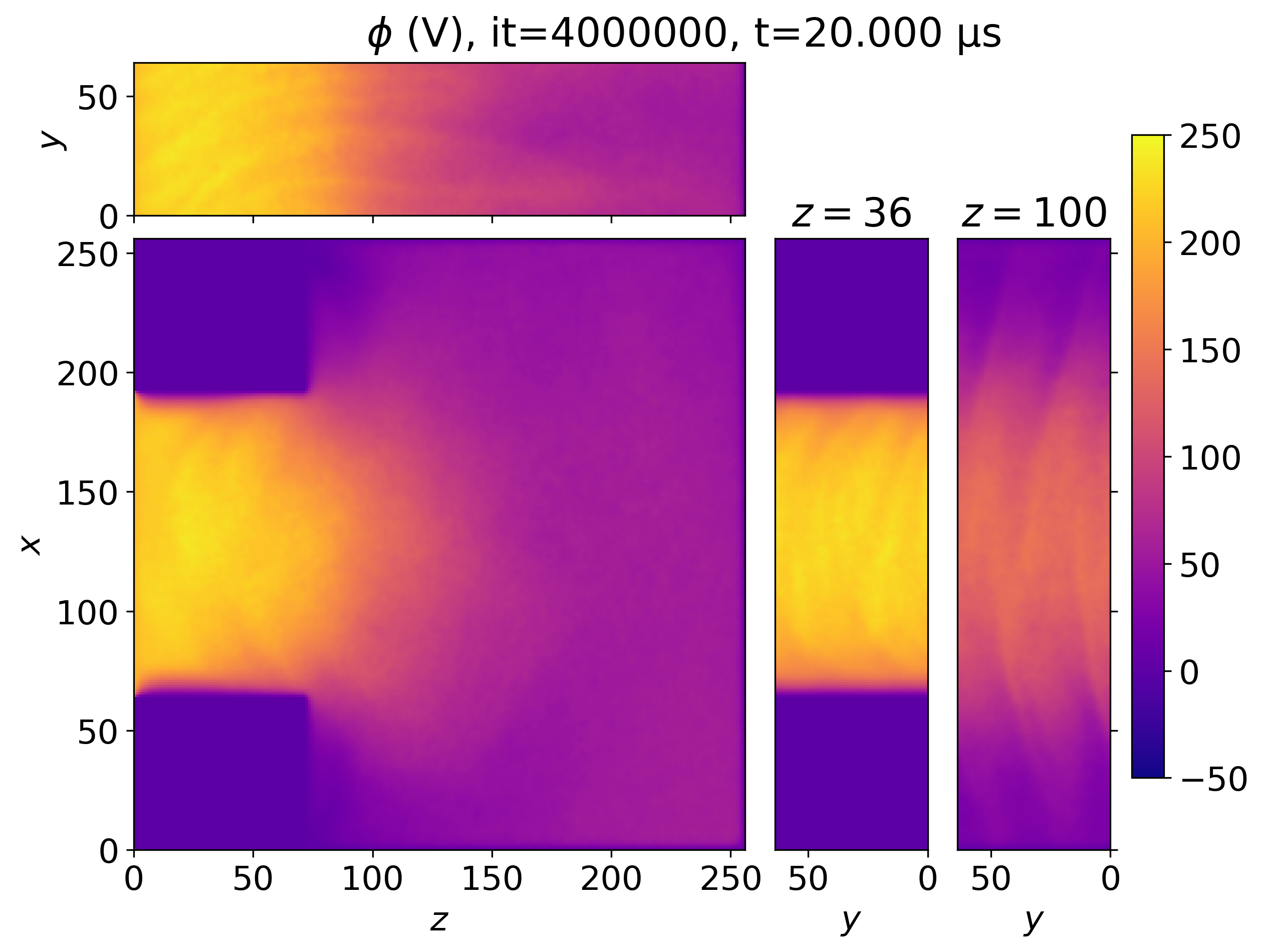}
    \includegraphics[width=0.32\linewidth]{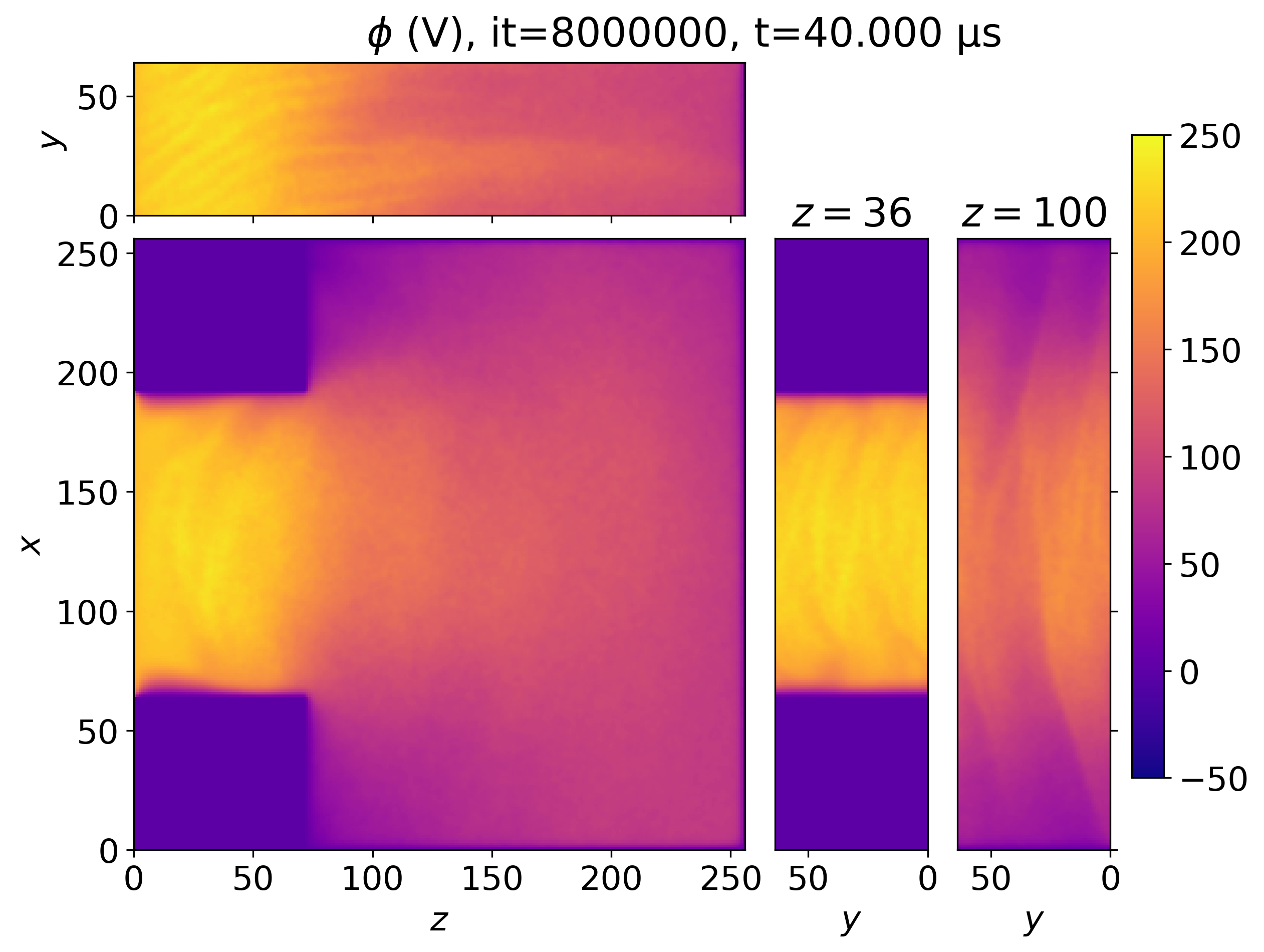}
    \includegraphics[width=0.32\linewidth]{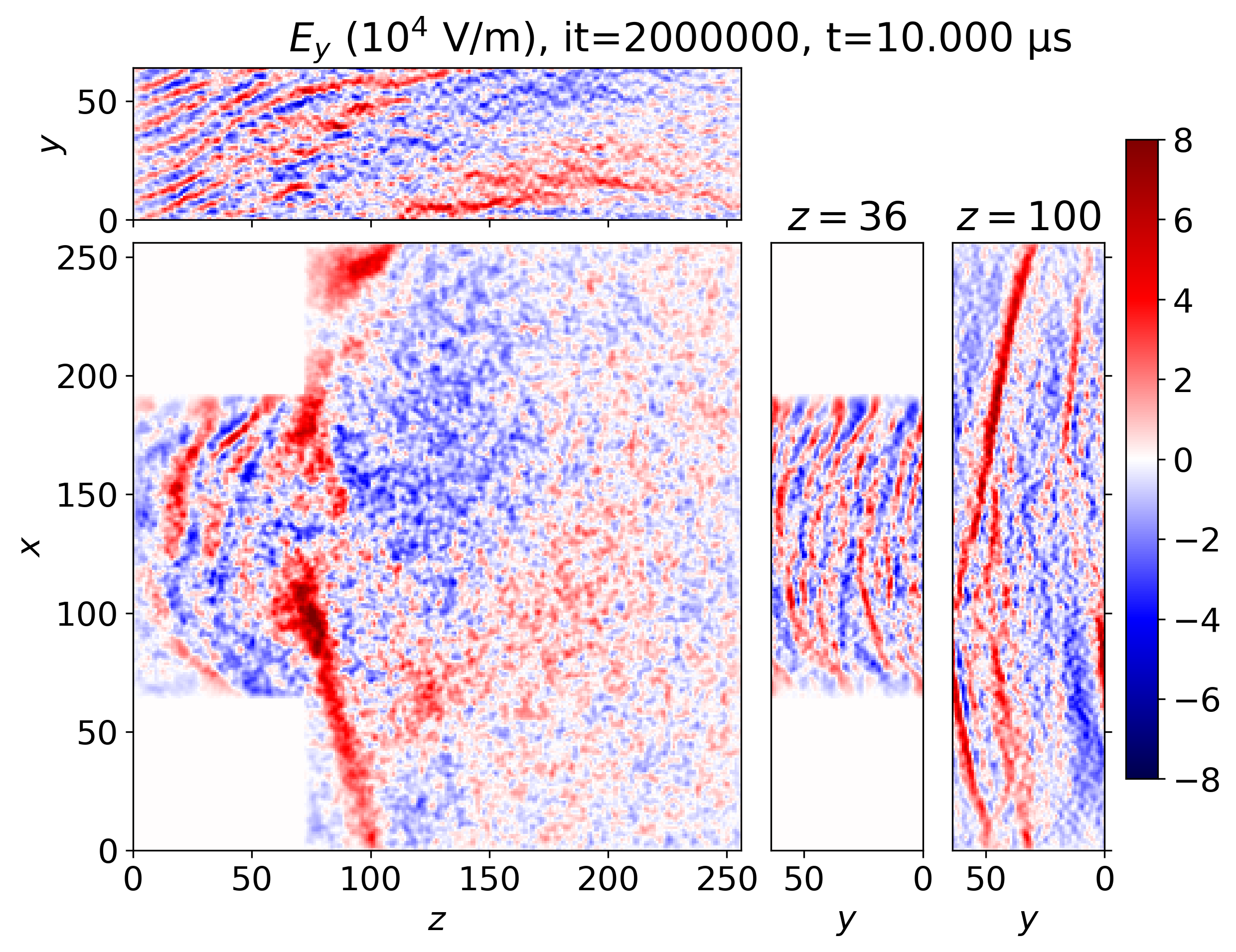}
    \includegraphics[width=0.32\linewidth]{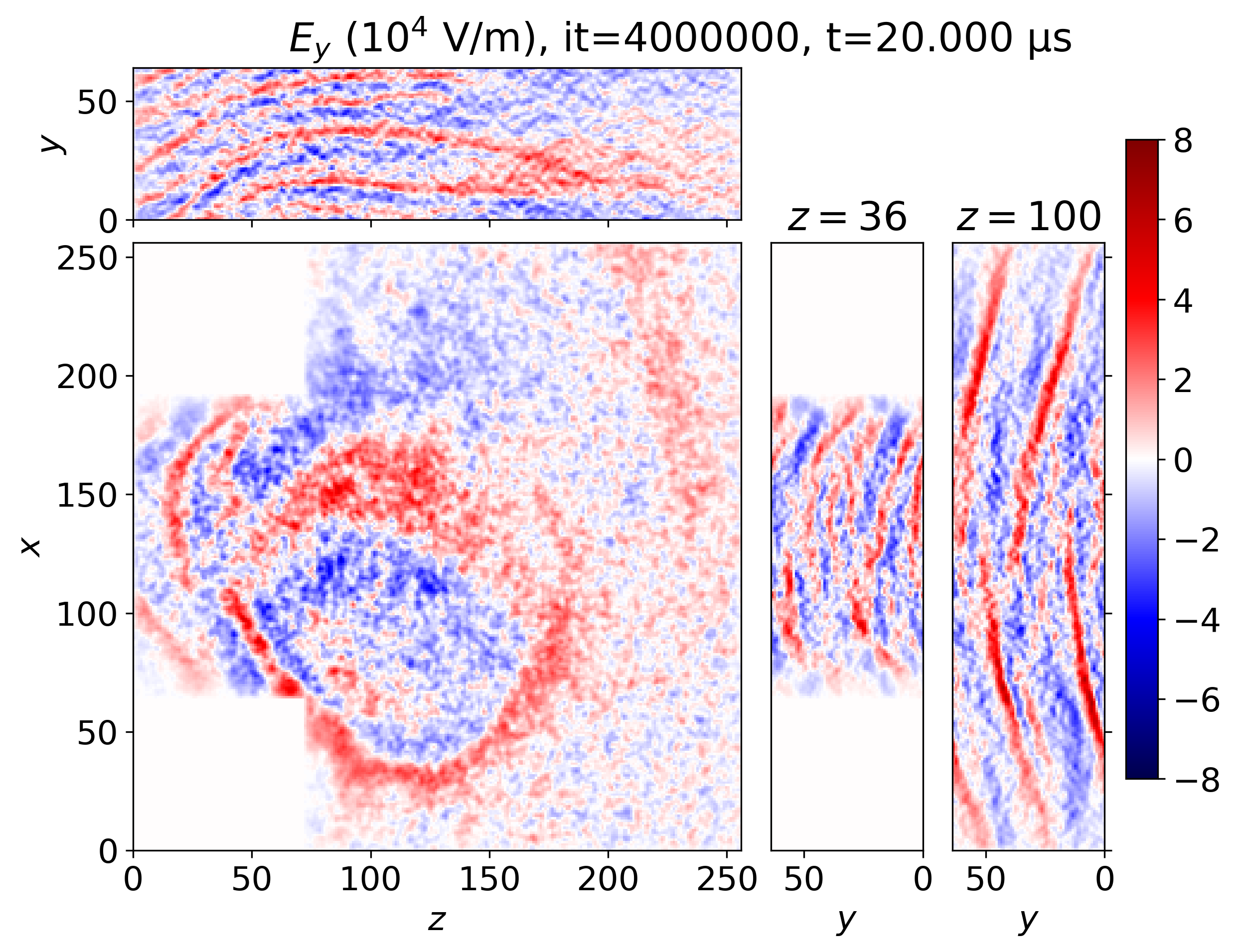}
    \includegraphics[width=0.32\linewidth]{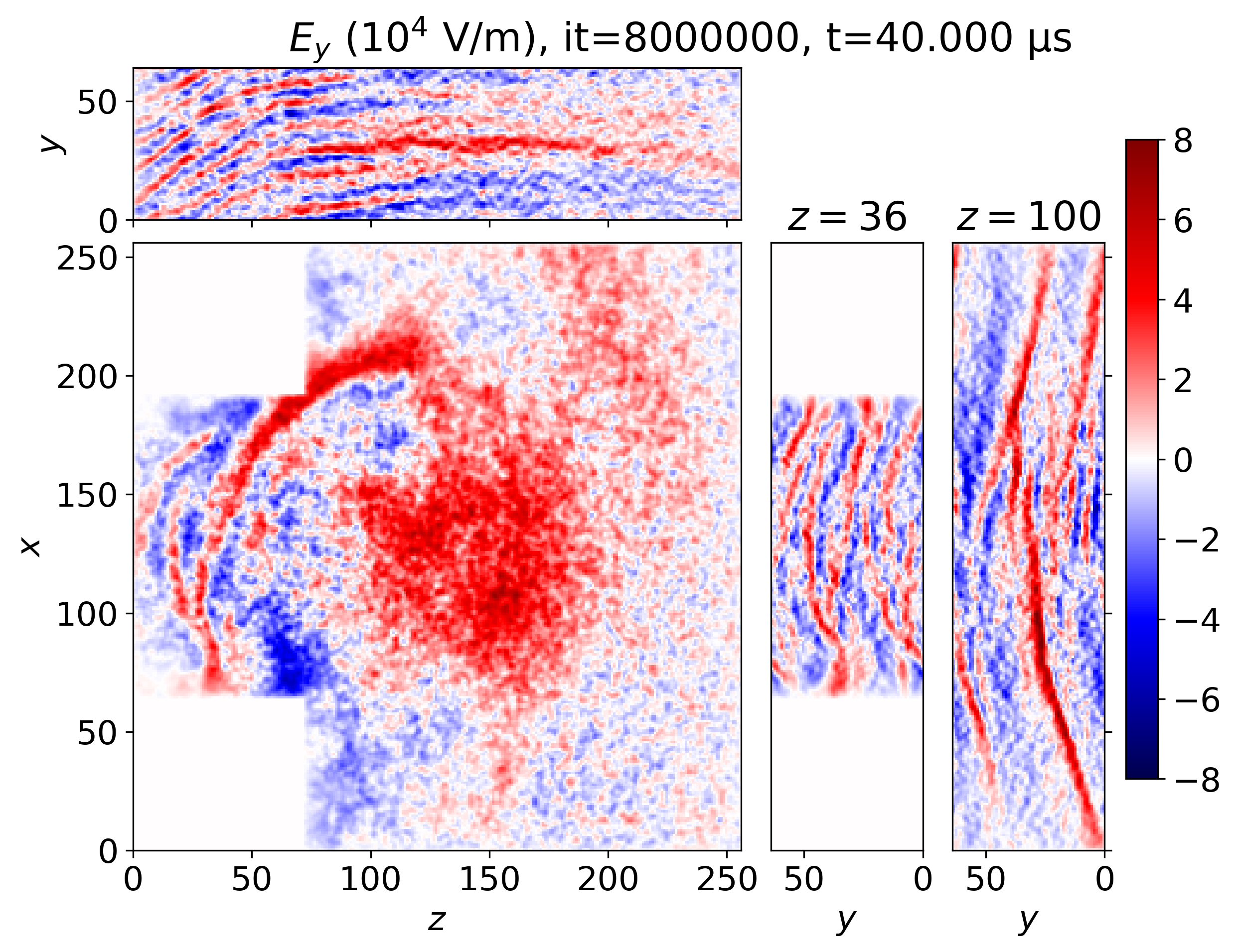}
    \includegraphics[width=0.32\linewidth]{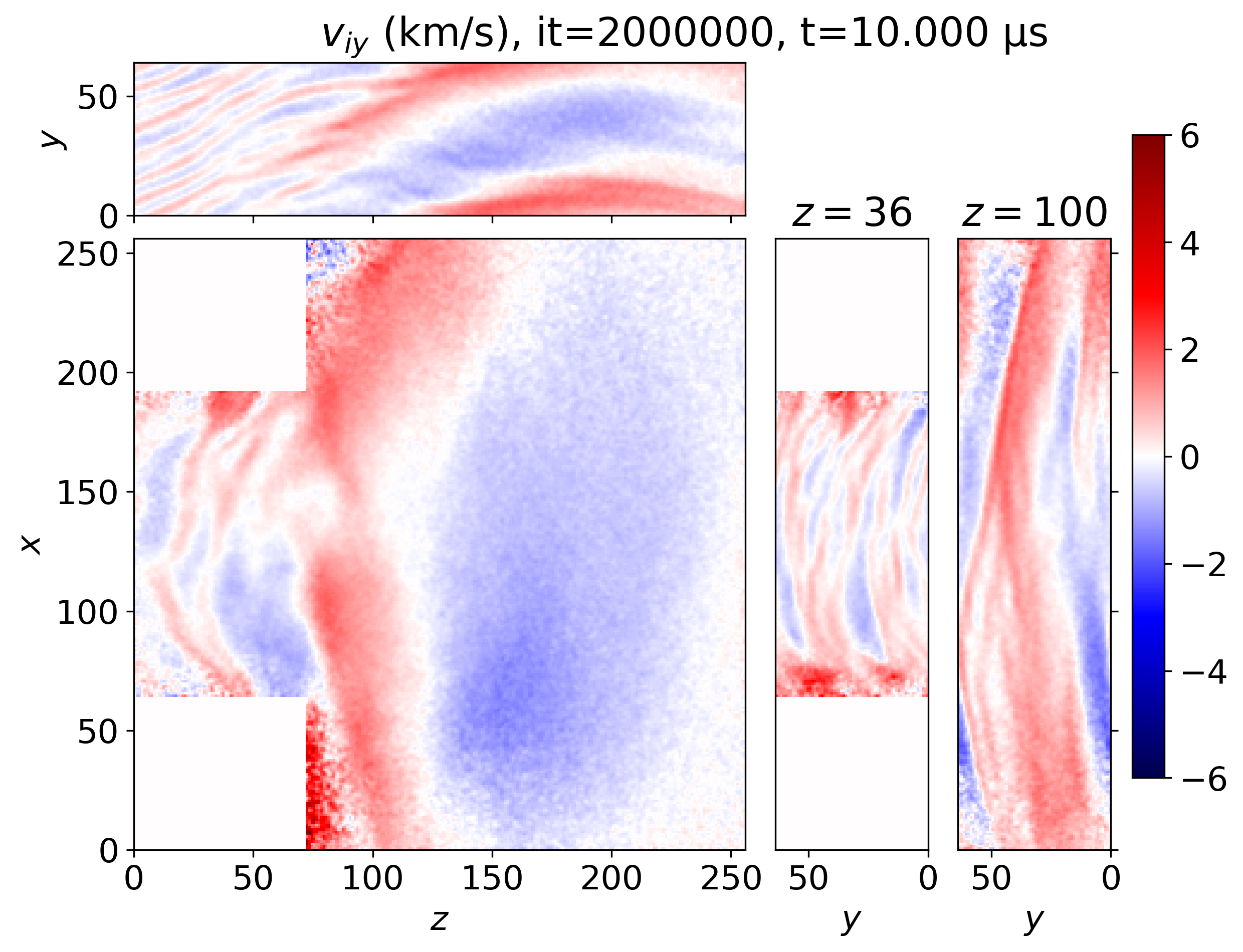}
    \includegraphics[width=0.32\linewidth]{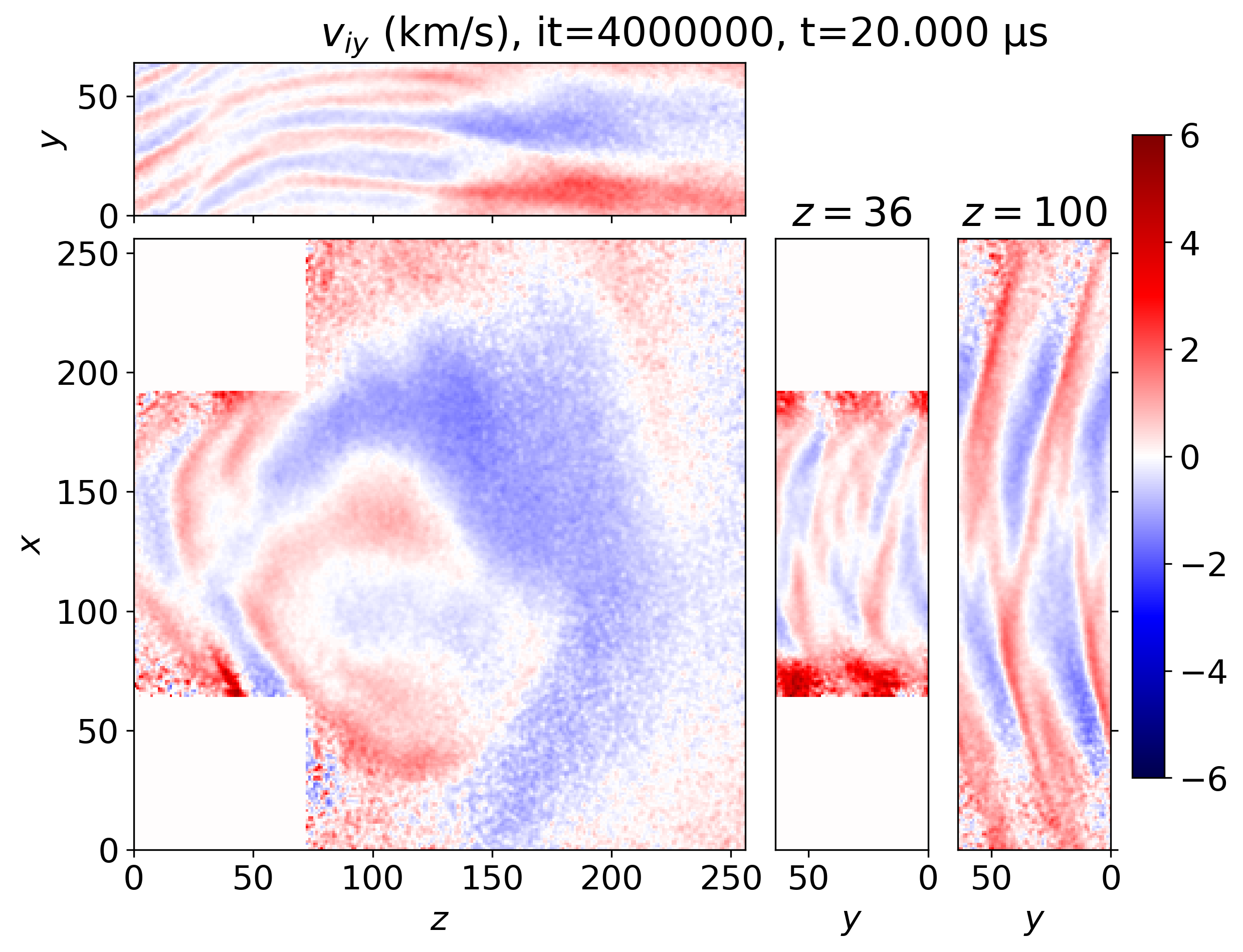}
    \includegraphics[width=0.32\linewidth]{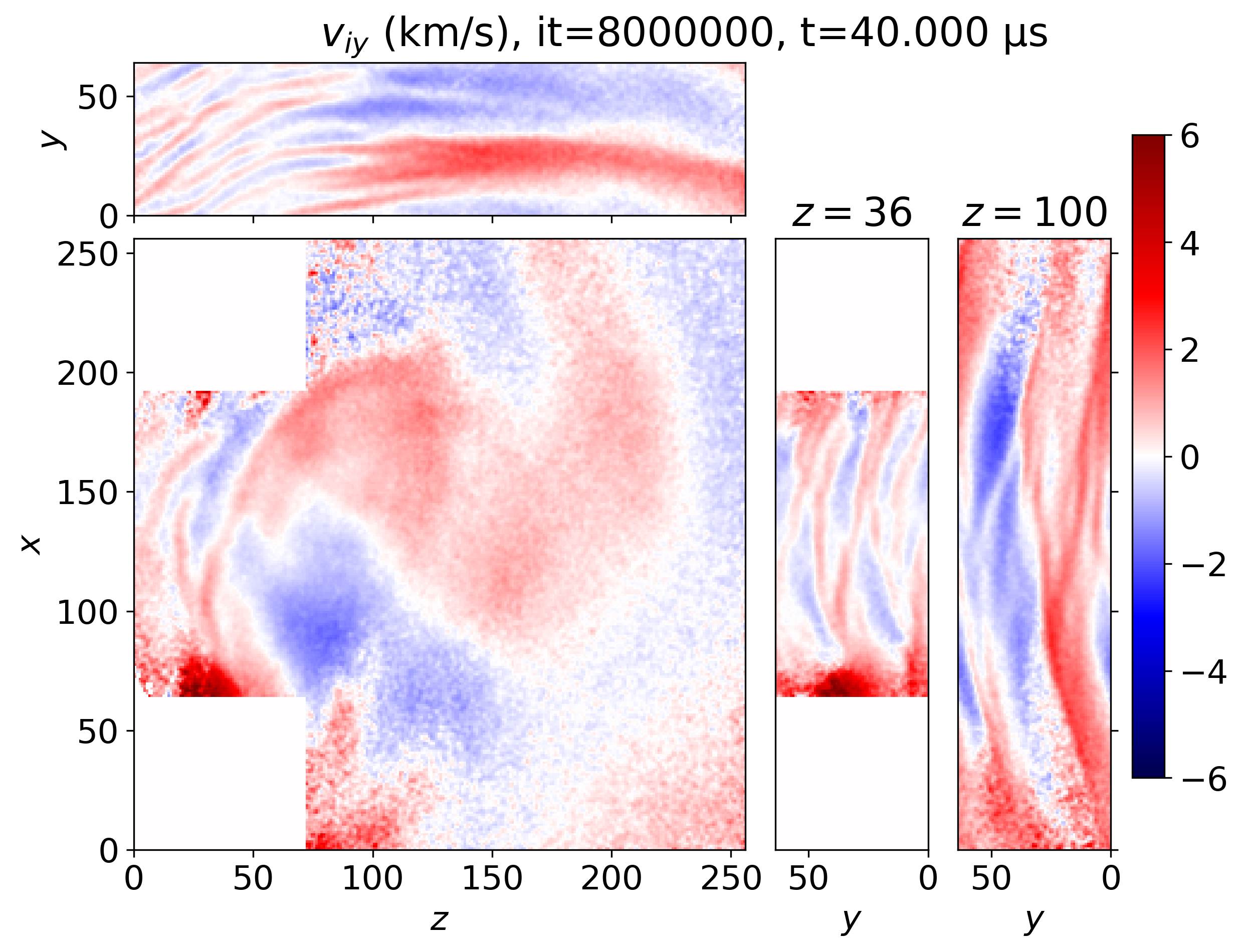}
    \includegraphics[width=0.32\linewidth]{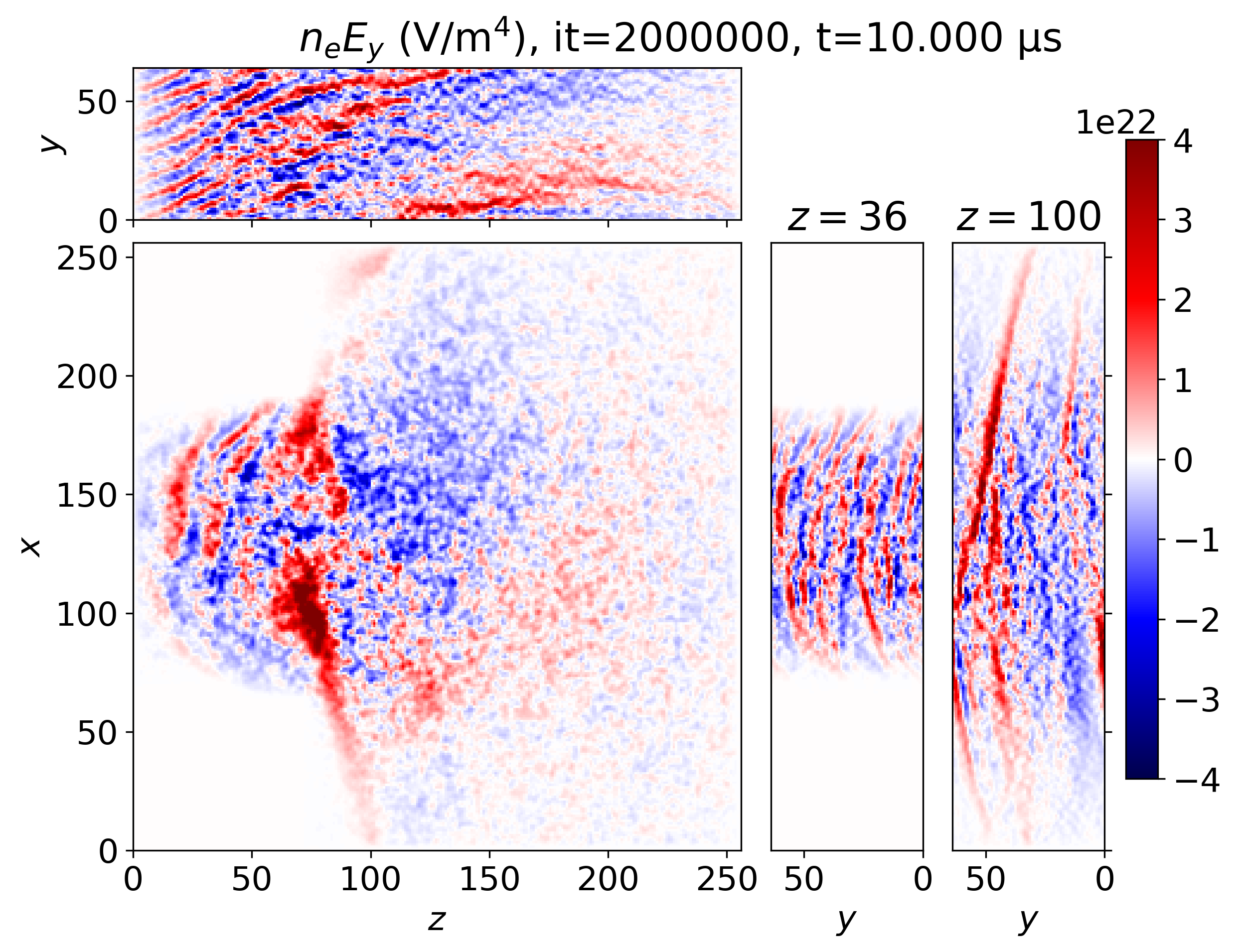}
    \includegraphics[width=0.32\linewidth]{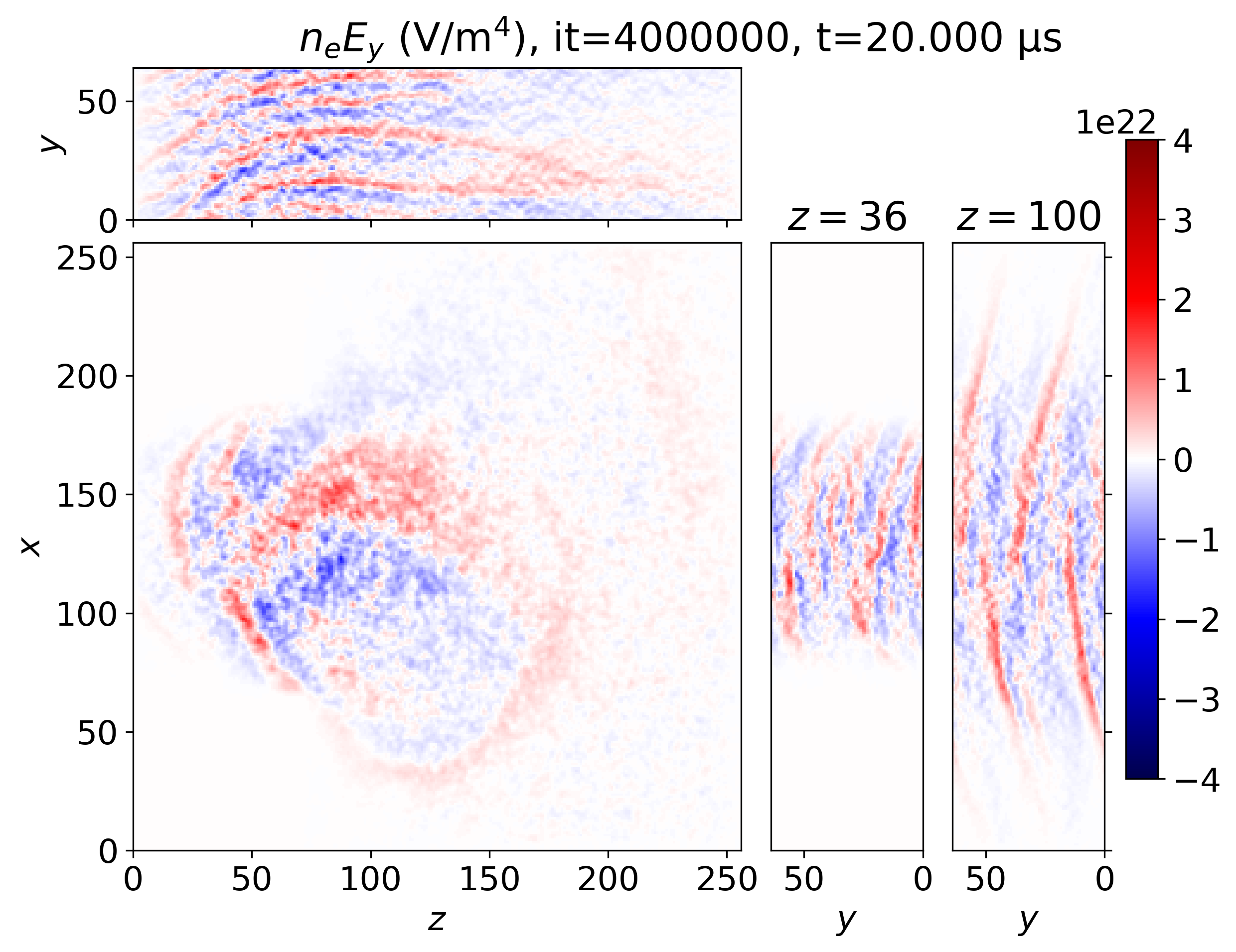}
    \includegraphics[width=0.32\linewidth]{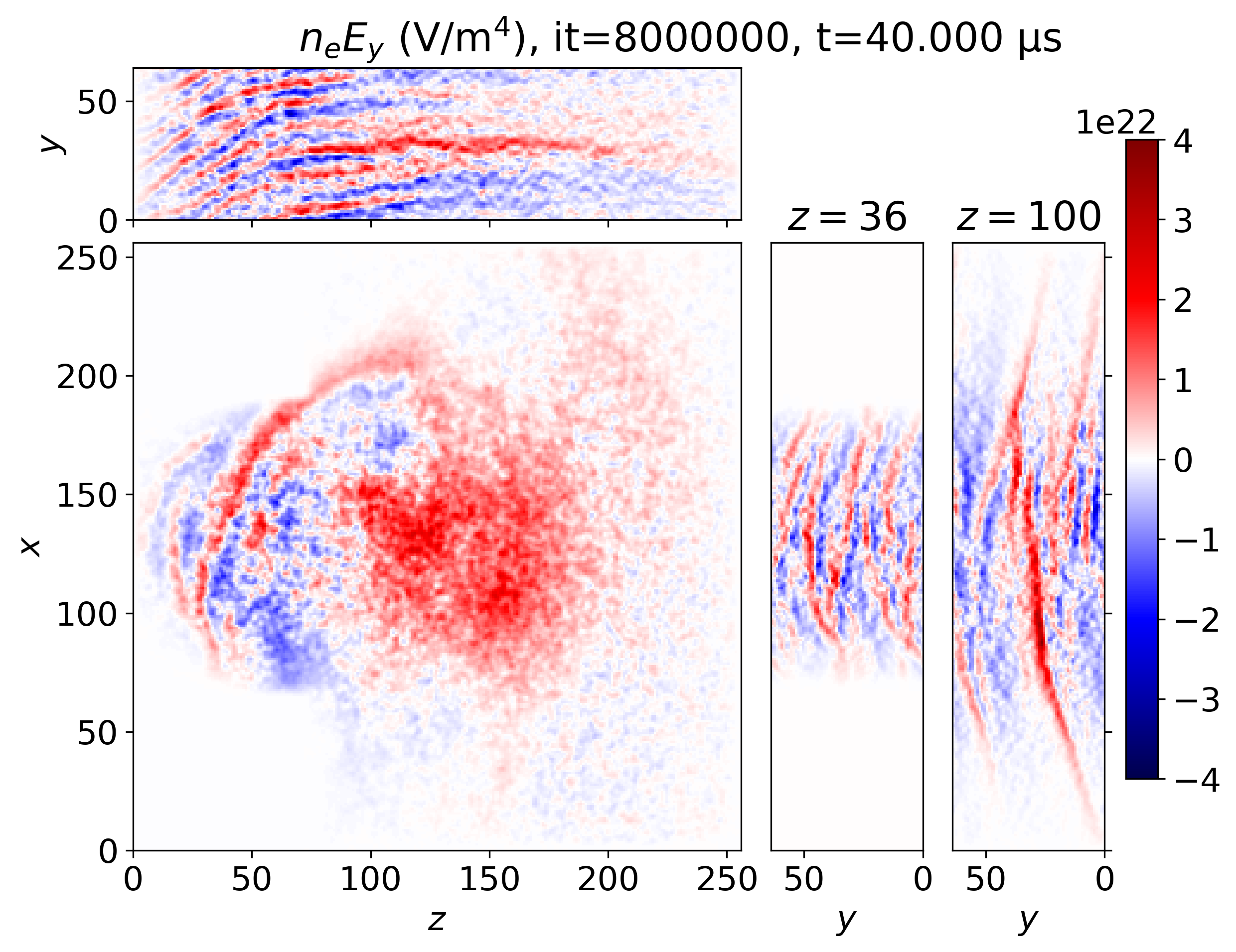}
    \caption{
Instantaneous three-dimensional field structure of Case~D at $t=10$, 20, and $40~\mu\mathrm{s}$ (left to right).
From top to bottom, the panels show ion number density $n_i$, electric potential $\phi$, azimuthal electric field $E_y$, azimuthal ion velocity $v_{iy}$, and the instantaneous correlation term $n_eE_y$.
Each snapshot includes mid-plane $z$--$y$ and $z$--$x$ cuts, together with $y$--$x$ slices at $z=36$ (inside the channel) and $z=100$ (near-exit plume).
Coordinates are reported in grid-index units.
}
\label{fig:Tripanel_t}
\end{figure*}

\subsection{Instantaneous Three-Dimensional EDI Structures Underlying the Pathway}
\label{sec:instantaneous_edi}

We now return to the instantaneous fields in order to clarify what three-dimensional structures underlie that averaged pathway.
The key point is that the transport map in Fig.~\ref{fig:nE} is not an independent background quantity; rather, it is the net remnant of strongly oscillatory, spatially coherent, three-dimensional EDI structures.
Fig.~\ref{fig:Tripanel_t} therefore provides the instantaneous counterpart to the averaged transport picture.

The ion-density snapshots show that the discharge remains strongly three-dimensional at all three representative stages.
At $t=10~\mu\mathrm{s}$, corresponding to the early high-density stage, $n_i$ forms a dense core in the downstream half of the channel and expands into the near-field plume, while clear azimuthal nonuniformity is visible in both the $z$--$y$ cut and the $y$--$x$ slices.
At $t=20~\mu\mathrm{s}$, corresponding to the depleted stage, the overall density level decreases substantially and the azimuthal modulation becomes weaker, but it does not disappear.
By $t=40~\mu\mathrm{s}$, the discharge recovers to a later, more repeatable state, yet appreciable three-dimensional structure remains visible both inside the channel and in the near-exit plume.
This persistent nonuniformity is important because it confirms that the later transport pathway is extracted from a genuinely three-dimensional fluctuating state, rather than from a nearly axisymmetric background perturbed only weakly by noise.

The potential snapshots retain the same global organization identified in the time-averaged fields of Fig.~\ref{fig:Tripanel_tavg}: the main axial potential drop remains localized near the exit and extends into the near-field plume.
Superimposed on this slowly varying structure, however, are small-scale ripples and distortions associated with the high-frequency instability.

Among the instantaneous quantities, $E_y$ most directly reveals the EDI wave pattern.
The $E_y$ panels in Fig.~\ref{fig:Tripanel_t} exhibit the characteristic stripe-like and wavefront-like structures associated with azimuthally propagating instability modes.
At $t=10~\mu\mathrm{s}$, the fluctuations are relatively strong and dominated by shorter-wavelength, tightly spaced fronts in the downstream channel and near-exit region.
At $t=20~\mu\mathrm{s}$, the fluctuation amplitude weakens and the characteristic wavelength becomes longer, with broader and more widely spaced patterns.
By $t=40~\mu\mathrm{s}$, the instability strengthens again and returns to a shorter-wavelength state.
Thus, both the fluctuation amplitude and the instantaneous wavelength are modulated over the slow discharge evolution.

The azimuthal ion velocity $v_{iy}$ shows that the ions respond coherently to these EDI fields.
Its wave-like modulation closely follows the spatial organization seen in $E_y$, demonstrating strong coupling between the azimuthal electric-field oscillations and the ion response.
At $t=10~\mu\mathrm{s}$ and $t=40~\mu\mathrm{s}$, the patterns are again comparatively fine-scaled, whereas at $t=20~\mu\mathrm{s}$ they are broader and longer-wavelength.
Unlike $E_y$, however, the overall amplitude of the instantaneous $v_{iy}$ fluctuations does not appear to decrease as markedly at $t=20~\mu\mathrm{s}$.
In addition, the instantaneous $v_{iy}$ field is biased toward a net positive direction, consistent with the preferred azimuthal sense set by the electron $\mathbf{E}\times\mathbf{B}$ drift and the dominant propagation direction of the instability.

Most directly relevant to anomalous transport is the instantaneous correlation term $n_eE_y$ shown in the bottom row of Fig.~\ref{fig:Tripanel_t}.
Its spatial pattern closely follows that of $E_y$, indicating that the instantaneous transport-driving correlation is carried by the same three-dimensional EDI structures that dominate the azimuthal electric field.
At $t=10~\mu\mathrm{s}$, $n_eE_y$ forms relatively strong, short-wavelength bands concentrated in the downstream channel and near the exit.
At $t=20~\mu\mathrm{s}$, the pattern weakens and shifts toward longer-wavelength modulation.
By $t=40~\mu\mathrm{s}$, it strengthens again and returns to a finer structure.
This close correspondence between $E_y$ and $n_eE_y$ demonstrates that the EDI is not merely present as an oscillatory field fluctuation; it directly carries the correlation responsible for the net anomalous electron transport after averaging.

This interpretation also helps clarify the role of averaging.
The instantaneous $n_eE_y$ field oscillates rapidly in sign and amplitude, so at any given time it is dominated by stripe-like wave structures rather than by a smooth transport channel.
However, when the oscillations are averaged over time and over the azimuthal direction, the fluctuating fine-scale pattern does not vanish completely.
Instead, it leaves behind the persistent two-band structure seen in Fig.~\ref{fig:nE}.
The averaged near-wall pathway can therefore be understood as the net transport footprint of a large ensemble of instantaneous three-dimensional EDI structures.

From the perspective of the paper's main claim, Fig.~\ref{fig:Tripanel_t} provides the dynamical foundation for the transport map.
The value of the 3D PIC approach is not only that it resolves the existence of EDI, but that it resolves how the instability occupies space in three dimensions and how its repeated action produces a nonuniform, wall-localized transport pathway after averaging.
Without access to these instantaneous three-dimensional structures, the origin of the near-wall transport topology would remain hidden.

\subsubsection{Potential Crest Surfaces and Three-Dimensional Wavefronts}
\label{sec:crest}

To further visualize the full three-dimensional wavefront geometry of the instantaneous EDI, we additionally extract crest isosurfaces from the electrostatic potential field $\phi(x,y,z,t)$.
The motivation is that, in the presence of non-negligible $E_x$ and $E_z$ components, the azimuthal electric field $E_y$ alone does not fully represent the wavefront structure, whereas the electrostatic potential is better suited to recover coherent equal-phase surfaces.
The detailed extraction procedure is described in Apx.~\ref{apx:wave_crest}.

\begin{figure}[htbp!]
    \centering
    \includegraphics[width=0.9\linewidth]{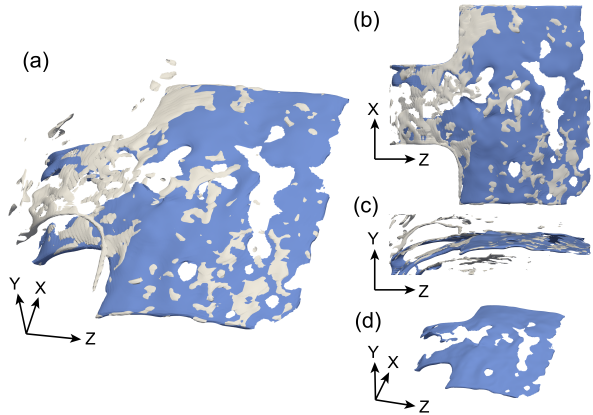}
    \caption{
    Electrostatic-potential crest isosurfaces at $t=29~\mu$s of Case~D.
    (a) Fragmented crest structure extracted directly from $\phi$, together with a representative crest isosurface obtained after optimized filtering;
    (b) radial--axial view;
    (c) azimuthal--axial view; and
    (d) the representative crest isosurface.
    }
    \label{fig:wave crest}
\end{figure}

As shown in Fig.~\ref{fig:wave crest}(a), the white fragments denote crest isosurfaces extracted directly from $\phi$ at $t=29~\mu$s of Case~D.
Although most of these fragments appear visually disconnected, they collectively outline a larger coherent wavefront.
This fragmentation is mainly attributed to the relatively high noise level in the raw potential field.
The blue surface in Fig.~\ref{fig:wave crest}(a) and (d) denotes a representative crest isosurface obtained with an optimized set of processing parameters $(\sigma_s,\sigma_b,q,\delta)$, followed by Connectivity and Threshold filtering to isolate the dominant coherent structure.

The extracted crest surface is approximately shell-like and convex toward the positive $y$ direction, consistent with propagation in the $\mathbf{E}\times\mathbf{B}$ direction.
Fig.~\ref{fig:wave crest}(b) shows that this coherent crest spans the radial--axial extent of the discharge region, while Fig.~\ref{fig:wave crest}(c) further indicates that successive crests are arranged along the azimuthal direction.
Approximately four to six crest surfaces can be identified, consistent with the instantaneous wave patterns observed previously in the field snapshots.

This visualization complements the $E_y$-based diagnostics by showing that the EDI is not merely an azimuthal stripe pattern on a two-dimensional cut, but a genuinely three-dimensional wavefront structure with non-negligible axial and radial extent.
In this sense, the potential crest surfaces provide an additional geometric view of the same instantaneous EDI dynamics that ultimately give rise to the averaged near-wall transport pathway discussed above.


\begin{figure*}[!htbp]
    \centering
    \includegraphics[width=0.9\linewidth]{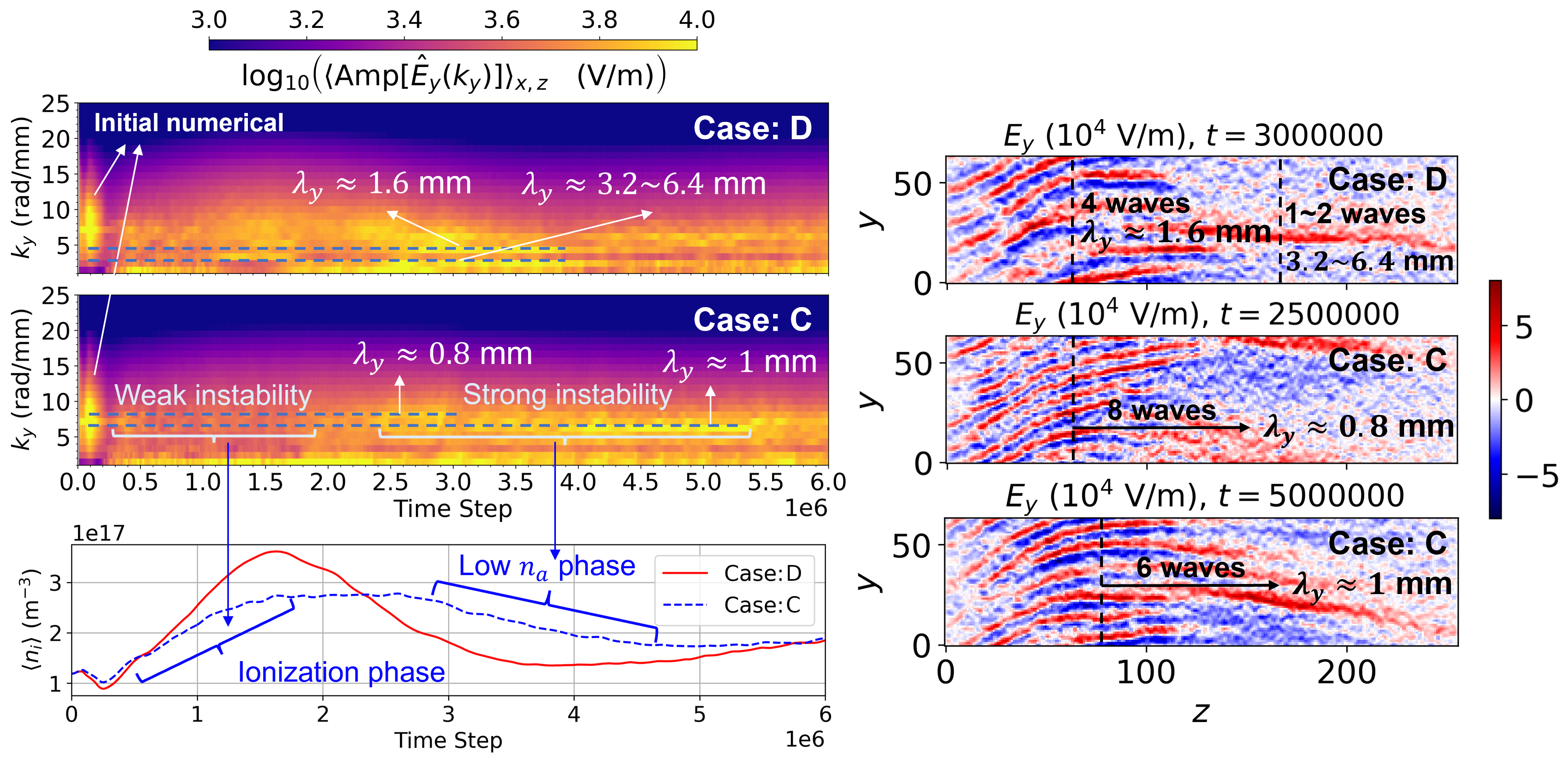}
    \caption{
Azimuthal wavenumber--time diagnostics of the $E_y$ fluctuations for Cases~D and C.
Left: evolution of the one-sided azimuthal spectrum, where $\hat{E}_y$ is obtained by FFT along the periodic $y$ direction and the amplitude is averaged over $x$ and $z$.
Right: representative $E_y$ snapshots on the $z$--$y$ plane at selected times, with $y$ and $z$ in grid-index units.
Bottom left: time evolution of the domain-averaged ion density $\langle n_i\rangle$ for both cases.
}
    \label{fig:Ey_ky_t}
\end{figure*}

\begin{figure}[htbp]
    \centering
    \includegraphics[width=0.9\linewidth]{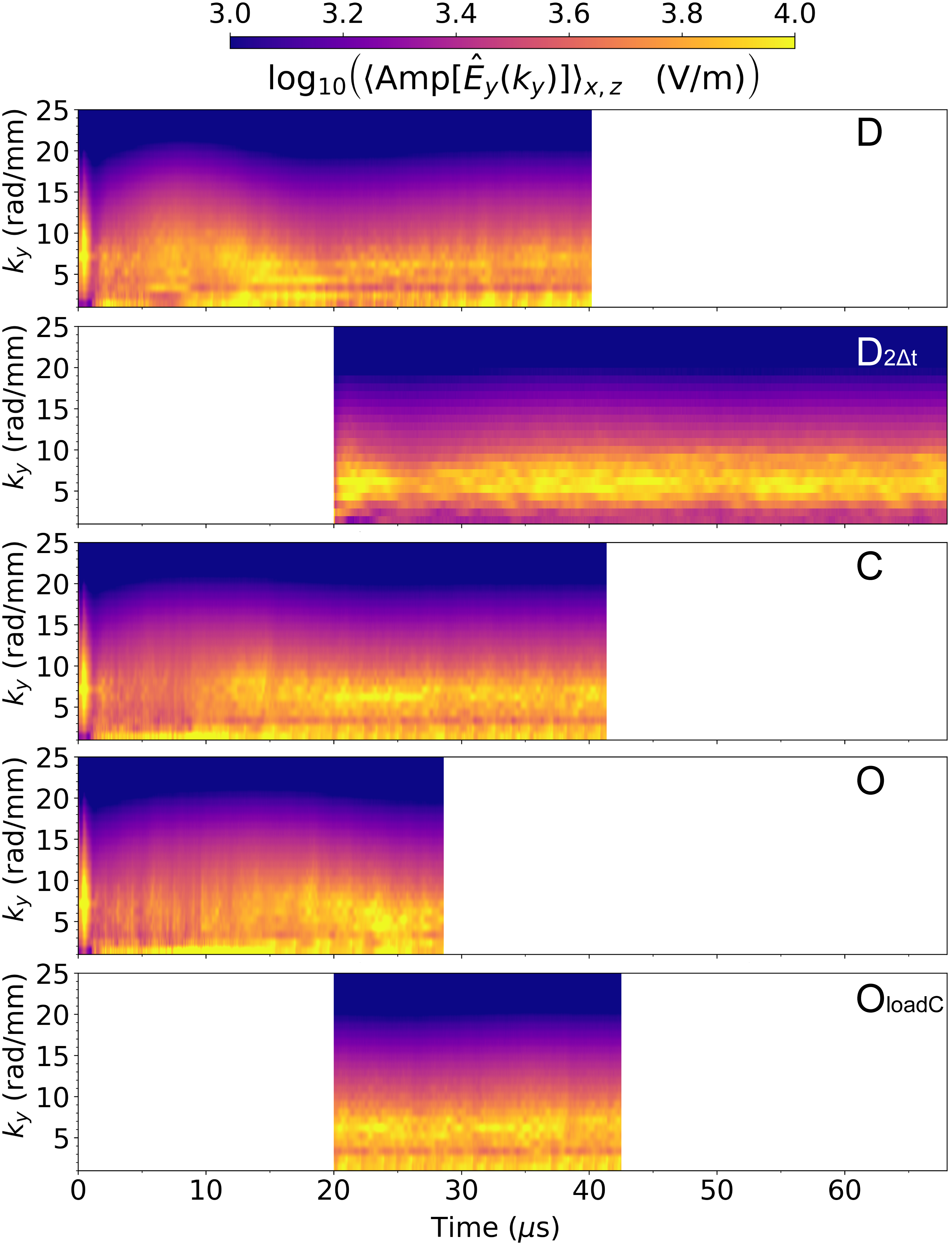}
\caption{
Azimuthal wavenumber--time spectrograms of the $E_y$ fluctuations for Cases~D, D$_{2\Delta t}$, C, O, and O$_{\mathrm{loadC}}$, constructed using the same procedure as in Fig.~\ref{fig:Ey_ky_t}.
White regions indicate time intervals not available for a given run.
}
    \label{fig:Eyky}
\end{figure}

\subsection{Spectral Characteristics of the EDI across Boundary Treatments}
\label{sec:spectral_boundary}

The near-wall transport pathway identified in Sec.~\ref{sec:transport_pathway} is extracted from time-averaged fields, but it is generated by the underlying high-frequency EDI dynamics.
It is therefore useful to characterize the spectral behavior of the instability and to determine how that behavior changes, or remains unchanged, under different boundary treatments.

\subsubsection{Azimuthal Wavelength Evolution in the Baseline Cases}
\label{sec:diag_ey_ky}

We first examine the temporal evolution of the azimuthal wavelength in the two baseline cases, D and C.
The detailed procedure used to construct the azimuthal wavenumber diagnostics is described in Apx.~\ref{apx:Ey-k}.
Fig.~\ref{fig:Ey_ky_t} shows the corresponding $k_y$--$t$ maps of the $E_y$ fluctuations, together with representative instantaneous $E_y$ patterns and the time history of the domain-averaged ion number density $\langle n_i\rangle$.

For both cases, the earliest stage contains a short-lived burst of strong broadband fluctuations during roughly the first $0.2\times 10^6$ time steps.
This initial burst is attributed primarily to the start-up procedure and does not represent the mature EDI state of interest here.
After this stage, the instability enters a weaker interval while $\langle n_i\rangle$ continues to increase.
In this interval, a significant fraction of the electron energy is consumed by ionization, which appears to hinder the development of stronger EDI activity.
At later times, both cases transition into a more developed ``strong-instability'' stage, during which the spectral intensity becomes larger and more persistent.
Finally, after this strong stage, the fluctuation level decreases again as the discharge moves into a later recovery stage.

During the stronger-instability interval, the dominant azimuthal wavelengths inferred from Fig.~\ref{fig:Ey_ky_t} remain on the order of $\sim 1~\mathrm{mm}$, consistent with the characteristic scale commonly associated with Hall-thruster EDI.
For Case~D, a representative snapshot around $t=3\times 10^6$ time steps shows approximately four wave periods across the simulated azimuthal extent near $z\simeq 80$, corresponding to $\lambda_y\approx 1.6~\mathrm{mm}$ and thus $k_y\sim 4$--$5~\mathrm{rad/mm}$.
Longer waves are also visible in the plume, with one to two periods across the domain and correspondingly smaller $k_y$.
For Case~C, representative snapshots indicate a modest wavelength shift over time, from roughly $\lambda_y\approx 0.8~\mathrm{mm}$ in an earlier strong stage to roughly $\lambda_y\approx 1~\mathrm{mm}$ at a later stage.
Thus, although the fluctuation intensity and preferred wavelength are modulated over the slow discharge evolution, the EDI remains confined to a relatively narrow characteristic azimuthal scale.

These baseline diagnostics support two conclusions relevant to the main transport result.
First, the EDI reaches a locally mature spectral state well before the global low-frequency envelope becomes strictly stationary, which justifies the late-time analysis window defined in Sec.~\ref{sec:analysis_window}.
Second, the near-wall transport pathway identified in Fig.~\ref{fig:nE} is not associated with an isolated or anomalous spectral event, but rather with a persistent EDI regime whose dominant azimuthal scale remains in the expected Hall-thruster range.

\subsubsection{Boundary-Condition Comparison of the Spectral Behavior}
\label{sec:diag_ey_ky_cases}

We next compare the spectral evolution across the different simulation cases.
Fig.~\ref{fig:Eyky} compiles the $k_y$--$t$ spectrograms of the azimuthal electric-field fluctuations for Cases~D, D$_{2\Delta t}$, C, O, and O$_{\mathrm{loadC}}$, using the same FFT-based diagnostics as in Fig.~\ref{fig:Ey_ky_t}.
The white segments indicate time intervals not covered by a given run.

The most important observation is that Cases~D, C, O, and O$_{\mathrm{loadC}}$ exhibit broadly similar spectral behavior after the initial transient.
In all four cases, the dominant fluctuation power remains concentrated within a comparable range of azimuthal wavenumbers, and the temporal evolution of the main spectral band follows the same general pattern of early development, stronger fluctuation stage, and later-time persistence.
These spectral similarities are particularly important in light of the transport maps in Fig.~\ref{fig:nE}: they indicate that the robust near-wall transport pathway identified there is generated under closely related EDI spectral conditions, despite the substantial changes in wall and outflow boundary closure.

Among these cases, O$_{\mathrm{loadC}}$ is particularly useful because it begins from a later-time state loaded from Case~C and therefore reaches a mature stage of the outflow-boundary evolution at lower additional computational cost.
Its spectrogram reproduces the same dominant wavelength content over its available interval and remains close to that of Case~O in the portion of the evolution that overlaps.
This agreement supports the use of O$_{\mathrm{loadC}}$ as a practical late-time proxy for Case~O in the transport analysis.

The differences among Cases~D, C, O, and O$_{\mathrm{loadC}}$ are mainly secondary.
Changing from the conducting Dirichlet walls of Case~D to the ceramic wall treatment of Case~C produces only modest modifications in the spectral intensity and preferred wavelength.
Likewise, introducing the outflow treatment in Case~O and Case~O$_{\mathrm{loadC}}$ does not shift the dominant spectral band qualitatively, even though it does affect the detailed amplitude distribution and the downstream coupling of the fluctuations.
This spectral robustness is consistent with the transport results obtained earlier: the boundary treatments redistribute the detailed strength and near-plume extension of the anomalous transport pathway, but they do not eliminate the near-wall pathway itself or replace it with a fundamentally different instability regime.

Case~D$_{2\Delta t}$ is included in Fig.~\ref{fig:Eyky} for reference, although it is primarily intended for the numerical-effects discussion rather than for the physical boundary-condition comparison.
Relative to the baseline Case~D, its spectrogram shows a weaker contribution from the low-$k_y$ portion of the spectrum, i.e., a reduction of the longer-wavelength components, together with a comparatively stronger concentration of power at intermediate $k_y$.
This suggests that increasing the timestep can bias the EDI spectrum toward shorter wavelengths.
Because this effect is numerical in origin rather than associated with a change in wall or outflow closure, a more detailed comparison between Cases~D and D$_{2\Delta t}$ is deferred to the later numerical-effects section, where the timestep influence will be discussed explicitly.

\subsubsection{Local Dispersion Characteristics}
\label{sec:dispersion}

\begin{figure}[htbp]
    \centering
    \includegraphics[width=\linewidth]{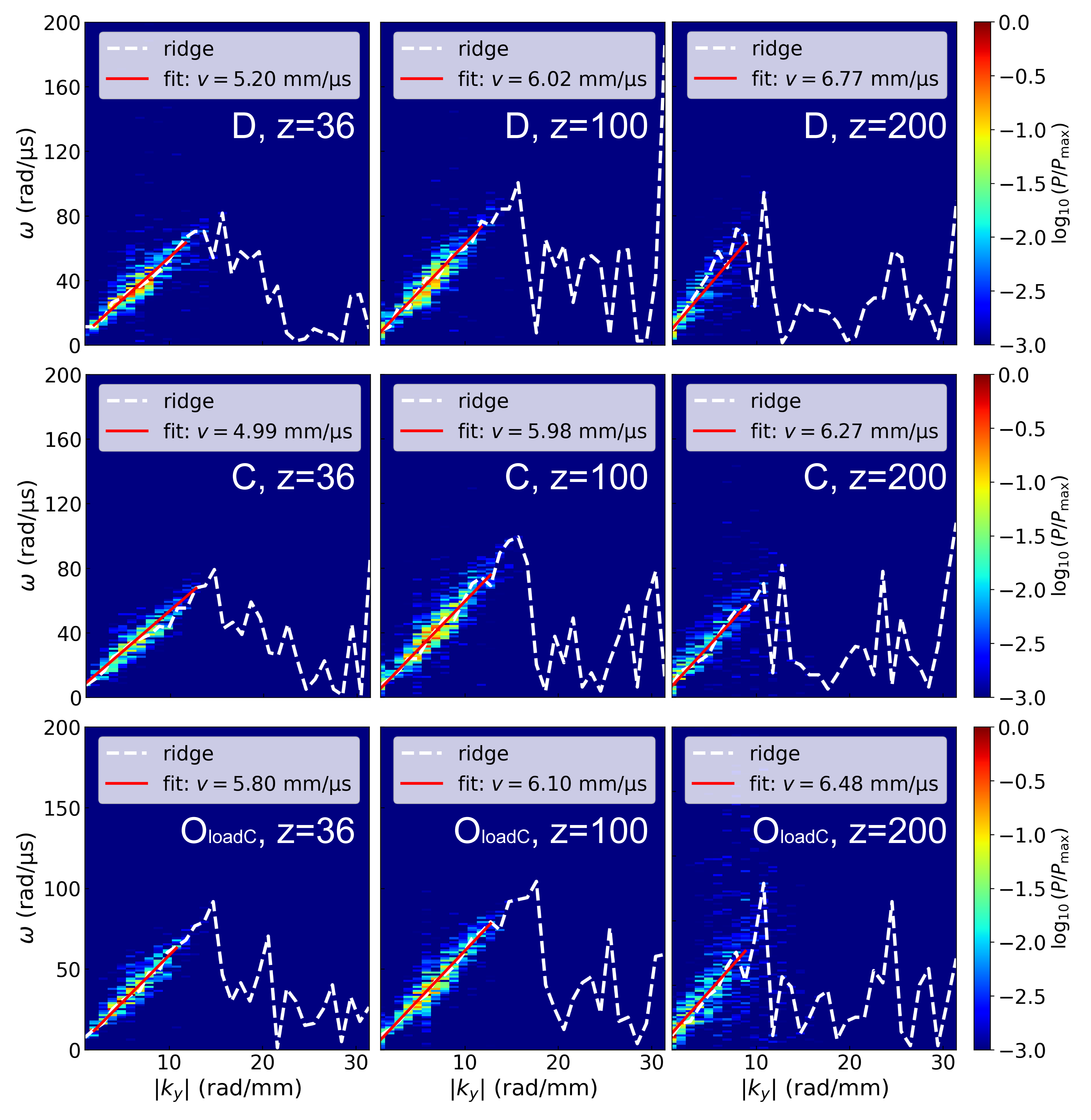}
    \caption{
    For Case~D, the fitted red-line slope is also comparable to the local ion acoustic speed, with $c_s=5.61$, 5.82, and $4.33~\mathrm{mm/\mu s}$ at $z=36$, 100, and 200, respectively. The agreement is closest in the in-channel and near-exit regions, whereas a larger deviation appears farther downstream, where the branch becomes broader and less coherent.
    }
    \label{fig:ky_omega_cases}
\end{figure}

To further characterize the local wave dynamics in the region of the transport pathway, Fig.~\ref{fig:ky_omega_cases} shows dispersion maps of the azimuthal electric-field fluctuations in the $(k_y,\omega)$ plane for Cases~D, C, and O$_{\mathrm{loadC}}$ at three representative axial locations, $z=36$, 100, and 200.
These diagnostics complement the $k_y$--$t$ analysis by resolving the frequency content associated with each azimuthal wavenumber.

For all three cases, the spectral power is dominated by a low-$k_y$ branch that is approximately linear over its most energetic region.
This branch is most clearly defined at $z=36$ and $z=100$, corresponding to the in-channel and near-exit regions where the coherent EDI signatures are strongest in the earlier field diagnostics.
At the farther downstream location, $z=200$, the branch becomes broader and less sharply defined, indicating that the wave activity is weaker and less coherent in the downstream plume.
At the same time, the high-$k_y$ portion of the spectrum is noticeably reduced there, so that the downstream fluctuation field is increasingly dominated by longer-wavelength components while the short-wavelength content becomes weaker.
This trend is consistent with the earlier observations from the instantaneous fields and the $k_y$--$t$ diagnostics: the clearest instability signatures and strongest transport activity remain concentrated inside the channel and near the exit, whereas the downstream plume is more diffuse and contains fewer short-wavelength structures.

The comparison among Cases~D, C, and O$_{\mathrm{loadC}}$ shows that the dominant low-$k_y$ dispersion branch is broadly preserved across boundary treatments.
In each case, the ridge follows a similar low-$k_y$ trend, and a linear fit to its energetic segment yields an apparent phase speed of order $5$--$7~\mathrm{mm/\mu s}$.
For Case~D, the fitted red-line slope in Fig.~13 is also comparable to the local ion acoustic speed, with $c_s=5.61$, 5.82, and $4.33~\mathrm{mm/\mu s}$ at $z=36$, 100, and 200, respectively, showing the closest agreement in the in-channel and near-exit regions, while the somewhat larger downstream deviation remains within the same order and is consistent with the broader, less coherent branch there.
For Case~D, the fitted phase speed is $5.20$, 6.02, and $6.77~\mathrm{mm/\mu s}$ at $z=36$, 100, and 200, respectively, whereas the corresponding local ion acoustic speeds are $5.61$, 5.82, and $4.33~\mathrm{mm/\mu s}$, so that the difference is small at $z=36$ and 100 ($\sim 0.4$ and $0.2~\mathrm{mm/\mu s}$) but becomes noticeably larger in the downstream plume at $z=200$ ($\sim 2.4~\mathrm{mm/\mu s}$).
The main boundary induced differences therefore appear in the sharpness and spectral spread of the branch, rather than in the emergence of a distinct new dominant mode.

Taken together, Figs.~\ref{fig:Ey_ky_t}, \ref{fig:Eyky}, and \ref{fig:ky_omega_cases} indicate that the near-wall anomalous transport pathway identified earlier is embedded in a robust EDI regime with broadly similar azimuthal scales and local dispersion characteristics across Cases~D, C, and O$_{\mathrm{loadC}}$.
The spectral diagnostics therefore support the transport-based interpretation by showing that the pathway is associated with a common underlying EDI dynamics, while the boundary treatment mainly modulates the coherence, spectral spread, and downstream persistence of that activity.


\begin{figure}
    \centering
    \includegraphics[width=0.9\linewidth]{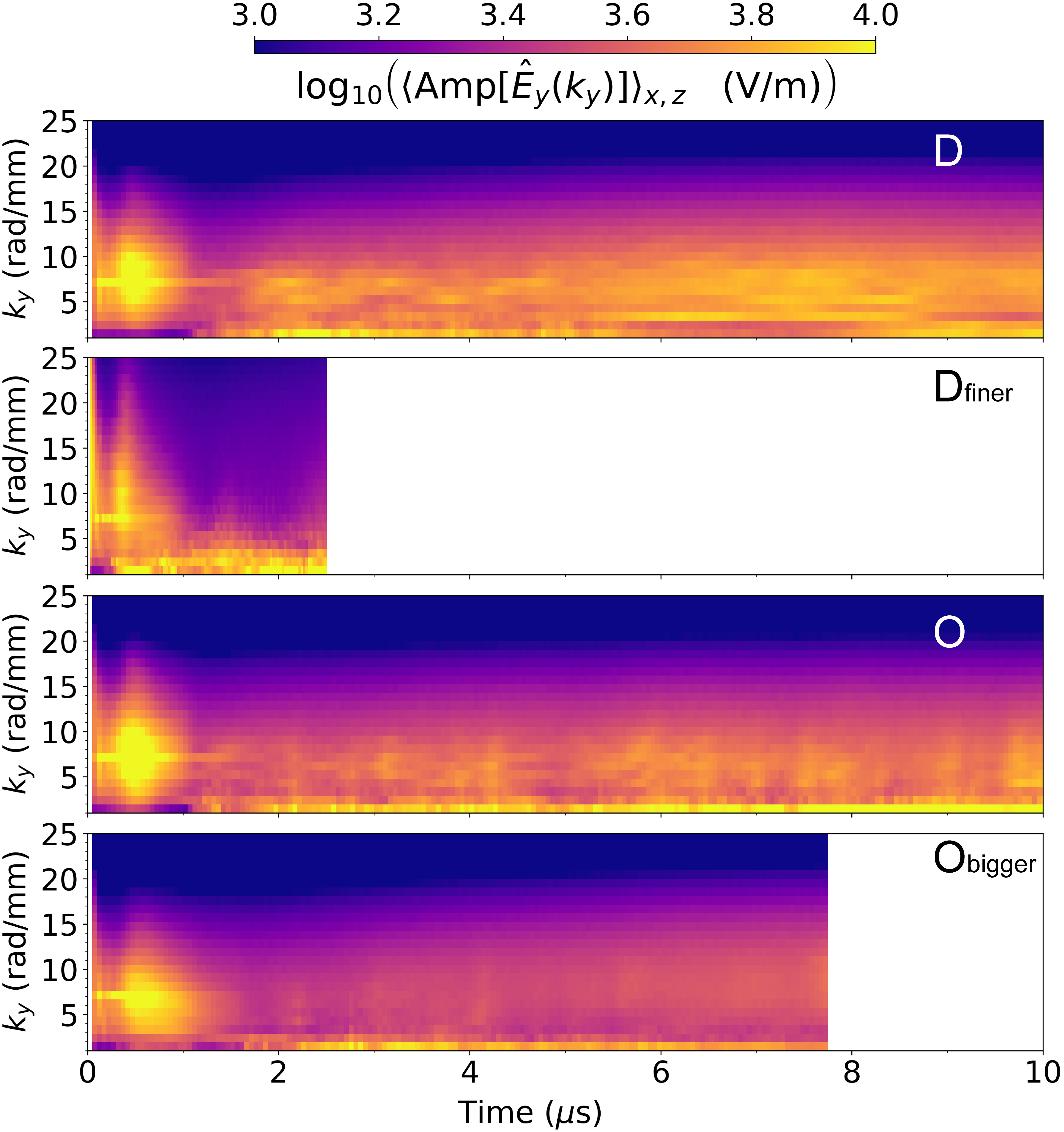}
    \caption{
Zoomed-in view of the $E_y$ azimuthal wavenumber--time spectrograms over the first $10~\mu$s for Cases~D, D$_{\mathrm{finer}}$, O, and O$_{\mathrm{bigger}}$, using the same diagnostics as in Fig.~\ref{fig:Ey_ky_t}.
}
    \label{fig:Eyky10us}
\end{figure}




\section{Numerical Sensitivities}
\label{sec:numerical}

After the extended presentation of the main simulation results, we now turn to a more focused discussion of numerical effects.
The following subsections assess how the resolved EDI dynamics and the associated anomalous electron transport depend on the timestep, grid resolution, and plume-domain size.

As a complement to the discussion in Sec.~\ref{sec:diag_ey_ky_cases}, Fig.~\ref{fig:Eyky10us} presents the corresponding zoomed-in $E_y$ $k_y$--$t$ spectrograms for the remaining numerical-variation comparisons, namely Case~D versus Case~$D_{\mathrm{finer}}$ and Case~O versus Case~$O_{\mathrm{bigger}}$.
The timestep-doubled case, Case~$D_{2\Delta t}$, is not repeated here because its spectrogram has already been shown in Fig.~\ref{fig:Eyky} and discussed in Sec.~\ref{sec:diag_ey_ky_cases}.
Over the limited overlapping interval, Case~$D_{\mathrm{finer}}$ appears to show a modest tendency toward lower $k_y$, corresponding to a somewhat longer azimuthal wavelength than in Case~D, although this trend should be interpreted cautiously because the refined simulation covers only a short early-time window and is not perfectly synchronized with Case~D.
By contrast, Cases~O and $O_{\mathrm{bigger}}$ exhibit broadly similar early-time spectral distributions, indicating that plume-domain enlargement does not produce a strong spectral reorganization; however, the overall fluctuation level in Case~$O_{\mathrm{bigger}}$ appears somewhat weaker.

\subsection{Timestep Sensitivity}
\label{sec:dt}

For timestep sensitivity, we compare the baseline Case~D with Case~D$_{2\Delta t}$, in which only the timestep is doubled, while the mesh and boundary treatment are kept unchanged.
As noted in Sec.~\ref{sec:results}, the ky--t spectrogram of Case~D$_{2\Delta t}$ already suggests a systematic spectral shift relative to Case~D: the low-$k_y$ part of the spectrum is weaker, whereas the fluctuation power is relatively more concentrated at intermediate $k_y$.
This indicates that increasing the timestep suppresses the longer-wavelength contribution and biases the resolved EDI toward shorter azimuthal wavelengths.

\begin{figure*}
    \centering
    \includegraphics[width=0.34\linewidth]{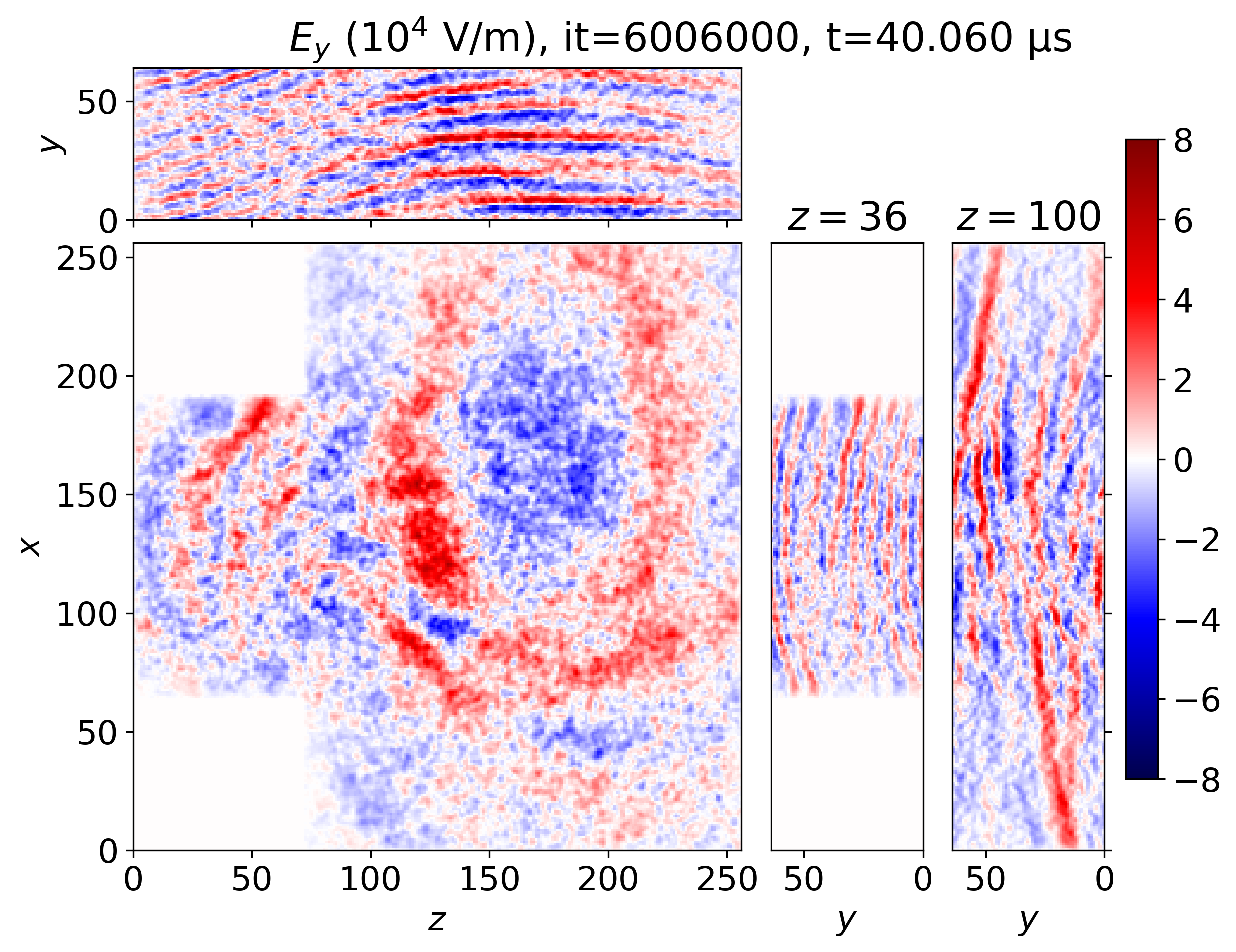}
    \includegraphics[width=0.34\linewidth]{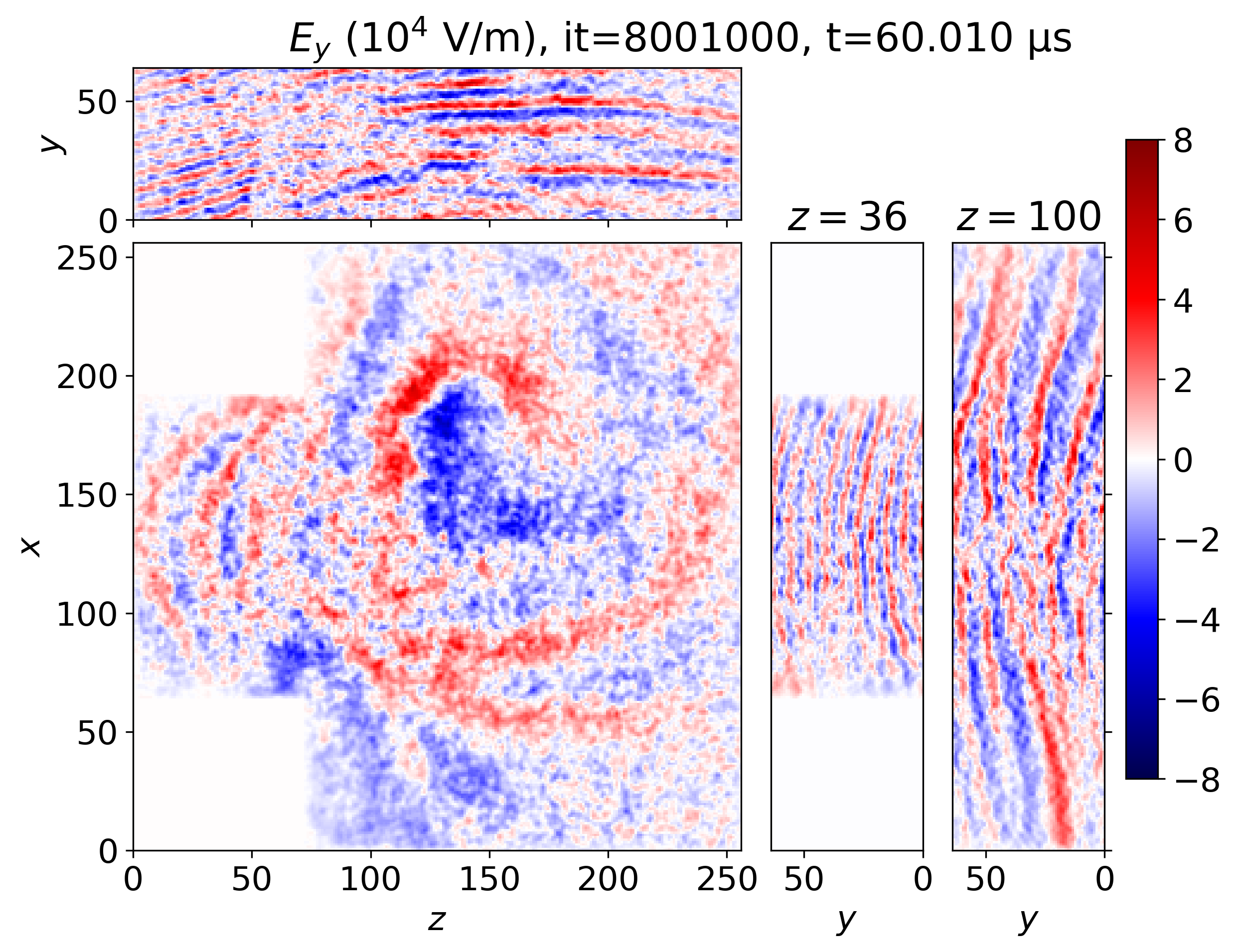}
    \includegraphics[width=0.28\linewidth]{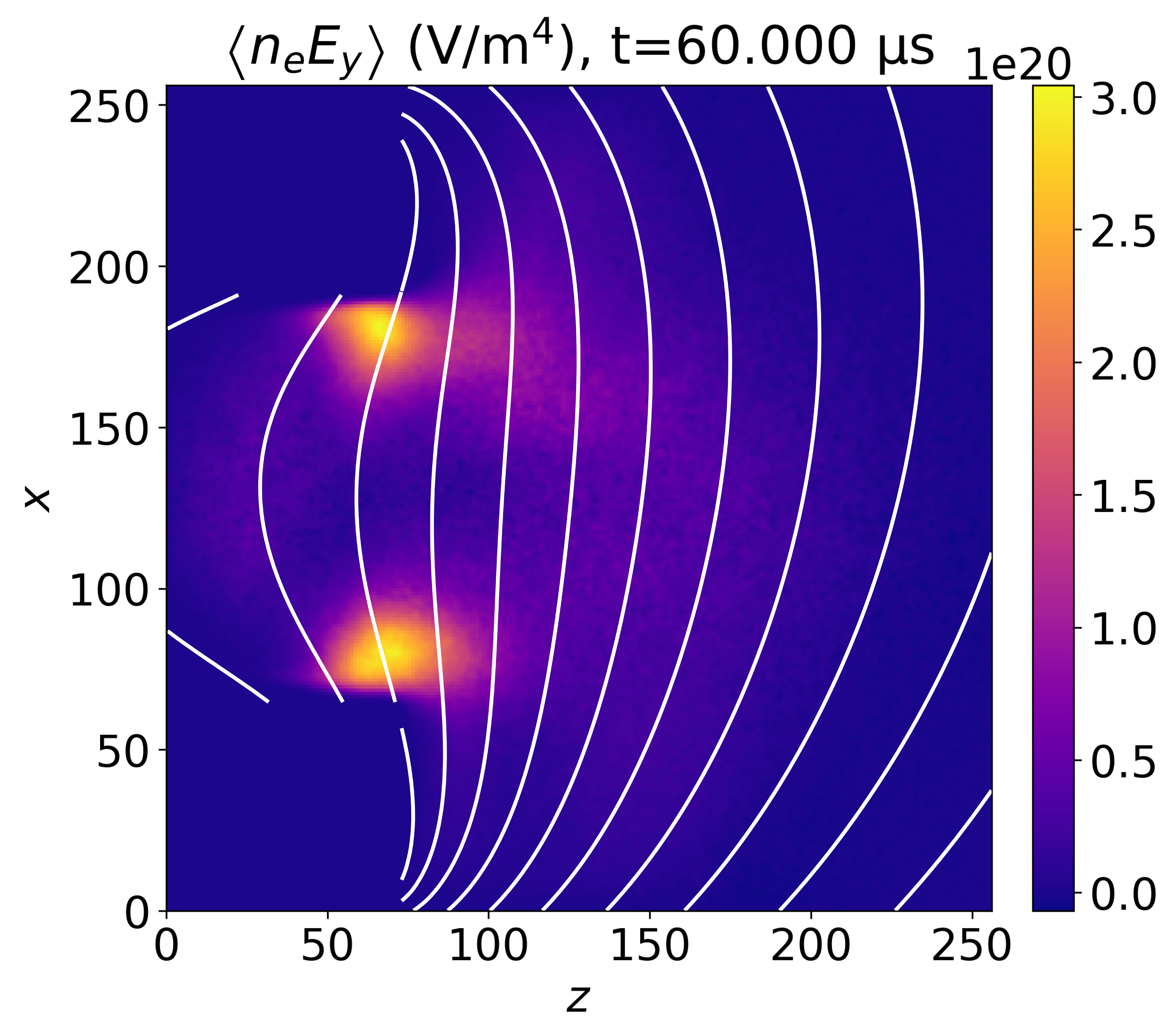}
    \caption{
    Timestep sensitivity of the EDI and the resulting transport organization.
    Left and middle: representative late-time instantaneous $E_y$ structures of Case~D$_{2\Delta t}$ at $t\approx 40~\mu$s and $60~\mu$s.
    Right: time- and azimuthally-averaged correlation term $\langle n_eE_y\rangle$ for Case~D$_{2\Delta t}$.
    }
    \label{fig:2dt}
\end{figure*}

\begin{figure}
    \centering
    \includegraphics[width=\linewidth]{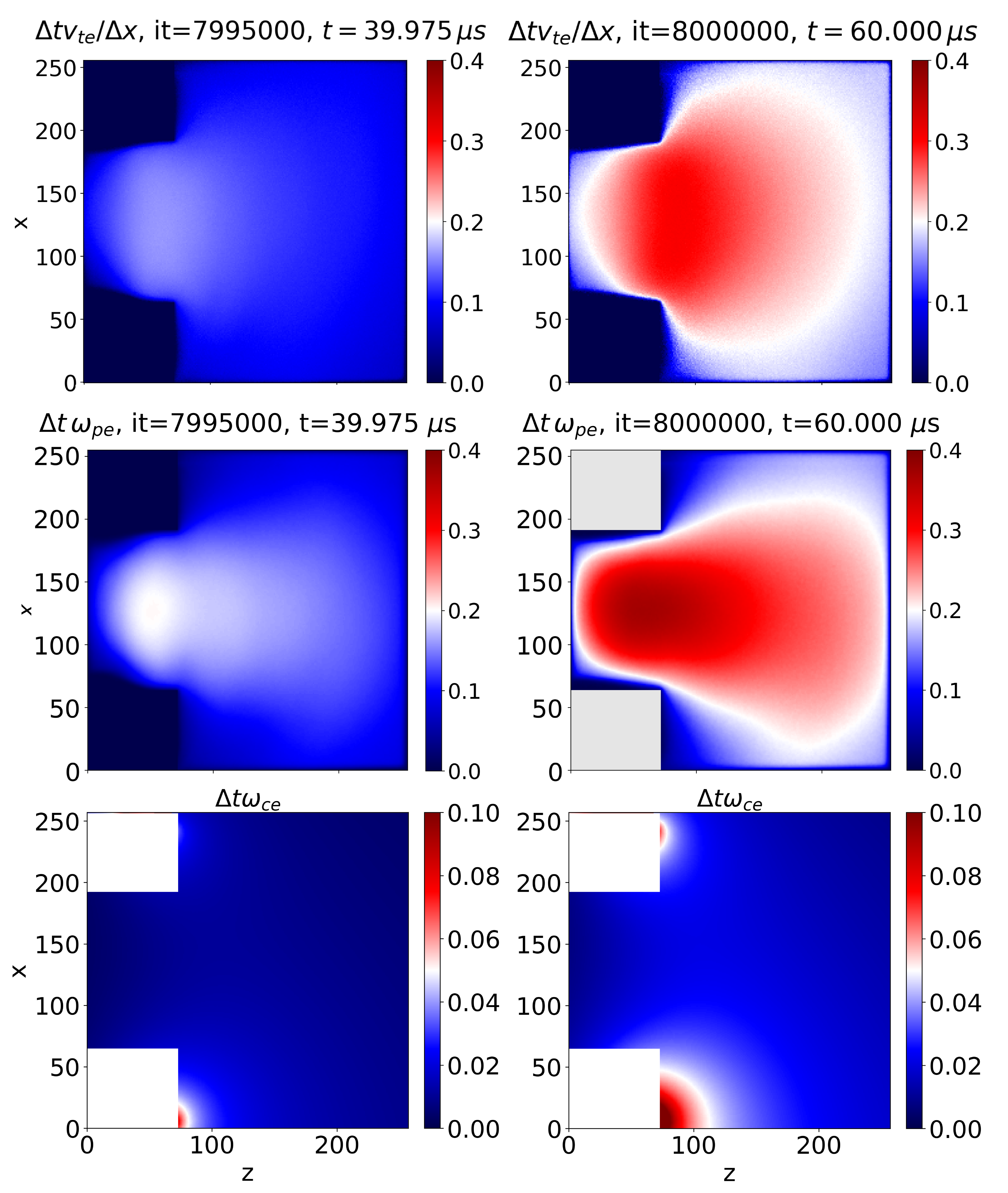}
    \caption{
    Local timestep-resolution diagnostics for Cases~D and D$_{2\Delta t}$ at representative late times.
    Left column: baseline Case~D.
    Right column: Case~D$_{2\Delta t}$.
    Top row: $\Delta t\,\omega_{pe}$.
    Middle row: $\Delta t v_{te} / \Delta x $.
    Bottom row: $\Delta t\,\omega_{ce}$.
    }
    \label{fig:dt_metrics}
\end{figure}

Fig.~\ref{fig:2dt} shows that the same tendency is clearly visible in physical space.
The left and middle panels present representative late-time instantaneous $E_y$ structures in Case~D$_{2\Delta t}$ at $t\approx 40~\mu$s and $60~\mu$s, respectively.
Compared with the corresponding late-stage structures of the baseline Case~D in Fig.~\ref{fig:Tripanel_t}, the plume-region pattern in Case~D$_{2\Delta t}$ is systematically finer, with more closely spaced stripe-like structures and a shorter apparent wavelength.
The same trend is also noticeable near the exit.
Thus, the timestep does not merely rescale the fluctuation amplitude; it modifies the wavelength content of the resolved EDI, preferentially weakening the long-wavelength, low-$k_y$ component.


The key point, however, is that this spectral and morphological sensitivity does not translate into a qualitative change of the net transport topology.
The right panel of Fig.~\ref{fig:2dt} shows the time- and azimuthally-averaged correlation term $\langle n_eE_y\rangle$ for Case~D$_{2\Delta t}$.
Despite the shorter-wavelength instantaneous structures, the averaged transport remains concentrated in two wall-adjacent bands located in the downstream channel and around the exit region, with only a weaker extension into the near plume.
This is the same spatial organization identified previously in Fig.~\ref{fig:nE} for the baseline transport analysis.
Therefore, doubling the timestep alters the detailed instantaneous realization of the instability, but it does not remove, shift, or qualitatively reorganize the dominant near-wall transport pathway.

In practical terms, this separation between spectral sensitivity and transport-topology robustness suggests a useful multistage workflow:
the doubled-timestep case can be used for accelerated exploratory simulations to identify overall trends and promising optimization directions at lower computational cost,
whereas the baseline timestep should be retained when the detailed wavelength content, fine-scale EDI morphology, and high-frequency plasma response need to be characterized more faithfully.

In addition, to provide a more direct measure of timestep adequacy, Fig.~\ref{fig:dt_metrics} presents three commonly used local timestep-resolution diagnostics, namely $\Delta t\,\omega_{pe}$, $\Delta t v_{te}/\Delta x$, and $\Delta t\,\omega_{ce}$, for Cases~D and D$_{2\Delta t}$.
The quantity $\Delta t\,\omega_{pe}$ reaches its largest values in the high-density region, as expected from the local increase of the electron plasma frequency.
For the baseline Case~D, $\Delta t\,\omega_{pe}$ remains below about $0.2$ throughout the domain.
By contrast, in Case~D$_{2\Delta t}$ it reaches values as high as $\sim 0.4$ in the dense core region, exceeding the commonly used guideline $\Delta t\,\omega_{pe}\lesssim 0.2$.
This indicates that the doubled timestep becomes noticeably less accurate for resolving the fastest plasma oscillations.
By comparison, the particle-transit indicator $\Delta t v_{te}/\Delta x$ remains below about $0.4$ even in Case~D$_{2\Delta t}$, while $\Delta t\,\omega_{ce}$ stays below $0.1$ in both cases and becomes appreciable only near the thruster end surfaces where the magnetic field is strongest.
Taken together, these diagnostics suggest that Case~D$_{2\Delta t}$ is not primarily limited by particle-flight or cyclotron-motion resolution, but rather by a coarser resolution of the electron plasma oscillation timescale.
This is likely the main reason why Case~D$_{2\Delta t}$ exhibits a systematic difference in the plume-region EDI wavelength relative to the baseline Case~D, even though the time-averaged near-wall transport topology remains robust.

\subsection{Grid-Resolution Sensitivity}
\label{sec:dx}

\begin{figure}
    \centering
    \includegraphics[width=\linewidth]{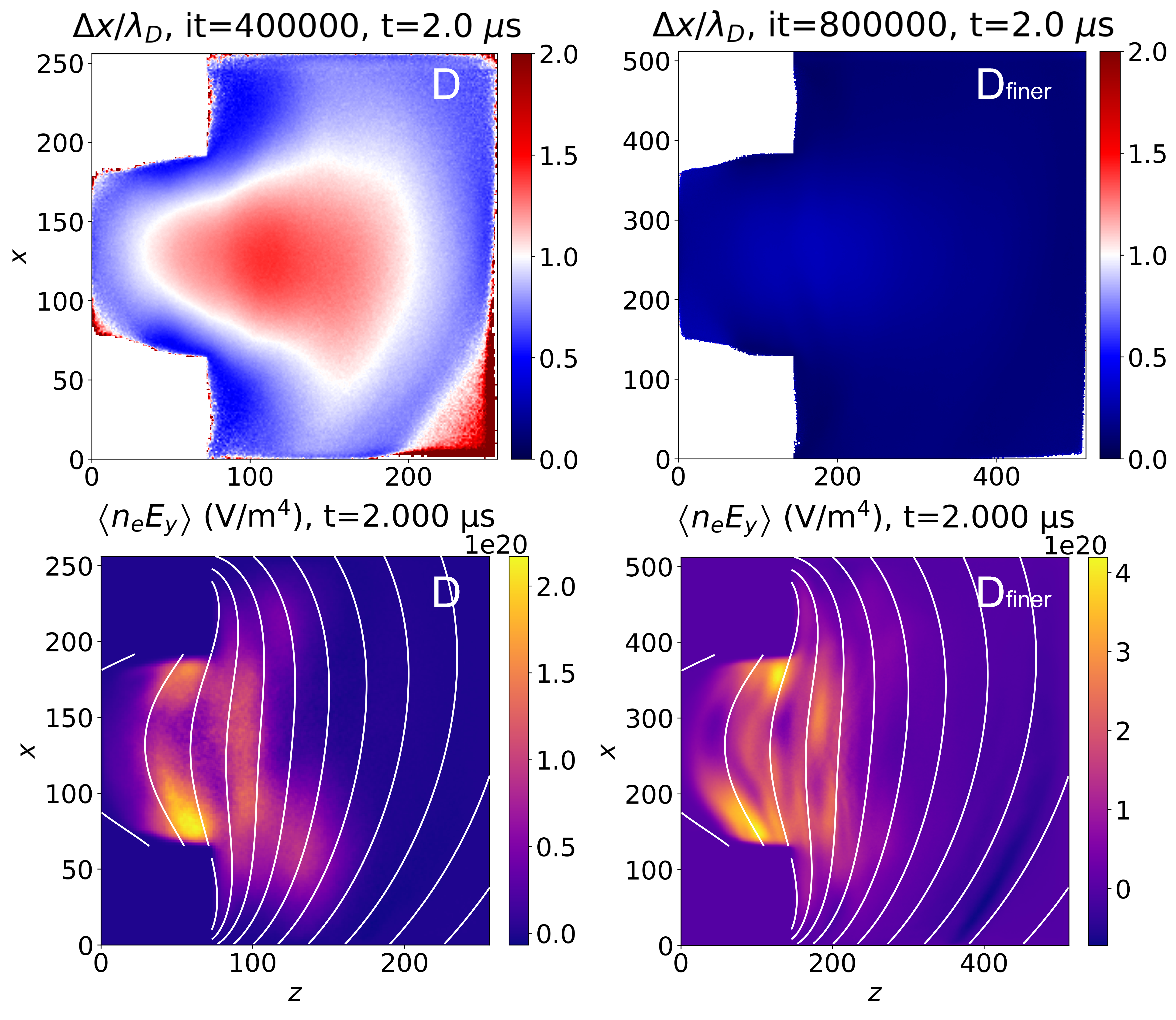}
    \caption{
    Comparison between Cases~D and $\rm D_{finer}$ at $t=2.0~\mu\mathrm{s}$.
    The upper row shows $\Delta x/\lambda_D$, and the lower row shows the corresponding transport-related structure.
    }
    \label{fig:Dfiner}
\end{figure}

To assess whether the baseline grid used in Case~D is sufficient to capture the near-wall pathways of anomalous electron transport, Fig.~\ref{fig:Dfiner} compares Cases~D and $D_{finer}$ at the same physical time, $t=2.0~\mu\mathrm{s}$. The left column corresponds to Case~D, and the right column to Case~$D_{finer}$. The upper row shows the local ratio between the grid spacing and the Debye length, $\Delta x/\lambda_D$, where $\lambda_D$ is evaluated from the local electron density and temperature. The lower row presents the corresponding two-dimensional transport-related structure used to diagnose the near-wall anomalous transport pathway. In the physically relevant dense-plasma region, the largest $\Delta x/\lambda_D$ values appear in the near-field plume core, around $x\approx120$--$150$ and $z\approx100$--$160$, with a peak of about $1.5$. Although larger values are found near the outer low-density boundaries, these occur outside the main transport-carrying region.
Thus, while Case~$D_{finer}$ satisfies the Debye-length resolution criterion more strictly, the baseline Case~D does not fully meet the most stringent Debye-length requirement everywhere, but it still appears to resolve the main near-wall transport region sufficiently to recover the same overall anomalous-transport topology.

More importantly, the transport-related structures obtained in Cases~D and $D_{finer}$ remain broadly consistent at the global level.
In both cases, the enhanced region is concentrated primarily near the downstream part of the channel and around the exit, exhibits clear near-wall intensification, and then extends into the near-plume region. Thus, both simulations recover the same overall picture: the anomalous electron transport is preferentially enhanced near the wall and forms a pathway that connects the near-wall channel region to the downstream plume.
This overall agreement suggests that the existence, location, and large-scale topology of the near-wall transport pathway are not strongly altered by the present level of grid refinement.
Although the coarser grid in Case~D smooths local features and cannot be regarded as fully Debye-resolved everywhere, it still captures the dominant structure, while Case~$D_{finer}$ mainly adds finer spatial detail.

The difference in absolute intensity between the two cases should not be interpreted simply as a grid-induced error. As shown by the time evolution of the particle number in Fig.~\ref{fig:ni_t_cases}, the two simulations are not perfectly synchronized in their slow global evolution. In particular, Case~$D_{finer}$ exhibits an evident delay relative to Case~D, with an offset of approximately $2~\mu\mathrm{s}$ in the particle-evolution history. As a result, even when compared at the same nominal physical time, the two cases do not correspond to exactly the same stage of the slowly evolving background discharge state. This phase lag naturally leads to a difference in amplitude.


Taken together, these results indicate that although Case~$D_{finer}$ provides stricter Debye-length resolution and reveals more local detail, the baseline Case~D still reproduces the same large-scale near-wall pathways of anomalous electron transport.
Given the substantially higher computational cost of Case~$D_{finer}$, which makes long simulations over several tens of microseconds difficult in practice, Case~D is therefore used as the baseline case for the systematic analysis in this work, while Case~$D_{finer}$ serves as a refinement check showing that the main transport topology is preserved and that the finer grid primarily contributes additional local detail.

\subsection{Plume-Domain Sensitivity}
\label{sec:domain size}

\begin{figure}
    \centering
    \includegraphics[width=\linewidth]{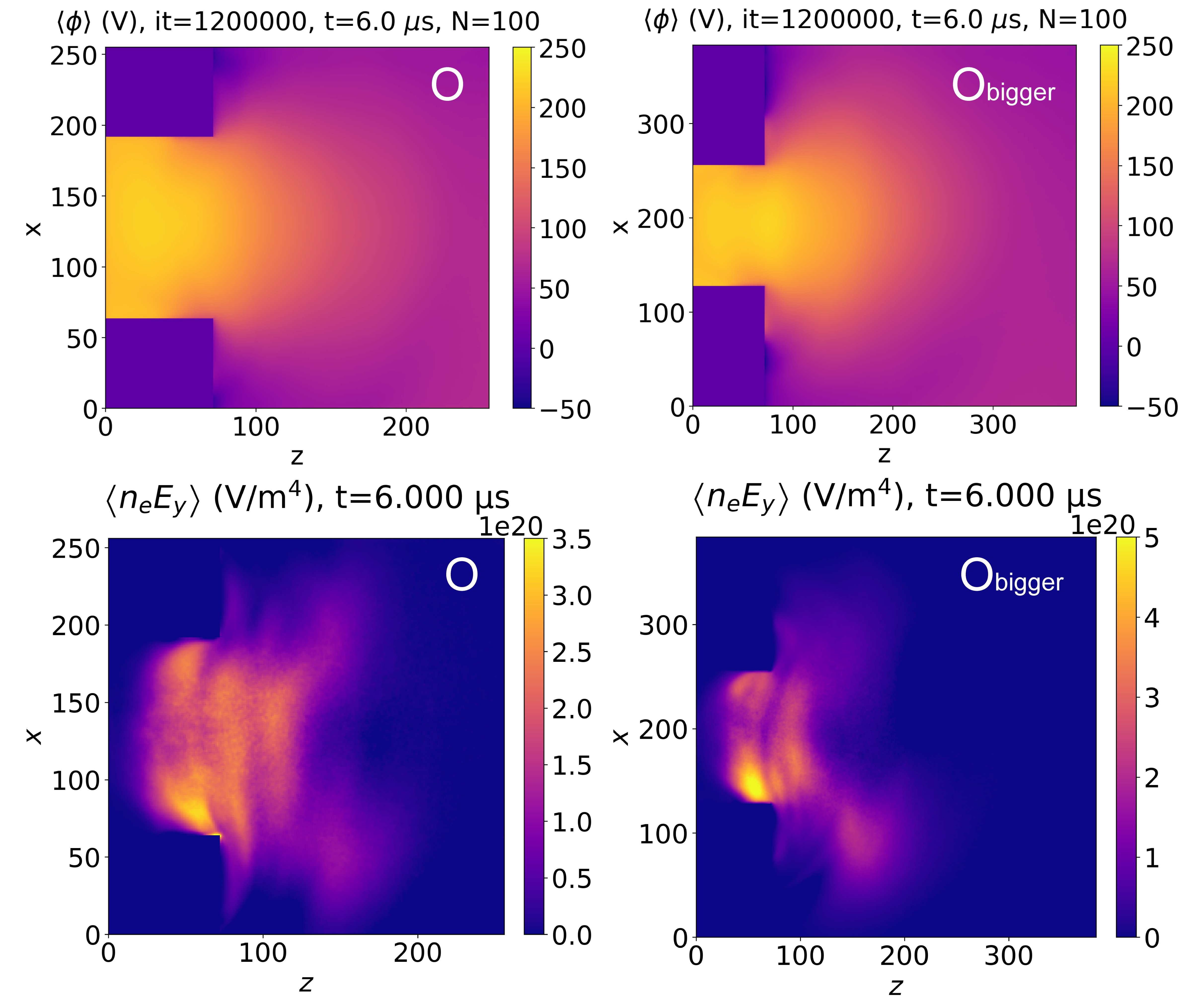}
\caption{
Sensitivity to plume-domain size.
Left: Case~O.
Right: Case~O$_{\mathrm{bigger}}$.
Top row: time-averaged potential $\langle \phi \rangle$.
Bottom row: time- and azimuthally-averaged correlation term $\langle n_e E_y\rangle$ with representative magnetic-field lines overlaid.
}
    \label{fig:Obigger}
\end{figure}

To further examine the sensitivity of the transport diagnostics to the
downstream plume extent, Fig.~\ref{fig:Obigger} compares Case~O with
Case~O$_{\mathrm{bigger}}$, in which the plume region is extended
substantially while keeping the same grid spacing and physical models.
This comparison is motivated by the role of the plume boundary in the
global electrostatic field solve: because the potential is determined
over the full computational domain, the downstream truncation can affect
how the potential relaxes in the plume and can thereby influence the
electric-field distribution near the exit. As already suggested by the
slow global evolution in Fig.~\ref{fig:ni_t_cases} and by the time-averaged field geometry
in Fig.~\ref{fig:Tripanel_tavg}, the near-wall anomalous transport pathway is organized around
the downstream half of the channel, the exit region, and the near plume.
The plume extent could therefore influence how this pathway couples to the
downstream plasma, even if the dominant transport structure itself is
formed closer to the channel and exit.

The top row of Fig.~\ref{fig:Obigger} shows that enlarging the plume
region mainly changes the downstream relaxation of the time-averaged
potential, while the overall potential structure inside the channel and
around the exit remains similar. In both cases, the dominant axial
potential drop is still localized near the downstream half of the
channel and the exit region, consistent with the averaged field geometry
previously shown in Fig.~\ref{fig:Tripanel_tavg}. The larger plume domain mainly allows a more
gradual potential relaxation in the downstream plume, without producing
a qualitative reorganization of the near-exit field geometry that
governs the transport pathway.

The bottom row of Fig.~\ref{fig:Obigger} shows that the transport maps
retain the same overall topology under plume-domain enlargement.
In both Case~O and Case~$O_{\mathrm{bigger}}$, the strongest
$\langle n_e E_y \rangle$ signal remains concentrated in band-like
regions adjacent to the inner and outer channel walls, with the dominant
signature located around the exit region and extending into the near
plume.
Thus, the near-exit transport structure is broadly the same in the two
cases: the correlation is strongest near the walls and then extends
toward the middle of the near-field plume.

At the same time, the enlarged plume domain provides a larger downstream
region over which the transport signature can develop and extend.
Accordingly, in Case~$O_{\mathrm{bigger}}$, the
$\langle n_e E_y \rangle$ structure continues farther into the plume,
whereas in Case~O the downstream part is more strongly truncated by the
limited computational extent.
This difference should therefore be interpreted mainly as a domain-size
effect on the available downstream development length, rather than as a
qualitative change in the transport pathway formed near the channel exit.

Taken together, Fig.~\ref{fig:Obigger}, together with the earlier
diagnostics in Figs.~\ref{fig:ni_t_cases}--\ref{fig:nE}, shows that the downstream plume extent
mainly affects how far the anomalous transport signature can persist and
extend into the plume, while the dominant transport topology near the
channel exit remains essentially unchanged.
The near-wall pathway therefore appears to be a robust feature of the
discharge structure, whereas the plume-domain size primarily controls
the degree to which its downstream continuation is retained or
artificially truncated in the simulation.

\section{Discussion and conclusions}
\label{sec:conclusion}

The present work was motivated by a longstanding question in Hall-thruster physics:
although instability-driven anomalous electron transport has been widely invoked to explain the observed cross-field conduction,
its net spatial pathway has remained much less clearly resolved than the instability itself.
By performing instability-resolving 3D PIC simulations with a highly integrated and boundary-aware model,
we have shown that the anomalous transport is not distributed uniformly across the channel cross section.
Instead, after time and azimuthal averaging of the fully three-dimensional oscillatory fields,
the net transport self-organizes into persistent near-wall pathways connected to the near-exit region.
In this sense, the main contribution of the present paper is not only to confirm the importance of EDI-driven transport,
but to reveal its spatial topology in a form that could not be obtained from
lower dimensional models.

A second major contribution of this work is methodological.
The simulations were carried out using what is, to our knowledge, one of the most physically complete 3D PIC frameworks yet applied to Hall-thruster EDI studies within a transport-resolving campaign.
The model combines a realistic magnetic-field configuration, electron-impact ionization treated by MCC,
a self-consistent continuum neutral-gas evolution model,
dielectric wall charging with SEE,
and an open near-plume outflow treatment, together with direct comparisons against simpler conducting-wall and truncated-boundary closures.
This combination is important because the wall response, the
near-plume truncation, and the neutral depletion dynamics are
not merely numerical details:
they strongly influence the slowly varying discharge structure and
the low-frequency evolution that set the background on which the
instability develops\cite{boeuf_garrigues_1998,choueiri_2001,fabris_2015}.
The present results nevertheless suggest that, while these effects
modify the quantitative distribution and downstream persistence of
the averaged transport signature, they do not qualitatively change
its dominant near-wall topology.

A related methodological point concerns the interpretation of convergence.
In a problem of this type, the slow global discharge envelope and the fast EDI dynamics evolve on clearly separated timescales.
The present results support the view that strict convergence of a slowly varying global quantity such as the spatially averaged ion density is not a prerequisite for meaningful instability diagnostics.
What is required instead is the existence of a local quasi-steady interval in which the high-frequency instability has become established and its spectral content and spatial organization vary only weakly over the averaging window.
This distinction is practically important, because it makes
transport-resolving analysis feasible even in simulations for
which the full low-frequency envelope is too expensive to follow
to completion, or may not become perfectly stationary at all due
to the persistent breathing-mode oscillation.
For 3D Hall-thruster PIC studies, this separation between global convergence and local instability maturity should therefore be regarded as a central element of the analysis strategy.

On the physics side, the most important conclusion is the robustness of the near-wall transport topology.
The transport maps extracted from the time- and azimuthally-averaged correlation term $\langle n_e E_y \rangle$
show that the dominant anomalous electron transport is concentrated in two wall-adjacent bands near the downstream channel and exit region, rather than in the channel core.
This result persists across substantially different boundary treatments.
When the conducting-wall baseline is replaced by dielectric ceramic walls with self-consistent charging and SEE,
and when an open-outflow treatment is further introduced in the plume,
the detailed strength and downstream extension of the transport pathway are modified,
but the pathway itself is not removed or replaced by a fundamentally different organization.
The simulations therefore indicate that near-wall localization is a robust property of the instability-driven transport, rather than an artifact of one specific electrostatic closure.

The numerical-sensitivity study adds an additional and, in our view, highly practical layer to these conclusions.
The results suggest that different simulation outputs have different levels of numerical robustness.
Quantities such as the detailed EDI wavelength content, fine-scale instantaneous morphology, and fastest plasma response are relatively sensitive to timestep and grid resolution.
By contrast, the large-scale topology of the time- and azimuthally-averaged transport pathway appears substantially more robust.
For example, doubling the timestep biases the resolved EDI spectrum toward shorter azimuthal wavelengths and alters the detailed instantaneous field structure,
yet the averaged $\langle n_e E_y \rangle$ map still recovers the same dominant near-wall pathway.
Likewise, the baseline grid does not satisfy the most stringent Debye-resolution criterion everywhere,
but it still reproduces the same large-scale near-wall transport topology as the refined case, with the finer grid contributing mainly additional local detail.
Enlarging the plume domain mainly affects the downstream relaxation of the potential and the retained downstream extension of the transport signature,
while leaving the dominant near-exit pathway essentially unchanged.

This hierarchy of robustness leads to an important practical recommendation.
When the cost of a fully resolved 3D PIC campaign exceeds the available budget or acceptable turnaround time,
simulations with moderately relaxed $\Delta t$ and $\Delta x$ should not be regarded as automatically useless or completely unphysical.
Rather, they should be interpreted as reduced-fidelity calculations:
they may distort wave details, spectral peaks, and some local amplitudes,
but they can still preserve the main topology of the averaged transport organization.
In particular, the present results suggest that the wall-localized $\langle n_e E_y \rangle$ pathway may remain a meaningful diagnostic even when the most conservative PIC resolution requirements are not satisfied everywhere.
This makes such calculations valuable for rapid parameter scans, trend identification, and early-stage design exploration.
A natural workflow for future studies is therefore hierarchical:
first perform lower-cost exploratory simulations to map broad tendencies in parameter space;
then apply more expensive, fully refined simulations only to a smaller set of selected cases for which detailed wave physics and quantitative convergence are essential.
In our view, this is not merely a compromise forced by limited resources,
but a practically important strategy for making large-scale 3D PIC studies scientifically productive.

The present study also suggests several natural directions for future investigation.
On the modeling side, it will be important to examine more systematically how the transport pathway depends on the details of the SEE model,
the emitted-electron energy distribution,
the magnetic-field topology,
the cathode position and injection characteristics,
the discharge voltage,
the propellant mass flow rate,
etc.
It will also be valuable to extend the present analysis toward broader operating-condition scans and toward closer connections with experimentally inferred transport and wave diagnostics.
At the same time, the strongest future direction, in our opinion, is not simply to add more numerical sophistication in isolation,
but to build a Hall-thruster prototype specifically intended for one-to-one comparison with transport-resolving 3D simulations.

This point deserves emphasis.
Historically, experiments have often been regarded as the more expensive part of the workflow, with simulations mainly expected to adapt themselves to existing experimental conditions.
For state-of-the-art 3D PIC Hall-thruster calculations, however, that assumption is becoming less valid.
Using the computational cost summarized in Tab.~I, the total campaign represented in the present work amounts to roughly $1.12\times10^{6}$ core-hours.
When multiplied by a realistic charge per core-hour, this already places the computational cost in the same order of magnitude as a targeted laboratory experiment, and in some cases potentially beyond it.
This changes the philosophy of model validation.
Rather than asking only that simulations be adjusted to whatever experiment is already available,
it becomes increasingly reasonable to design experiments and simulations together from the outset,
so that each constrains the other.
Indeed, for certain questions, it may be advantageous to design the experimental configuration in a way that is intentionally closer to the assumptions and diagnostics of the simulation,
thereby making the comparison sharper, cleaner, and more informative for both sides.

In summary, the present work establishes that instability-driven anomalous electron transport in a Hall thruster can organize into robust near-wall pathways that are directly revealed only after resolving the fully three-dimensional fluctuating dynamics.
At the same time, the study provides a practical set of lessons for conducting large-scale 3D PIC Hall-thruster simulations, including the importance of physically informed initialization,
hybrid neutral modeling,
careful interpretation of local quasi-steady instability states,
and a hierarchical view of numerical fidelity.
We therefore hope that this paper can serve both as a physics result and as a methodological reference.
More broadly, we expect that future progress in this area will depend not only on more complete 3D kinetic models,
but also on a tighter co-development of simulation and experiment aimed specifically at resolving the spatial structure of anomalous transport in Hall thrusters.

\section*{Acknowledgment}

The authors acknowledge the support from National Natural Science Foundation of China (Grant No. 52472403).

\section*{Conflict of interest}
The authors have no conflicts to disclose.

\section*{Data Availability}

The data that support the findings of this study are available from
the corresponding author upon reasonable request.

\appendix

\section{Additional Implementation Details for Boundary Treatments}
\label{apx:boundary_treatments}

\subsection{Discrete implementation of the dielectric-wall boundary condition}
\label{apx:dielectric_bc_discretization}

The electrostatic potential is solved on a cell-centered mesh using a seven-point stencil\cite{coco2013finite}.
For an interior cell $(i,j,k)$,
\begin{equation}
6\phi_{i,j,k} - \sum_{\ell\in\mathcal{N}(i,j,k)} \phi_{\ell}
\;=\;
h^2\frac{\rho_{i,j,k}}{\epsilon_0},
\label{eq:poisson_interior_apx}
\end{equation}
where $\mathcal{N}(i,j,k)$ is the set of the six nearest neighbors and $h$ is the uniform grid spacing.

At a planar boundary, the wall lies midway between an interior cell center and an outside ghost cell\cite{hara2017radial}.
For a face perpendicular to the $x$ direction, the boundary field is discretized using a one-sided difference across the two cell centers.
At the $x$-minimum boundary, the prescribed boundary field $E_{bc}$ gives
\begin{equation}
E_{bc} \;=\; -\,\frac{\phi_{i,j,k}-\phi_{i-1,j,k}}{h},
\qquad
\phi_{i-1,j,k} \;=\; \phi_{i,j,k} + hE_{bc},
\end{equation}
so eliminating the ghost value in Eq.~\eqref{eq:poisson_interior_apx} yields
\begin{equation}
\begin{aligned}
5\phi_{i,j,k}
&-\Big(
\phi_{i+1,j,k}
+\phi_{i,j-1,k}
+\phi_{i,j+1,k}
+\phi_{i,j,k-1}
+\phi_{i,j,k+1}
\Big) \\
&=
h^2\frac{\rho_{i,j,k}}{\epsilon_0}
+
hE_{bc}.
\end{aligned}
\label{eq:neumann_xmin_stencil_apx}
\end{equation}

At the $x$-maximum boundary,
\begin{equation}
E_{bc} \;=\; -\,\frac{\phi_{i+1,j,k}-\phi_{i,j,k}}{h},
\qquad
\phi_{i+1,j,k} \;=\; \phi_{i,j,k} - hE_{bc},
\end{equation}
which gives
\begin{equation}
\begin{aligned}
5\phi_{i,j,k}
&-\Big(
\phi_{i-1,j,k}
+\phi_{i,j-1,k}
+\phi_{i,j+1,k}
+\phi_{i,j,k-1}
+\phi_{i,j,k+1}
\Big) \\
&=
h^2\frac{\rho_{i,j,k}}{\epsilon_0}
-
hE_{bc}.
\end{aligned}
\label{eq:neumann_xmax_stencil_apx}
\end{equation}

The same construction applies on the $y$ and $z$ faces using the corresponding field component and index direction.
Thus, for a node adjacent to a dielectric boundary, the outside unknown is eliminated,
the diagonal coefficient is reduced from $6$ to $5$,
and the right-hand side receives the appropriate $\pm hE_{bc}$ contribution.
At edges and corners, contributions from multiple eliminated ghost neighbors are accumulated.

\subsection{Stochastic sampling of SEE and emitted-electron initialization}
\label{apx:see_sampling}

The continuous SEE yield $\sigma$ is converted into an integer number of emitted secondary macroparticles while preserving the correct mean emitted charge\cite{Taccogna2022CouplingPICChemistry}.
For $\sigma\le 1$, the emission is treated as a Bernoulli event,
\begin{equation}
N=
\begin{cases}
1, & r<\sigma,\\
0, & r\ge\sigma,
\end{cases}
\label{eq:see_bernoulli_apx}
\end{equation}
where $r\sim\mathcal{U}(0,1)$.
For $\sigma>1$, stochastic rounding is applied around $n=\lfloor \sigma \rfloor$ such that $N=n+1$ with probability $\sigma-n$ and $N=n$ otherwise.
This preserves $\mathbb{E}[N]=\sigma$ without introducing fractional macroparticles.

Emitted electrons are launched into the plasma half-space using a low-energy distribution representative of true secondaries.
Let $\hat{\mathbf{n}}$ denote the inward unit normal pointing from the wall into the plasma.
The normal component is sampled from the Maxwellian flux distribution,
\begin{equation}
v_n = \sqrt{-\frac{2{\mathrm e}T_{\mathrm{SEE}}}{m_e}\ln(1-r_1)},
\qquad
\mathbf{v}_n=v_n\hat{\mathbf{n}},
\label{eq:see_vn_apx}
\end{equation}
while the two tangential components are sampled from a Maxwellian at the same temperature,
\begin{equation}
\begin{aligned}
v_{t,1}
&=
\sqrt{\frac{{\mathrm e}T_{\mathrm{SEE}}}{m_e}}
\sqrt{-2\ln(1-r_2)}
\cos(2\pi r_3),\\
v_{t,2}
&=
\sqrt{\frac{{\mathrm e}T_{\mathrm{SEE}}}{m_e}}
\sqrt{-2\ln(1-r_2)}
\sin(2\pi r_3),
\end{aligned}
\label{eq:see_vt_apx}
\end{equation}
so that
\begin{equation}
\mathbf{v}_{\mathrm{emit}}=\mathbf{v}_n+\mathbf{v}_t .
\end{equation}

Emitted particles are initialized at the impact location and then advanced for the remaining fraction of the time step after impact.
This ensures that they are placed on the plasma side of the interface and avoids immediate artificial reintersection with the wall.

\subsection{Discrete implementation of the open Robin outflow boundary}
\label{apx:open_bc_discretization}

For a cell-centered discretization on a uniform grid with spacing $h$,
consider a cell adjacent to an open boundary with outward normal aligned with a coordinate direction\cite{Arias2018RobinPoisson}.
Let $\phi_P$ denote the potential at the interior cell center and $\phi_G$ the ghost value at the outside cell center.
Approximating the normal derivative and boundary value by
\begin{equation}
\left.\frac{\partial \phi}{\partial n}\right|_{bc} \approx \frac{\phi_G-\phi_P}{h},
\qquad
\phi_b \approx \frac{\phi_P+\phi_G}{2},
\label{eq:open_discrete_defs_apx}
\end{equation}
the continuous Robin condition becomes
\begin{equation}
\frac{\phi_G-\phi_P}{h}
+\kappa_b\left(\frac{\phi_P+\phi_G}{2}-\phi_\infty\right)=0.
\label{eq:open_robin_discrete_apx}
\end{equation}
Solving for the ghost value gives
\begin{equation}
\phi_G
=
\frac{(2-\kappa_b h)\phi_P + 2\kappa_b h\,\phi_\infty}{2+\kappa_b h}.
\label{eq:open_ghost_apx}
\end{equation}

Substituting Eq.~\eqref{eq:open_ghost_apx} into the seven-point Poisson stencil eliminates the ghost unknown and modifies the discrete linear system locally.
In particular, the diagonal coefficient receives an additional contribution
\begin{equation}
\Delta a_P=\frac{\kappa_b h-2}{\kappa_b h+2},
\label{eq:open_diag_update_apx}
\end{equation}
and the right-hand side acquires
\begin{equation}
\Delta b_P=\frac{2\kappa_b h}{\kappa_b h+2}\,\phi_\infty.
\label{eq:open_rhs_update_apx}
\end{equation}

The geometric factor $\kappa_b$ is computed locally from the boundary-face position.
For example, on the $z$-maximum face,
$\mathbf{x}_b=(x_i,y_j,z_{\max})$ and $\hat{\mathbf{n}}_b=(0,0,1)$,
so $\kappa_b=(\hat{\mathbf{n}}_b\cdot\mathbf{r}_b)/(\mathbf{r}_b\cdot\mathbf{r}_b)$ is evaluated for each boundary-adjacent cell and then inserted into Eqs.~\eqref{eq:open_diag_update_apx} and \eqref{eq:open_rhs_update_apx}.
This construction yields an open field boundary compatible with the cell-centered Poisson solver while reducing the truncation sensitivity associated with a clamped Dirichlet outflow.

\section{Wavelength Evolution}
\label{apx:Ey-k}

The simulation domain is periodic in $y$ with $N_y$ cells and cell size $\Delta y$, so that the periodic length is $L_y=N_y\Delta y$\cite{OppenheimSchafer2009DTSP}.
At each output time $t$, we assemble the full 3D field and restrict the analysis to a region of interest in $x$ (here, a slab centered at mid-radius with thickness 100 cells) while retaining the full $y$--$z$ extent.

\paragraph{(i) One-sided Fourier transform in the periodic direction.}
For each $(x,z)$ we compute the one-sided discrete Fourier transform along $y$,
\begin{equation}
\begin{aligned}
\hat{E}_y(x,m,z,t)
&=\sum_{j=0}^{N_y-1}E_y(x,y_j,z,t)\,
\exp\!\left(-i\frac{2\pi m j}{N_y}\right),\\
&m=0,\ldots,\left\lfloor\frac{N_y}{2}\right\rfloor .
\end{aligned}
\end{equation}
where $y_j=j\Delta y$.
The corresponding physical azimuthal wavenumber and wavelength are
\begin{equation}
k_y(m)=\frac{2\pi m}{L_y},
\qquad
\lambda_y(m)=\frac{L_y}{m}\quad (m\ge 1).
\end{equation}

\paragraph{(ii) Conversion to spectral amplitude and averaging over $(x,z)$.}
Because the FFT is one-sided for a real-valued signal, we convert the unnormalized coefficients to a one-sided amplitude (in V/m) via
\begin{equation}
A(x,m,z,t)=
\begin{cases}
\displaystyle \frac{1}{N_y}\,|\hat{E}_y(x,0,z,t)|, & m=0,\\[6pt]
\displaystyle \frac{2}{N_y}\,|\hat{E}_y(x,m,z,t)|, & 1\le m \le \frac{N_y}{2}-1,\\[6pt]
\displaystyle \frac{1}{N_y}\,|\hat{E}_y(x,N_y/2,z,t)|, & m=N_y/2~(N_y~\text{even}),
\end{cases}
\end{equation}
and then form an averaged amplitude spectrum (RMS over $x,z$)
\begin{equation}
\bar{A}(m,t)=\left\langle A^2(x,m,z,t)\right\rangle_{x,z}^{1/2}.
\end{equation}
In practice we exclude $m=0$ (infinite wavelength) and optionally the Nyquist mode $m=N_y/2$ to avoid numerical artifacts at the one-sided boundary.

\paragraph{(iii) Time--wavelength spectrogram.}
Finally, we visualize the evolution of azimuthal structure using a time--wavelength map
\begin{equation}
S(\lambda_y,t)=\log_{10}\!\left[\bar{A}\!\left(m(\lambda_y),t\right)\right],
\end{equation}
where $m(\lambda_y)=L_y/\lambda_y$.

\section{Extraction of Wave Crest Surfaces}
\label{apx:wave_crest}

\paragraph{Wall and sheath masking.}
The wall/solid region is identified once, from the first reconstructed full-domain snapshot, as the set of grid points where the stored diagnostic field is exactly zero,
\begin{equation}
M_{\rm wall}(x,y,z) \equiv \mathbb{I}\left[\phi(x,y,z,t_0)=0\right],
\end{equation}
where $\mathbb{I}[\cdot]$ is the indicator function.

To further exclude the near-wall sheath region, a wall-boundary mask is constructed on the non-periodic faces, and the Euclidean distance-to-wall field $d(x,y,z)$ is then computed using a distance transform applied to the complement of the wall mask.
The sheath exclusion mask is defined as
\begin{equation}
M_{\rm sheath}(x,y,z) \equiv \mathbb{I}\!\left[d(x,y,z)\le n_{\rm sheath}\right]\wedge \neg M_{\rm wall},
\end{equation}
and the total exclusion mask is given by
\begin{equation}
M_{\rm ign}\equiv M_{\rm wall}\vee M_{\rm sheath}.
\end{equation}
In this work, $n_{\rm sheath} = 15$ grid cells.
Since the $y$ direction is periodic, no boundary masking is applied on the corresponding boundary faces.

\paragraph{Denoising and background removal.}
To isolate wave-like fluctuations from the large-scale potential variation (e.g.\ the sheath-related and quasi-static gradients), a band-limited fluctuation field is constructed by subtracting a smoothed background:
\begin{align}
\phi_s &= G_{\sigma_s}\ast \phi, \\
\phi_{\rm bg} &= G_{\sigma_b}\ast \phi_s, \\
\phi' &= \phi_s - \phi_{\rm bg},
\end{align}
where $G_\sigma$ is a 3D Gaussian kernel, $\ast$ denotes convolution, and the kernel widths satisfy $\sigma_b > \sigma_s$; here, $(\sigma_s,\sigma_b) = (2,5)$ grid cells.
For subsequent spectral processing, $\phi'(x,y,z)$ is set to zero within the excluded region, i.e.\ in $M_{\rm ign}$, to avoid contamination from masked values.

\paragraph{3D monogenic signal and local phase.}
Since the dominant wave does not propagate along a fixed direction and the wavefronts are generally curved, the local phase should be defined without reference to any prescribed orientation. 
The 3D monogenic signal is therefore employed, which extends the 1D analytic signal to three-dimensional fields through the Riesz transform \cite{felsberg2001monogenic}.

Given the detrended scalar field $\phi'(\mathbf{x})$, let $\widehat{\phi'}(\mathbf{k})$ denote its Fourier transform, where $|\mathbf{k}|=\sqrt{k_x^2+k_y^2+k_z^2}$.
The corresponding quadrature components are then obtained via the 3D Riesz transform:
\begin{equation}
\widehat{R_i}(\mathbf{k})
= -\,i\,\frac{k_i}{|\mathbf{k}|}\,\widehat{\phi'}(\mathbf{k}),
\quad i\in\{x,y,z\},
\label{eq:riesz}
\end{equation}
with the multiplier set to zero at $\mathbf{k}=\mathbf{0}$. 
The factor $-i$ introduces the quadrature phase shift, whereas $k_i/|\mathbf{k}|$ makes the construction rotation-covariant.

Transforming back to physical space yields the Riesz vector field $\mathbf{R}(\mathbf{x})=(R_x,R_y,R_z)$, whose magnitude is
\begin{equation}
q(\mathbf{x})=\|\mathbf{R}(\mathbf{x})\|=\sqrt{R_x^2+R_y^2+R_z^2}.
\end{equation}
The local monogenic amplitude and phase are then defined as
\begin{equation}
A(\mathbf{x})=\sqrt{\phi'(\mathbf{x})^2+q(\mathbf{x})^2},
\quad
\theta(\mathbf{x})=\arctan2\!\big(q(\mathbf{x}),\phi'(\mathbf{x})\big),
\label{eq:mono_phase}
\end{equation}
so that $\theta\in[0,\pi]$ since $q(\mathbf{x})\ge 0$ by construction.  
Constant $\theta$ surfaces therefore define local equal-phase surfaces independent of the instantaneous propagation direction. 

Under the usual locally narrowband, approximately plane-wave assumption, $\phi'(\mathbf{x})\approx A(\mathbf{x})\cos\theta(\mathbf{x})$ and $q(\mathbf{x})\approx A(\mathbf{x})|\sin\theta(\mathbf{x})|$, so that $(\phi',q)$ forms an orientation-independent local phasor pair. 
Additionally, the normalized Riesz vector $\mathbf{R}/(q+\varepsilon)$ can be interpreted as an estimate of the local wavefront normal, although it is not used explicitly here.

\paragraph{Amplitude gating and crest/trough isosurfaces.}
To suppress weak-signal regions and residual noise, an amplitude gate is applied using a high quantile threshold,
\begin{equation}
A_{\rm thr} = Q_{q}(A),\quad q=0.9,
\end{equation}
where $Q_q(\cdot)$ denotes the $q$-quantile evaluated over the computational domain.
Locations with $A<A_{\rm thr}$ are excluded by treating $\theta$ as invalid there. In addition, $\theta$ is also marked invalid in $M_{\rm ign}$.
Additionally, for visualization we set $\theta$ invalid on $M_{\rm ign}$ (wall and sheath).
The crest and trough surfaces are then identified as the isosurfaces
\begin{align}
\mathcal{S}_{\rm crest}  &= \{\mathbf{x}:\ \theta(\mathbf{x})=\delta\}, \\
\mathcal{S}_{\rm trough} &= \{\mathbf{x}:\ \theta(\mathbf{x})=\pi-\delta\},
\end{align}
where $\delta=0.25~\mathrm{rad}$ is introduced to avoid numerical sensitivity near the endpoints $0$ and $\pi$.
The resulting isosurfaces are extracted from the structured grid using a marching-cubes algorithm \cite{lorensen1987marching} and exported as VTK PolyData (\texttt{.vtp}) for interactive and time-resolved visualization.

\paragraph{Connectivity filtering and region selection.}
When visualized in ParaView \cite{ahrens2005paraview}, the extracted crest/trough isosurfaces typically appear as multiple fragmented surface patches.
The Connectivity filter assigns a \texttt{RegionId} to each connected component, thereby distinguishing independent regions. The Threshold filter can then be applied to retain selected \texttt{RegionId} values, allowing the largest connected wavefront patch to be isolated.

\nocite{*}
\bibliography{reference}

\end{document}